\documentclass[11pt]{article}
\pdfoutput=1

\usepackage{fullpage,amsfonts,amsmath,amssymb,amsthm,bm,xspace,thmtools,thm-restate,asymptote,mathrsfs,appendix,array, comment}
\usepackage{multirow,graphicx, algorithm,algorithmic,caption,subcaption,url,cite,float}
\usepackage{mathtools}
\usepackage{hyperref}
\hypersetup{
	pdfpagelayout=TwoPageLeft,
	pdfstartview=Fit,
	bookmarks=true,
	bookmarksopen=true,
	bookmarksnumbered=true,
	breaklinks=true,
	colorlinks=true,
	linkcolor=blue,
	anchorcolor=black,
	citecolor=blue,
	filecolor=black,
	menucolor=red,
	urlcolor=blue
}

\usepackage{todonotes}
\usepackage{array} 
\usepackage[noabbrev]{cleveref}
\usepackage{listings}
\usepackage{xpatch}
\usepackage{thmtools}
\usepackage{apptools}

\graphicspath{{NumericalExperiment_simulated/figs_asy/}{NumericalExperiment_simulated/other/}}

\usepackage{changepage}

\makeatletter
\xpatchcmd{\thmt@restatable}{\csname #2\@xa\endcsname\ifx\@nx#1\@nx\else[{#1}]\fi}%
{\IfAppendix{\csname #2\@xa\endcsname}{\csname #2\@xa\endcsname[{#1}]}}
\makeatother

\newcolumntype{C}{ >{\centering\arraybackslash} m{7cm} }
\newcolumntype{D}{ >{\centering\arraybackslash} m{2cm} }
\newcolumntype{E}{ >{\centering\arraybackslash} m{1.5cm} }
\newcolumntype{F}{ >{\centering\arraybackslash} m{8.3cm} }
\newcolumntype{G}{ >{\centering\arraybackslash} m{8cm} }

\newtheorem{theorem}{Theorem}[section]

\newtheorem{lemma}[theorem]{Lemma}

\theoremstyle{remark}

\numberwithin{equation}{section}

\newcommand{\wt}[1]{\widetilde{ #1 } }

\newcommand{\C}{\mathbb{C}}

\newcommand{\R}{\mathbb{R}}

\newcommand{\MAT}[1]{\begin{bmatrix} #1 \end{bmatrix}}

\newcommand{\abs}[1]{\left| #1 \right|}

\newcommand{\sqbr}[1]{\left[ #1 \right]}
\newcommand{\brac}[1]{\left( #1 \right) }
\newcommand{\ml}[1]{\mathcal{ #1 } }
\newcommand{\op}[1]{ \operatorname{#1} }

\newcommand{\normF}[1]{\left\| #1 \right\| _{F}}

\newcommand{\PROD}[2]{\left \langle #1, #2\right \rangle}

\usepackage{booktabs}
\newcommand{\scatterAllPlotScale}{0.3}

\newcommand{\scatterAllGTSize}{2pt}
\newcommand{\subScatterPlotScale}{1}
\newcommand{\scatterTitleFont}{\normalsize}
\newcommand{\scatterTitleYshift}{-5}

\newcommand{\scatterToneMin}{.05}
\newcommand{\scatterToneMax}{3}
\newcommand{\scatterTtwoMin}{.004}
\newcommand{\scatterTtwoMax}{1}
\newcommand{\scatterTickFont}{\normalsize}
\newcommand{\scatterTickShift}{1}
\newcommand{\scatterXTickRotate}{45}
\newcommand{\scatterYTickRotate}{45}
\newcommand{\scatterYLabelShift}{-10}
\newcommand{\scatterXLabelShift}{-5}
\newcommand{\scatterMajorTickLen}{2}

\newcommand{\scatterLabelFont}{\normalsize}

\newcommand{\heatSubPlotWidth}{4}
\newcommand{\heatSubPlotWidthSmall}{3}

\usepackage{pgfplots}
\pgfplotsset{compat=1.14,
	colormap={bwr}{
		rgb255=(0,0,255)
		rgb255=(255,255,255)
		rgb255=(255,0,0)}
}
\usetikzlibrary{spy}
\usepgfplotslibrary{external}
\usepgfplotslibrary{fillbetween}
\usepgfplotslibrary{groupplots}
\usetikzlibrary{arrows,automata}
\usepackage{blkarray}
\usepackage[hang,flushmargin,multiple]{footmisc}

\tikzsetexternalprefix{figs_tikz/}
\usepackage{authblk}

\title{Multicompartment Magnetic Resonance Fingerprinting}

\author[1]{Sunli Tang\thanks{Sunli Tang and Carlos Fernandez-Granda contributed equally to this paper.}}
\author[1,2]{Carlos Fernandez-Granda\textsuperscript{*}}
\author[2,3]{Sylvain Lannuzel}
\author[1]{Brett Bernstein}
\author[4,5]{Riccardo Lattanzi}
\author[4,5]{Martijn Cloos}
\author[4,5]{Florian Knoll}
\author[4,5]{Jakob Assl\"ander}

\affil[1]{\small Courant Institute of Mathematical Sciences, New York University}
\affil[2]{\small Center for Data Science, New York University}
\affil[3]{\small \'Ecole CentraleSup\'elec}
\affil[4]{\small Center for Biomedical Imaging, Department of Radiology, New York University School of Medicine}
\affil[5]{\small Center for Advanced Imaging and Innovation Research (CAI$^2$R), Department of Radiology, New York University School of Medicine}

\date{February 2018}

\begin{document}

\maketitle

\vspace{-0.3in}

\begin{abstract}
\noindent
Magnetic resonance fingerprinting (MRF) is a technique for quantitative estimation of spin-relaxation parameters from magnetic-resonance data. Most current MRF approaches assume that only one tissue is present in each voxel, which neglects the tissue's microstructure, and may lead to artifacts in the recovered parameter maps at boundaries between tissues. In this work, we propose a multicompartment MRF model that accounts for the presence of multiple tissues per voxel. % Fitting the model requires solving a sparse linear inverse problem at each voxel, in order to express the magnetization signal as a linear combination of a few fingerprints in the precomputed dictionary. 
The model is fit to the data by iteratively solving a sparse linear inverse problem at each voxel, in order to express the magnetization signal as a linear combination of a few fingerprints in the precomputed dictionary. Thresholding-based methods commonly used for sparse recovery and compressed sensing do not perform well in this setting due to the high local coherence of the dictionary. Instead, we solve this challenging sparse-recovery problem by applying reweighted-$\ell_1$-norm regularization, implemented using an efficient interior-point method. The proposed approach is validated with simulated data at different noise levels and undersampling factors, as well as with a controlled phantom imaging experiment on a clinical magnetic-resonance system. 
\end{abstract}


\newpage
\section{Introduction}

\subsection{Quantitative magnetic resonance imaging}
\label{sec:qmri}
Magnetic resonance imaging (MRI) is a medical-imaging technique that measures the response of the atomic nuclei in biological tissues to high-frequency radio waves when placed in a strong magnetic field~\cite{zhi2000principles}. Due to its superior soft-tissue contrast, MRI has become a key technology for non-invasive diagnostics. %MRI sequences induce complex physical processes on the spins of hydrogen atoms in the human body and capture the resulting signal averaged over all spins within a voxel. 
In contrast to other medical imaging modalities such as computed tomography or positron emission tomography, MR images are usually qualitative in nature. They consist of gray values that capture relative signal intensity changes between certain tissues, which are then interpreted by radiologists to detect pathologies. %Traditionally, the output of a typical clinical MRI exam is a set of image volumes with specific pre-defined image contrasts that depend on a particular clinical question. 

Due to their qualitative nature, MR images are subject to variations between different MRI systems and particular data-acquisition settings, hampering reproducibility. This makes it very challenging to use data from current clinical MRI systems for longitudinal studies, early detection and progress-tracking of disease, as well as computer-aided diagnosis. The goal of quantitative MRI is to measure physical parameters \emph{quantitatively}, in a way that is reproducible across different MRI systems ~\cite{giedd1996quantitative,suddath1989temporal,tofts2005quantitative,Deoni2010}. Some of these quantities, such as spin-relaxation times, have great potential as quantitative biomarkers for various pathologies~\cite{reich2007multiparametric,matzat2013quantitative,neema2007t1} and allow to synthesize images with standardized image contrasts~\cite{noth2015improved,Deshmane2016}. Unfortunately, data-acquisition protocols for quantitative MRI often lead to long measurement times that are challenging to integrate in the clinical work-flow and as a consequence are not widely deployed.

\subsection{Magnetic resonance fingerprinting}
\label{sec:mrf}
Magnetic resonance fingerprinting (MRF) \cite{Ma2013} is a recent approach to quantitative MRI designed to measure tissue-relaxation parameters within clinically-feasible scan times. MRF departs from traditional MRI acquisition by deliberately avoiding a steady state of the magnetization. For a fixed value of the parameters of interest, the signal evolution of the magnetization at a given voxel (which results from the train of radiofrequency pulses and gradient waveforms used to acquire the data) can be simulated by solving a system of differential equations, called the Bloch Equations~\cite{bloch1946nuclear}. This makes it possible to precompute a dictionary of possible signals or \emph{fingerprints} associated with different values of the relaxation parameters. An estimate of the actual values consistent with the data can then be obtained by finding the fingerprint that best matches the observed measurements. This approach has been shown to be robust to technical imperfections of the scanner hardware, such as inhomogeneities of the magnetic fields used to generate the MR signal~\cite{Cloos2016}.

\pgfplotsset{compat = 1.3}
\begin{figure}[tp]
\begin{tikzpicture}
\begin{axis}[
	width=0.5\textwidth, 
	height=0.4\textwidth, 
	xmin=0, 
	xmax=850,
	ymin=-0.29,
	ymax=0.15,
	legend pos=south east, 
	legend columns = 2,
	xlabel = {time (s)}, 
	ylabel = {signal (a.u.)}, 
	y dir=reverse, 
	scaled y ticks={real:-1}, 
	ytick scale label code/.code={}, 
	scaled x ticks={real:222.222}, 
	xtick scale label code/.code={}, 
	xtick distance=222.222,
	name=signal]
\addplot[thick,cyan] file {dat_files/single_comp/sig1_single_comp.dat};
\addplot[thick,orange] file {dat_files/single_comp/sig2_single_comp.dat};
\addplot[thick,blue] file {dat_files/single_comp/sig_res_single_comp.dat};
\addplot[thick, red] file {dat_files/single_comp/sig_mix.dat};
\legend{I/E water, Myelin water, SC model, Data}
\end{axis}

\begin{axis}[
	at=(signal.right of north east),
	anchor=left of north west,
	xshift=0.2cm,
	height=0.4\textwidth, 
	xlabel={$T_1$ (s)}, 
	ylabel={$T_2 $ (ms)},
	ylabel shift = 2 pt]
\addplot[scatter, only marks, scatter src=\thisrow{class},scatter/classes={0={mark=square*,cyan}, 1={mark=square*,orange},2={mark=square*,blue}}]
      table[x=x,y=y] {
    x       y        class
    1.0840  69.0   0
    0.5000  12.0   1
    0.8557  40.7   2
};
\end{axis}
\end{tikzpicture}
\caption{The left image shows the simulated signals corresponding to two different tissues: intra/extra (I/E) axonal water (light blue) and myelin water (orange). The signal in a voxel containing both tissues corresponds to the sum of both signals (red).  A single-compartment (SC) model (blue) is not able to approximate the data (left) and results in an inaccurate estimate of the relaxation parameters $T_1$ and $T_2$ even in the absence of noise (right). The MRF dictionary is generated using the approach described in~\cite{asslander2016pseudo}.}
\label{fig:sc_error}
\end{figure}
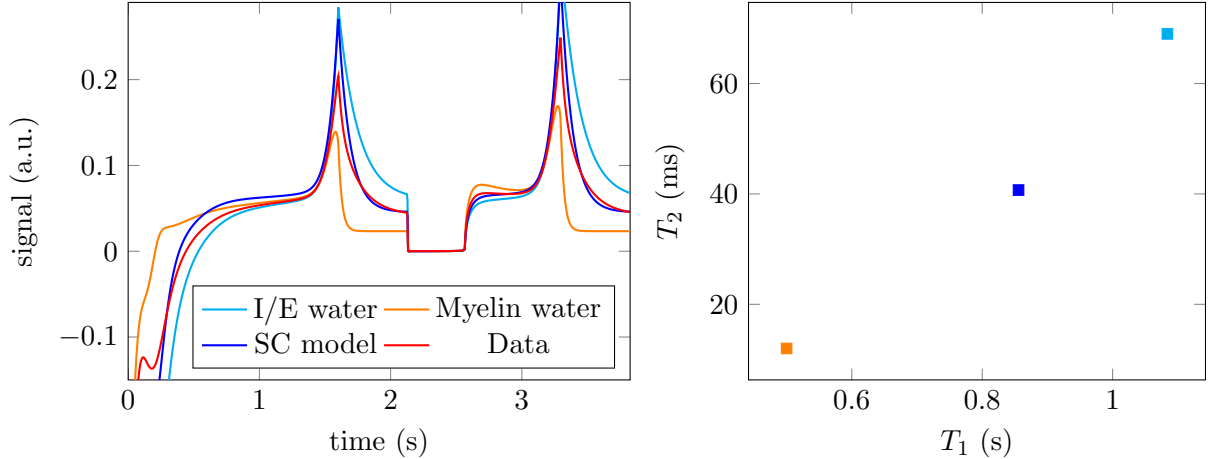

Current MRF methods usually assume that only one type of tissue with a well-defined set of relaxation times is present in each voxel, which degrades their accuracy when this \emph{single-compartment} assumption is violated. Fitting a single-compartment model to data obtained from a voxel with several tissues yields erroneous parameter estimates that may not correspond to any of the contributing compartments, as illustrated in Figure~\ref{fig:sc_error}. As a result of this \emph{partial-volume} effect~\cite{tohka2014partial}, single-compartment MRF methods do not accurately characterize boundaries between tissues, as discussed by the original developers of MRF~\cite{Deshmane2016}. This limitation is particularly problematic in applications that require geometric measurements, such as the diagnostic of Alzheimer’s disease, where cortical thickness provides a promising biomarker for early detection~\cite{Han2006}. 

Single-compartment models also fail to account for other intra-voxel structure. Due to the limited spatial resolution of MRI (on the order of a millimeter), the cellular microstructure of biological tissues results in multiple-tissue compartments being present even in voxels that are not necessarily at tissue boundaries. An important example is white matter in the brain, which consists of signal contributions of extra- and intra-axonal water with similar relaxation times, as well as water trapped between the myelin sheets, with substantially shorter relaxation times~\cite{Laule2006} (Figure~\ref{fig:sc_error} is based on this example). If a particular disease causes demyelination, this leads to a corresponding reduction of one of these compartments (the so-called myelin-water fraction) \cite{Laule2004}. Multicompartment models make it possible to evaluate the myelin-water fraction within a voxel, which is therefore useful for the diagnosis of neurodegenerative diseases such as multiple sclerosis~\cite{Laule2006,Laule2004}. 

It is important to note that the cellular structure of biological tissue generally does not result in entirely separated compartments within each voxel. In fact, one can observe chemical exchange of water molecules between compartments \cite{Allerhand1964}. However, in the case of tissue boundaries, these effects play a subordinate role due to the macroscopic nature of the interfaces. Similarly, in the case of myelin-water imaging, neglecting chemical exchange is a reasonable and commonly-used assumption in the literature~\cite{Laule2004}. In this work, we consider a multicompartment model that incorporates this assumption. As a result, the measured signal at a given voxel corresponds to the sum of the signals corresponding to the individual tissues present in the voxel~\cite{McGivney2017}. 

\subsection{Related work}
\label{sec:related_work}
% Quantitative multicompartment models have a long history. For example, 
Previous works applying multicompartment estimation in the context of myelin-water imaging, applied bi-exponential fitting to data obtained from a multi-echo experiment~\cite{Laule2004,Laule2006}. Other works used non-convex optimization to fit a multicompartment model to measurements from a multi-steady-state experiment~\cite{Deoni2008,Deoni2013}. In the context of MR fingerprinting, the multicompartment problem has been tackled by exhaustive search in~\cite{Hamilton2015}. The work that lies closer to our proposed approach is~\cite{McGivney2017}, which applies a reweighted-least-squares method based on a Bayesian framework to fit the multicompartment parameters to MRF data. This method produces a dense solution for each compartment in the parameter space, which must then be post-processed to produce the parameter estimates. In contrast, our proposed method yields sparse solutions, from which the parameters can usually be estimated directly.

Sparse linear models are a fundamental tool in statistics~\cite{hastie2015statistical,friedman2001elements}, signal processing~\cite{mallat2008wavelet} and machine learning~\cite{mairal2014sparse}, as well as across many applied domains~\cite{Wright2010,Lustig2008a}. Most works that apply these models to recover sparse signals focus on incoherent measurements~\cite{donoho2003optimally,gribonval2003sparse,tropp2004greed,donoho2006compressed,needell2009cosamp,candes2005decoding,bickel2009simultaneous,candes2008restricted}. As explained in more detail in Section~\ref{sec:sparse_estimation_l1}, this setting is not relevant to MRF where dictionary columns can be highly correlated. Instead, the present work is more related to recent advances in optimization-based methods for sparse recovery from coherent dictionaries~\cite{candes2014towards,fernandez2016super,bernstein_deconvolution,esser2013method,malioutov2005sparse}. Related applied work includes the application of these techniques to source localization in electroencephalography~\cite{silva2004evaluation,xu2007lp}, analysis of positron-emission tomography data~\cite{heins2014locally,gunn2002positron} and radar imaging~\cite{potter2010sparsity}. In addition, our work combines insights from recovery methods based on reweighting convex sparsity-inducing norms~\cite{candes2008enhancing,wipf2010iterative} and techniques for distributed optimization~\cite{boyd2011distributed}.    

\subsection{Contributions}

%As described in Section~\ref{eq:model}, fitting such a model requires solving sparse-recovery problem a. Our main contribution an optimization-based approach to tackle this problem that is shown to be both accurate and computationally efficient. 
% In this section, we describe a method to achieve this based on the following insights: 
The main contribution of this work is a method for fitting multicompartment MRF models. These models explain the signal evolution of a voxel using a small number of precomputed basis functions or \emph{fingerprints}, which requires solving a sparse linear inverse problem. We combine the following insights to solve this challenging inverse problem both accurately and efficiently:
\begin{itemize}
\item Fitting the multicompartment model can be decoupled into multiple sparse-recovery problems-- one for each voxel-- using the alternating-direction of multipliers framework (see Section~\ref{sec:admm}). 
\item The correlations between atoms in the dictionary make it possible to compress the dictionary to decrease the computational cost (see Section~\ref{sec:compression}).
\item Thresholding-based algorithms, typically used for sparse recovery and compressed sensing, do not perform well on the multicompartment MRF problem, due to the high correlation between dictionary atoms. However, $\ell_1$-norm minimization does succeed in achieving exact recovery in the absence of noise as long as the problem is solved using higher-precision second-order methods (see Section~\ref{sec:sparse_estimation_l1}). 
\item The sparse-recovery approach can be made robust to noise and model imprecisions by using reweighted-$\ell_1$-norm methods (see Section~\ref{sec:reweightedl1}).
\item These reweighted methods can be efficiently implemented using a second-order interior-point solver described in Section~\ref{sec:interior_point}.
\end{itemize}
We validate our proposed method via numerical experiments on simulated data (Section~\ref{sec:simulated_phantom}), as well as on experimental data from a custom-built phantom using a clinical MR system (see Section~\ref{sec:real_phantom}).

\section{Methods}
\label{sec:methods}

\subsection{Multicompartment model}
\label{eq:model}

In this work, we consider the problem of estimating the relaxation times $T_1$ and $T_2$ from magnetic-resonance data. $T_1$ is the time constant that governs the relaxation of the components of the nuclear-spin magnetization vector that are parallel to the external magnetic field. $T_2$ relaxation affects the components that are perpendicular to the external field~\cite{nishimura1996principles}. 

Magnetic-resonance fingerprinting (MRF)~\cite{Ma2013} is based on the following insight. For a fixed data-acquisition scheme, the mapping between the time-relaxation parameters $T_1$ and $T_2$ of a particular tissue and the corresponding magnetization signal measured by the MRI system is not known explicitly, but it can be \emph{simulated numerically} by solving a system of differential equations, called the Bloch Equations~\cite{bloch1946nuclear}. This makes it possible to construct a dictionary of possible signal evolutions or \emph{fingerprints}, each corresponding to a specific value of the relaxation times. 

Let us denote the signal measured at a voxel $j$ by the vector
\begingroup
 \renewcommand{\arraystretch}{1.4}\begin{align}
 x^{\sqbr{j}} & := \MAT{ x_{t_1}^{\sqbr{j}} \\ x_{t_2}^{\sqbr{j}} \\ \cdots \\ x_{t_n}^{\sqbr{j}} },
\end{align}
\endgroup
where $t_1$, $t_2$, \ldots, $t_n$ are the times at which the signal is sampled. We assume that the signal is well described as an additive superposition of fingerprints in the precomputed dictionary $D \in \R^{n \times m} $ (there are $m$ possible fingerprints),
\begin{align}
\label{eq:x_Dc}
x^{\sqbr{j}} & = Dc^{\sqbr{j}}, \quad 1\leq j \leq N,
\end{align}
where $N$ is the number of voxels in the volume of interest. The vectors of coefficients $c^{\sqbr{1}}$, $c^{\sqbr{2}}$, \ldots, $c^{\sqbr{N}}$ are assumed to be sparse and nonnegative: each nonzero entry in $c^{\sqbr{j}}$ corresponds to the proton density of a tissue that is present in voxel $j$.

Raw MRI data sampled at a given time do not directly correspond to the magnetization signal of the different voxels, but rather to samples from the spatial Fourier transform of the magnetization over all voxels in the volume of interest~\cite{ljunggren1983simple}, a representation known as \emph{k-space} in the MRI literature~\cite{Twieg1983}. Staying within clinically-feasible scan times in MRF requires subsampling the image k-space and therefore violating the Nyquist-Shannon sampling theorem~\cite{Ma2013}. The k-space is typically subsampled along different trajectories, governed by changes in the magnetic-field gradients used in the measurement process. We define a linear operator $F_{t_i} \in \C^{d \times n}$ that represents the linear operator mapping the magnetization at time $t_i$ to the measured data $y_{\sqbr{t_i}} \in \C^{d}$, where $d$ is the number of k-space samples at $t_i$. This operator encodes the chosen subsampling pattern in the frequency domain. The model for the data is consequently of the form
\begingroup
 \renewcommand{\arraystretch}{1.4}
\begin{align}
\label{eq:y_Fx_ti}
 y_{t_i} & = F_{t_i} \MAT{ x_{t_i}^{\sqbr{1}} \\ x_{t_i}^{\sqbr{2}} \\ \vdots \\ x_{t_i}^{\sqbr{N}} }, \qquad 1 \leq i \leq n,
\end{align}
\endgroup
where we have ignored noise and model inaccuracies. If the data are acquired using multiple receive coils, the measurements can be modeled as samples from the spectrum of the pointwise product between the magnetization and the sensitivity function of the different coils~\cite{Sodickson1997,pruessmann1999sense}. In that case the operator $F_{t_i}$ also includes the sensitivity functions and $d$ is the number of k-space samples multiplied by the number of coils. 

% Our goal is to fit the multicompartment model enforcing this sparsity assumption in a computationally efficient fashion.
% In this section, we describe a method to achieve this based on the following insights: 
%\begin{itemize}
%\item Fitting the model can be decoupled into $N$ sparse-recovery problems-- one for each voxel-- using the alternating-direction of multipliers framework, as described in Section~\ref{sec:admm}. 
%\item The correlations between atoms in the dictionary make it possible to compress the dictionary to decrease the computational cost, as discussed in Section~\ref{sec:compression}.
%\item Thresholding-based algorithms typically used for sparse recovery and compressed sensing cannot be used to tackle the sparse-recovery problems at each voxel, due to the high correlation between dictionary atoms. However, $\ell_1$-norm minimization does succeed in achieving exact recovery in the absence of noise as long as the problem is solved using higher-precision second-order methods (see Section~\ref{sec:sparse_estimation_l1}). 
%\item The sparse-recovery approach can be made robust to noise and model imprecisions by using reweighted-$\ell_1$-norm methods described in Section~\ref{sec:reweightedl1}.
%\item The aforementioned reweighted-$\ell_1$-norm methods can be efficiently implemented using a second-order interior-point solver described in Section~\ref{sec:interior_point}.
%\end{itemize}
\subsection{Parameter Map Reconstruction via Alternating Minimization}
\label{sec:admm}
For ease of notation, we define the magnetization and coefficient matrices
%\begingroup
%\renewcommand{\arraystretch}{1.25}
\begin{align}
X & :=  \MAT{ x^{\sqbr{1}}, &x^{\sqbr{2}}, &\ldots, &x^{\sqbr{N}} }, \\
C & :=  \MAT{ c^{\sqbr{1}}, &c^{\sqbr{2}}, &\ldots, &c^{\sqbr{N}} },
%Y:= \MAT{ y_{\sqbr{1}} & y_{\sqbr{2}} & \cdots & y_{\sqbr{n}}}, \qquad X & :=  \MAT{ (x^{\sqbr{1}}^T \\ x_{\sqbr{2}}^T \\ \cdots \\ x_{\sqbr{N}}^T }, \qquad C  :=  \MAT{ c_{\sqbr{1}}^T \\ c_{\sqbr{2}}^T \\ \cdots \\ c_{\sqbr{N}}^T }, % \\
% c^{\sqbr{i}}_j & \geq 0, \quad 1 \leq j \leq m,
\end{align}
%\endgroup
so that we can write~\eqref{eq:x_Dc} as 
\begin{align}
\label{eq:x_Dc_mat}
X & = DC.
\end{align}
In addition we define the data matrix $Y \in \C^{d \times n}$ 
\begin{align}
Y:= \MAT{ y_{t_1}, &y_{t_2}, &\ldots, &y_{t_n}}
\end{align}
and a linear operator $\ml{F}$ such that 
\begin{align}
Y & = \ml{F}X.
\end{align}
Note that by~\eqref{eq:y_Fx_ti} the $i$th column of $Y$ is the result of applying $F_{t_i}$ to the $i$th \emph{row} of $X$. 

In principle, we could fit the coefficient matrix by computing a sparse estimate such that $Y \approx \ml{F}DC$. Unfortunately, this would require solving a sparse-regression problem of intractable dimensions. Instead we incorporate a variable $\wt{X}$ to represent the magnetization and formulate the model-fitting problem as
\begin{alignat}{2}
& \underset{\wt{X} ,\wt{C} }{\op{minimize}} \qquad && \normF{Y -  \ml{F} \wt{X} }^2 \\
& \text{subject to} \qquad && \wt{X} = D \, \wt{C}, \label{eq:constraint_M_DC}
\end{alignat}
with the additional constraint that the entries of $\wt{C}$ are sparse and nonnegative. In order to alternate between updating the magnetization and coefficient variables, we follow the framework of the alternative-direction method of multipliers (ADMM)~\cite{boyd2011distributed}. Consider the augmented Lagrangian with respect to the constraint~\eqref{eq:constraint_M_DC},
\begin{align}
\ml{L}_{\mu}\brac{\wt{X},\wt{C},\Lambda} := \normF{Y - \ml{F} \wt{X}  }^2 + \PROD{\Lambda}{\wt{X} - D \wt{C} } + \frac{\mu}{2}\normF{ \wt{X} -  D \, \wt{C} }^2,
\end{align}
where the constant $\mu >0$ is a parameter and the dual variable $\Lambda$ is an $N \times n$ matrix. ADMM alternates between minimizing the augmented Lagrangian with respect to each primal variable sequentially and updating the dual variable. We now describe each of the updates in more detail. We denote the values of the different variables at iteration $l$ by $C^{\brac{l}}$, $X^{\brac{l}}$ and $\Lambda^{\brac{l}}$. More details on the implementation can be found in Ref.~\cite{asslander2017low}.

\subsubsection*{Updating $\wt{X}$}

If we fix $\wt{C}$ and $\Lambda$, minimizing $\ml{L}_{\mu}$ over $\wt{X}$ is equivalent to solving the least-squares problem 
\begin{align}
X^{\brac{l+1}}:= \arg \min_{\wt{X}} %\underset{\wt{M} }{\op{minimize}} \qquad 
\normF{Y -  F\wt{X} }^2 + \frac{\mu}{2}\normF{ \wt{X} - D \, C^{\brac{l}} - \frac{1}{\mu} \Lambda^{\brac{l}} }^2.
\end{align}
This step amounts to estimating a magnetization matrix that fits the data while being close to the magnetization corresponding to the current estimate of dictionary coefficients. The optimization problem has a closed-form solution, but due to the size of the matrices it is more efficient to solve it using an iterative algorithm such as the conjugate-gradients method~\cite{shewchuk1994introduction}. 

\subsubsection*{Updating $\wt{C}$}

If we fix $\wt{X}$ and $\Lambda$, minimizing $\ml{L}_{\mu}$ over $\wt{C}$ decouples into $N$ subproblems. In more detail, the goal is to compute a sparse nonnegative vector of coefficients $ \tilde{c}^{\sqbr{j}}$ for each voxel $j$, $1 \leq j \leq N$, such that
\begin{align}
\brac{X^{\brac{l}} - \frac{1}{\mu} \Lambda^{\brac{l}} }_j  & \approx D \, \tilde{c}^{\sqbr{j}},
\label{eq:sparse_recovery_voxel}
\end{align}
where the left hand side corresponds to the $j$th column of $X^{\brac{l}} - \frac{1}{\mu} \Lambda^{\brac{l}} $. This collection of sparse-recovery problems is very challenging to solve due to the correlations between the columns of $D$. In Sections~\ref{sec:sparse_estimation_l1}, \ref{sec:reweightedl1} and \ref{sec:interior_point} we present an algorithm to tackle them efficiently.

\subsubsection*{Updating $\Lambda$}
The dual variable is updated by setting
\begin{align}
\Lambda^{\brac{l+1}} := \Lambda^{\brac{l}} +  \mu \brac{X^{\brac{l}}- \widehat{ D} \, C^{\brac{l}} }.
\end{align}
We refer to~\cite{boyd2011distributed} for a justification based on dual-ascent methods.

%\begin{figure}
%\begin{center}
%\begin{tikzpicture}
%\begin{axis}[legend pos=south west ,xmin=0,xmax=800,width=0.95\textwidth, height=\axisdefaultheight,ymax=0.05,ymin=-0.27,ticks=none]
%% \addlegendimage{empty legend}
%\addplot[thick,orange] file {dat_files/admm_phantom/mag_init.dat};
%\addplot[thick,cyan] file {dat_files/admm_phantom/mag_1.dat};
%\addplot[thick,blue] file {dat_files/admm_phantom/mag_10.dat};
%\addplot[thick,magenta] file {dat_files/admm_phantom/mag_signal.dat};
%\legend{Iteration 1, Iteration 2, Iteration 10, Ground truth}
%\end{axis}
%\end{tikzpicture}
%\end{center}
%\caption{Magnitude of the magnetization signal for a specific voxel}
%\end{figure}

\subsection{Dictionary compression}
\label{sec:compression}
\begin{figure}
\begin{tikzpicture}
\begin{groupplot}[group style={group size=2 by 1, horizontal sep=2cm, vertical sep=1cm},height=4cm,width=6cm]
\nextgroupplot[ymode=log,width=0.5\textwidth, height=\axisdefaultheight, title={Singular values},xmin=0,xmax=40.5]
\addplot[only marks, blue,thick] file {dat_files/svd/sv.dat};
\nextgroupplot[
width=0.5\textwidth, 
height=\axisdefaultheight, 
title={Singular vectors},
legend pos=south east,
legend columns = 2,
xmin=0,xmax=850,ymin=-0.14,ymax=0.14,
xlabel = {time (s)}, 
ylabel = {signal (a.u.)}, 
ytick distance = 0.1,
scaled x ticks={real:222.222}, 
xtick scale label code/.code={}, 
xtick distance=222.222,]
\addplot[thick,cyan] file {dat_files/svd/sve1.dat};
\addplot[ thick,red] file {dat_files/svd/sve2.dat};
\addplot[ thick,blue] file {dat_files/svd/sve3.dat};
\addplot[ thick,orange] file {dat_files/svd/sve4.dat};
%\addplot[ thick,magenta] file {dat_files/svd/sve5.dat};
\legend{1,2,3,4}
\end{groupplot}
\end{tikzpicture}
\caption{Singular values (left) and corresponding left singular vectors (right) of an MRF dictionary generated using the approach in~\cite{asslander2016pseudo,Asslander2017arxiv}.}% \textbf{FK TODO: Is this figure referenced correctly? It seems to me that Figure 3 is referenced before Figure 2 in the text.}}
\label{fig:svd}
\end{figure}
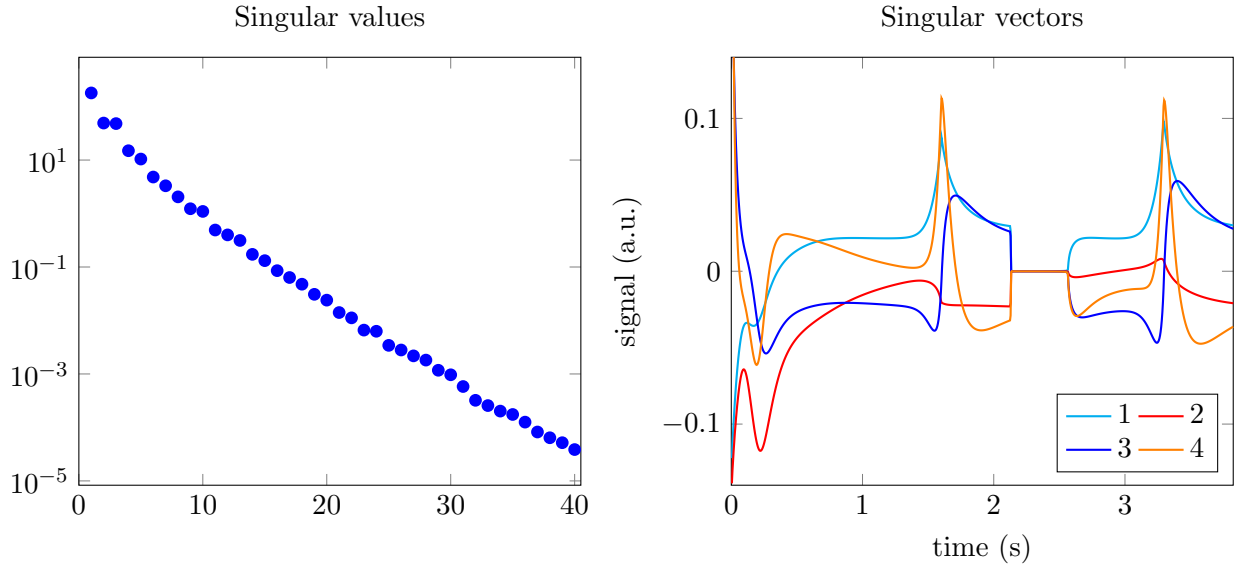

As explained in Section~\ref{sec:sparse_estimation_l1} the columns in the dictionary of fingerprints $D$ are highly correlated. Consequently, $D$ is approximately low rank: if we consider its singular value decomposition
\begin{align}
D = U\Sigma V^T,
\end{align}
where the superscript $T$ denotes the transpose. The singular values in $\Sigma$ present a rapid decay, so its column space is well approximated as the span of the first few left singular vectors (see Figure~\ref{fig:svd}). Following~\cite{McGivney2017,asslander2017low}, we exploit this low-rank structure to compress the dictionary. Let $U_{k}\Sigma_{k}V_k^T$ denote the rank-$k$ truncated SVD of $D$, where $U_{k}$ contains the first $k$ left singular vectors and $\Sigma_{k}$ and $V_k$ the corresponding singular values and right singular vectors. If the measured data follow the linear model $Y \approx \ml{F}DC$, then
\begin{align}
Y &  \approx \ml{F} U_{k}\Sigma_{k}V_k^T C \\
& = \overline{\ml{F}} \, \overline{D}C, \label{eq:compressed_model}
\end{align}
where $ \overline{\ml{F}} := \ml{F} U_{k}$ is a modified linear operator and
\begin{align}
\overline{D} & := \Sigma_{k}V_k^T \\
& = U_{k}^T D.
\end{align}
$\overline{D}$ can be interpreted as a compressed dictionary with dimensions $k \times m$ obtained by projecting each fingerprint onto the $k$-dimensional subspace spanned by the columns of $U_{k}$, which are depicted on the right image of Figure~\ref{fig:svd}. In our experiments, $D$ is well approximated by $U_{k}\Sigma_{k}V_k^T$ for values of $k$ between 10 and 15. As a result, replacing $\ml{F} $ by $ \overline{\ml{F}}$ and $D$ by $\overline{D}$ within the ADMM framework in Section~\ref{sec:admm} dramatically decreases the computational cost. % This is verified through computational experiments in Section~\ref{sec:numerical_single_voxel}. Since $k$ is at least an order of magnitude smaller than the length of the magnetization sequences $n$, dictionary compression significantly decreases the computational cost. 

\subsection{Sparse estimation via $\ell_1$-norm minimization}
\label{sec:sparse_estimation_l1}
In this section, we consider the problem of estimating the tissue parameters at a fixed voxel from the time evolution of its magnetization signal. As described in Section~\ref{sec:admm}, this is a crucial step in fitting the MRF multicompartment model. To simplify the notation, we denote the discretized magnetization signal at an arbitrary voxel by $x \in \R^{n}$ (where $n$ is the number of time samples) and the sparse vector of coefficients by $c \in \R^{m}$. In Eq.~\eqref{eq:sparse_recovery_voxel}, $x$ represents the column of $X^{\brac{l}} - \frac{1}{\mu} \Lambda^{\brac{l}}$ corresponding to the voxel at a particular iteration $l$ of the alternating scheme. 

Let us first consider the sparse recovery problem assuming that the magnetization estimate is exact. In that case there exists a sparse vector $c$ such that 
\begin{align}
\label{eq:xDc_noiseless}
x = D c.
\end{align} 
Even in this simplified scenario, computing $c$ from $x$ is very challenging. The dictionary is overcomplete: there are more columns than time samples because the number of fingerprints $m$ is larger than the number of time samples $n$. As a result the linear system is underdetermined: there are infinite possible solutions. However, we are not interested in arbitrary solutions, but rather in sparse solutions, where $c$ contains a small number of nonzero entries corresponding to the tissues present in the voxel. 

There is a vast literature on the recovery of sparse signals from underdetermined linear measurements. Two popular techniques are greedy methods that select columns from the dictionary sequentially~\cite{pati1993orthogonal,mallat1993matching} and optimization-based approaches that minimize a sparsity-promoting cost function such as the $\ell_1$ norm~\cite{chen2001atomic,donoho2006compressed,candes2006robust}. Theoretical analysis of these methods has established that they achieve exact recovery and are robust to additive noise if the overcomplete dictionary is \emph{incoherent}, meaning that the correlation between the columns in the dictionary is low~\cite{donoho2003optimally,gribonval2003sparse,tropp2004greed,donoho2006compressed,needell2009cosamp,candes2005decoding,bickel2009simultaneous,candes2008restricted}. Some of these works assume stronger notions of incoherence such as the restricted-isometry condition~\cite{candes2005decoding} or the restricted-eigenvalue condition~\cite{bickel2009simultaneous}.  

\begin{figure}[tp]
\begin{tikzpicture}[scale=1]
\begin{axis}[
width=0.5\textwidth, 
height=\axisdefaultheight, 
ymin=-0.1,ymax=0.1,
xmin=0,xmax=850,
legend pos=south east, 
xlabel={time (s)}, 
ylabel={signal (a.u.)}, 
y dir=reverse,
scaled y ticks={real:-1}, 
ytick scale label code/.code={}, 
ytick distance=0.1,
scaled x ticks={real:222.222}, 
xtick scale label code/.code={}, 
xtick distance=222.222,
title = MRF dictionary,
name=signal_MRF]
\addlegendimage{empty legend}
\addplot[thick,cyan] file {dat_files/dictionary_atoms/dic_t1_120_2.dat};
%\addplot[thick,teal] file {dat_files/dictionary_atoms/dic_t1_160.dat};
\addplot[thick,blue] file {dat_files/dictionary_atoms/dic_t1_300.dat};
\addplot[thick,magenta] file {dat_files/dictionary_atoms/dic_t1_720.dat};
\addplot[thick,violet] file {dat_files/dictionary_atoms/dic_t1_3200.dat};
\legend{\hspace{-.7cm} $T_1$ (ms),$120$,$300$,$720$,$3200$}
% \addplot[black] coordinates {(0,0) (0,1000)};
\end{axis}

\begin{axis}[
width=0.5\textwidth, 
height=\axisdefaultheight, 
ymin=-0.1,ymax=0.1,
xmin=0,xmax=850, 
legend pos=south east, 
xlabel={time (s)}, 
yticklabel=\empty,
ytick distance=0.1,
scaled x ticks={real:222.222}, 
xtick scale label code/.code={}, 
xtick distance=222.222,
name=signal_gauss,
at=(signal_MRF.right of north east),
anchor=left of north west,
title = Gaussian dictionary,
]
\addlegendimage{empty legend}
\addplot[thick,cyan] file {dat_files/dictionary_atoms/dic_g1.dat};
\addplot[thick,teal] file {dat_files/dictionary_atoms/dic_g2.dat};
\addplot[thick,blue] file {dat_files/dictionary_atoms/dic_g3.dat};
\addplot[thick,magenta] file {dat_files/dictionary_atoms/dic_g4.dat};
\addplot[thick,violet] file {dat_files/dictionary_atoms/dic_g5.dat};
% \addplot[black] coordinates {(0,0) (100,0)};
 \legend{\hspace{-.7cm} Atom,20,50,110,200,350}
\end{axis}

\begin{semilogxaxis}[
width=0.5\textwidth, 
height=\axisdefaultheight,  
xmin=.1, xmax = 5,
ymin=0,ymax=1.05,
xlabel= $T_1$ (s), 
ylabel = correlation coefficient, %ylabel shift = 5 pt,
%xlabel shift = 5 pt,
name=corr_MRF,
at=(signal_MRF.below south west),
anchor=above north west,
]
\addlegendimage{empty legend}
\addplot[very thick,cyan] file {dat_files/corr/cor_120.dat};
%\addplot[very thick,teal] file {dat_files/corr/cor_160.dat};
\addplot[very thick,blue] file {dat_files/corr/cor_300.dat};
\addplot[very thick,magenta] file {dat_files/corr/cor_720.dat};
\addplot[very thick,violet] file {dat_files/corr/cor_3200.dat};
 \addplot[black] coordinates {(0,0) (100,0)};
% \legend{\hspace{-.7cm} $T_1$ (ms),$120$,$300$,$720$,$3200$}
\end{semilogxaxis}

\begin{axis}[
width=0.5\textwidth, 
height=\axisdefaultheight,  
ymin=0,ymax=1.05,
xlabel = dictionary index, 
yticklabel=\empty,
at=(corr_MRF.right of north east),
anchor=left of north west]
\addlegendimage{empty legend}
\addplot[very thick,cyan] file {dat_files/corr/corg1.dat};
\addplot[very thick,teal] file {dat_files/corr/corg2.dat};
\addplot[very thick,blue] file {dat_files/corr/corg3.dat};
\addplot[very thick,magenta] file {dat_files/corr/corg4.dat};
\addplot[very thick,violet] file {dat_files/corr/corg5.dat};
% \addplot[black] coordinates {(0,0) (100,0)};
%   \legend{\hspace{-.7cm} Atom,20,50,110,200,350}
\end{axis}
\end{tikzpicture}
\caption{The figure shows a small selection of columns in a magnetic-resonance fingerprinting (MRF) dictionary (top left) and an i.i.d. Gaussian compressed-sensing dictionary (top right). In the MRF dictionary, generated using the approach in~\cite{asslander2016pseudo,Asslander2017arxiv}, each column corresponds to the time evolution of the magnetization for a value of the relaxation parameter $T_1$ (for simplicity $T_2$ is fixed to 62~ms). In the bottom row we show the correlation of the selected columns with every other column in the dictionary. This reveals the high local coherence of the MRF dictionary (left), compared to the incoherence of the compressed-sensing dictionary (right).}
\label{fig:dictionary_correlations}
\end{figure}
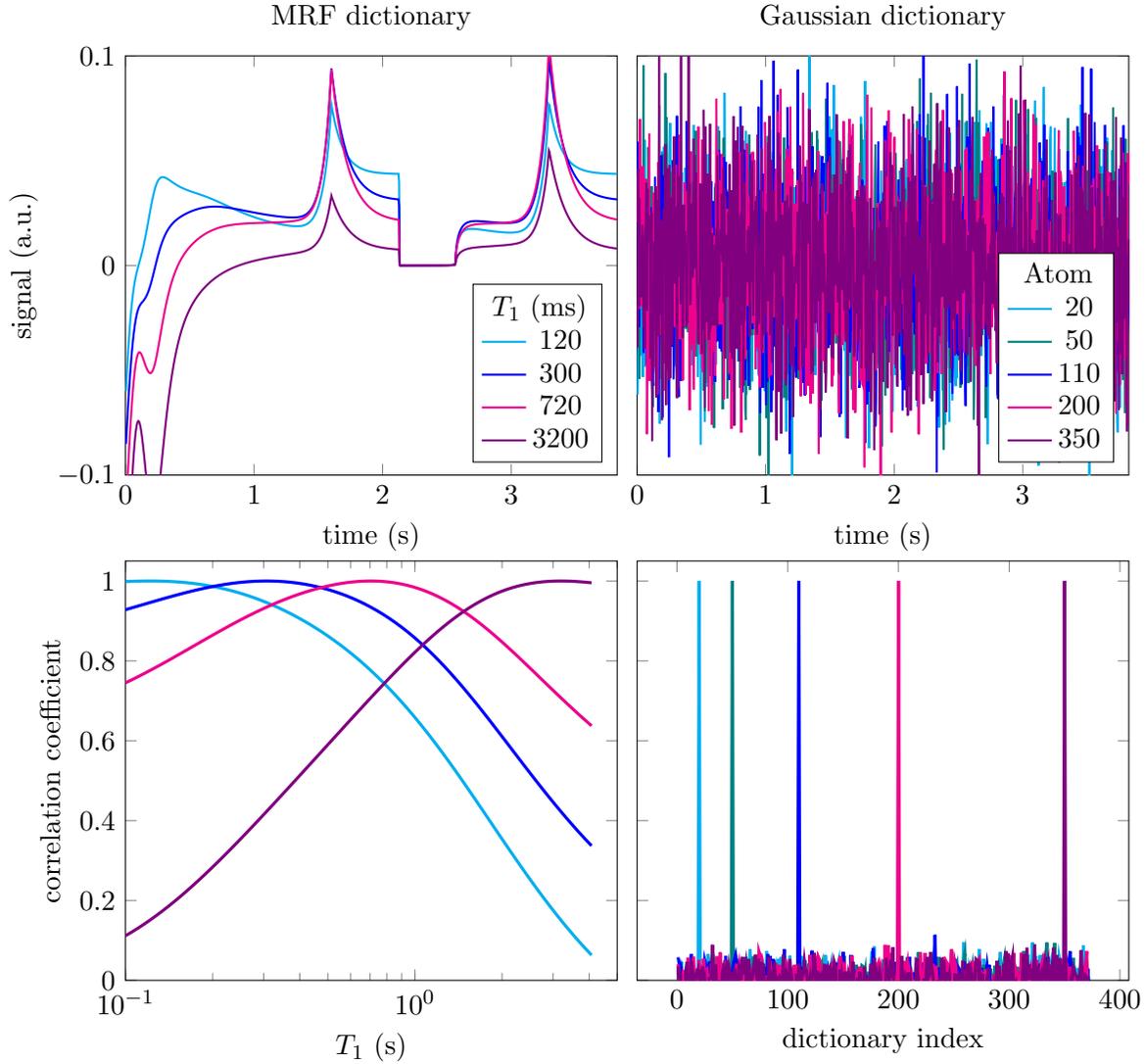

Unfortunately, MRF dictionaries are highly coherent: the correlation between fingerprints corresponding to similar parameters is very high. This is illustrated by Figure~\ref{fig:dictionary_correlations}, which shows the correlations between columns for a simple MRF dictionary simulated following~\cite{asslander2016pseudo,Asslander2017arxiv}. For ease of visualization, the only parameter that varies in the dictionary is $T_1$ (the value of $T_2$ is fixed to 62~ms), so that the correlation in the depicted dictionary has a very simple structure: neighboring columns correspond to similar values of $T_1$ and are, therefore, more correlated, whereas well-separated columns are less correlated. For comparison, the figure also shows the correlations between columns for a typical incoherent compressed-sensing dictionary generated by sampling each entry independently from a Gaussian distribution~\cite{donoho2006compressed,candes2006near}. 

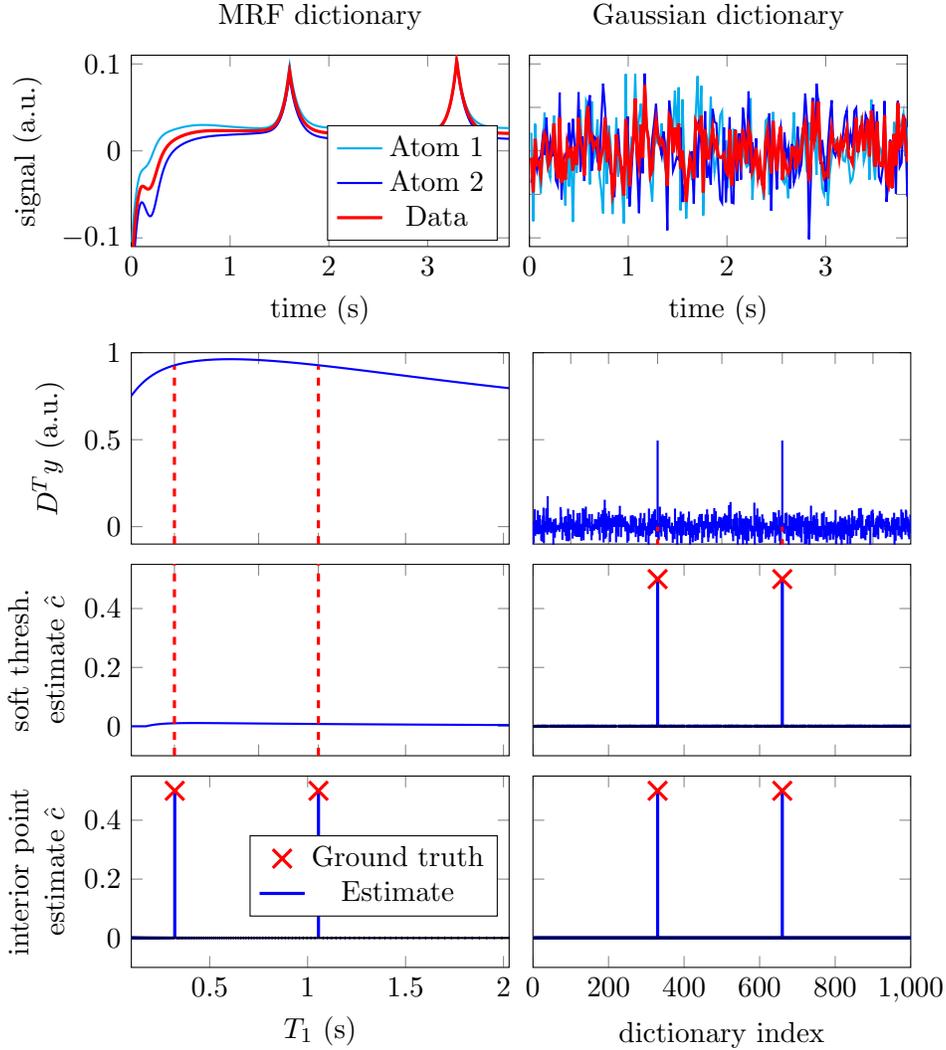
\begin{figure}[tp]
	\centering
\begin{tikzpicture}
\begin{axis}[
width=0.4\textwidth, 
height=0.25\textwidth,
ymin=-0.11,
ymax=0.11,
xmin=0,
xmax=850, 
xlabel={time (s)}, 
ylabel = signal (a.u.),
y dir=reverse,
scaled y ticks={real:-1}, 
ytick scale label code/.code={}, 
ytick distance=0.1,
scaled x ticks={real:222.222}, 
xtick scale label code/.code={}, 
xtick distance=222.222,
legend pos=south east,
name=signal_MRF,
title = MRF dictionary,
]
\addplot[ticks=none,thick,cyan] file {dat_files/figure_4by2/sig_mrf_1.dat};
\addplot[ticks=none,thick,blue] file {dat_files/figure_4by2/sig_mrf_2.dat};
\addplot[ticks=none,very thick,red] file {dat_files/figure_4by2/sig_mrf_mix.dat}; 
\legend{Atom 1,Atom 2,Data}

%\node[right, inner sep=0mm,minimum size=5mm] at (axis cs:  750, -0.08) {\textbf{a}};
\end{axis}

\begin{axis}[
width=0.4\textwidth, 
height=0.25\textwidth,
xmin=0,
xmax=200,
ymin = -0.22,
ymax=0.22, 
xlabel={time (s)},
yticklabel=\empty,
scaled x ticks={real:52.288}, 
xtick scale label code/.code={}, 
xtick distance=52.288,
name=signal_gauss,
at=(signal_MRF.right of north east),
anchor=left of north west,
title = Gaussian dictionary,
]
\addplot[thick,cyan] file {dat_files/figure_4by2/sig_gau_1.dat};
\addplot[ticks=none,thick,blue] file {dat_files/figure_4by2/sig_gau_2.dat};
\addplot[ticks=none,very thick,red] file {dat_files/figure_4by2/sig_gau_mix.dat}; 
%\legend{Atom 1,Atom 2,Data}

%\node[right, inner sep=0mm,minimum size=5mm] at (axis cs:  750/222.222*52.288, 0.16) {\textbf{b}};
\end{axis}

\begin{axis}[
width=0.4\textwidth, 
height=0.25\textwidth, 
xmin=1e-1,xmax=2.0287387e+00,
xtick={0.32,0.75,1.055,1.5,2.0},
xticklabels={Atom 1,0.75,Atom 2,1.5,2},
ymin=-0.1,
ymax=1, 
%xlabel=Dictionary entries (indexed by $T_1$ values),
xticklabel=\empty,
ylabel=$D^Ty$ (a.u.),
name=DTy_MRF,
at=(signal_MRF.below south west),
anchor=above north west,
]
\addplot[ticks=none,thick,blue] file {dat_files/figure_4by2/cor_mrf.dat};
\addplot[red, very thick, dashed, samples=2] coordinates {(1.055, -.1)(1.055, 0.925)};
\addplot[red, very thick, dashed, samples=2] coordinates {(0.32, -.1)(0.32, 0.925)};
%\legend{correlation, GT}
\end{axis}

\begin{axis}[
width=0.4\textwidth, 
height=0.25\textwidth,
xmin=0,xmax=1000, 
ymin=-0.1,ymax = 1,
xtick={0,100,200,330,500, 660,800,900},
xticklabel=\empty,
yticklabel=\empty,
name=DTy_gauss,
at=(DTy_MRF.right of north east),
anchor=left of north west,
]
\addplot[ycomb,ticks=none,thick,blue] file {dat_files/figure_4by2/gau_corr.dat};
\addplot[red, very thick, dashed, samples=2] coordinates {(330, -0.3)(330, 0)};
\addplot[red, very thick, dashed, samples=2] coordinates {(660, -0.3)(660, 0)};
\end{axis}

\begin{axis}[
width=0.4\textwidth, 
height=0.25\textwidth, 
legend pos = south east,
xmin=1e-1,xmax=2.0287387e+00,
xtick={0.32,0.75,1.055,1.5,2.0}, 
xticklabels={Atom 1,0.75,Atom 2,1.5,2},
ymin=-0.1, ymax=0.55, 
%ytick scale label code/.code={}, 
xticklabel=\empty,
ylabel style={align=center},
ylabel={soft thresh. \\ estimate $\hat{c}$},
name=soft_MRF,
at=(DTy_MRF.below south west),
anchor=above north west,
]
 \addplot[ticks=none,thick,blue] file {dat_files/figure_4by2/recon_mrf_tfocs.dat};
\addplot[red, very thick, dashed, samples=2] coordinates {(1.055, -.1)(1.055, 0.925)};
\addplot[red, very thick, dashed, samples=2] coordinates {(0.32, -.1)(0.32, 0.925)};
 
%\node[right, inner sep=0mm,minimum size=5mm] at (axis cs:  {750/850*(2.029-0.1)+0.1}, 0.425) {\textbf{c}};
\end{axis}

\begin{axis}[
width=0.4\textwidth, 
height=0.25\textwidth,
xticklabel=\empty,
yticklabel=\empty,
xmin=0,xmax=1000,
ymin=-0.1, ymax=0.55, 
name=soft_gauss,
at=(soft_MRF.right of north east),
anchor=left of north west,
]
\addplot+[only marks, mark=x, mark size=5pt, very thick, red] file {dat_files/figure_4by2/gt_gau.dat};
\addlegendimage{no markers,very thick,blue}
\addplot[ycomb,ticks=none,very thick,blue] file {dat_files/figure_4by2/recon_gauss_tfocs.dat};
\addplot[black, thick, samples=2] coordinates {(0, 0)(1e3, 0)};
%\legend{Signal, Estimate}

%\node[right, inner sep=0mm,minimum size=5mm] at (axis cs:  {750/850*1000}, 0.425) {\textbf{d}};
\end{axis}

\begin{axis}[
width=0.4\textwidth, 
height=0.25\textwidth,
ymin=-0.1, ymax=0.55, 
xlabel=$T_1$ (s),
ylabel style={align=center},
ylabel={interior point \\ estimate $\hat{c}$},
xmin=1e-1,xmax=2.0287387e+00,
legend pos=south east,
legend style = {yshift = 0.6cm},
name=interior_MRF,
at=(soft_MRF.below south west),
anchor=above north west,
]
\addplot+[only marks, mark=x, mark size=5pt, very thick, red] file {dat_files/figure_4by2/gt_mrf.dat};
\addlegendimage{no markers,very thick,blue}
\addplot[ycomb,ticks=none,very thick,blue] file {dat_files/figure_4by2/recon_mrf_cvx.dat};
\addplot[black, thick, samples=2] coordinates {(1e-1, 0)(3, 0)};
\legend{Ground truth, Estimate}

%\node[right, inner sep=0mm,minimum size=5mm] at (axis cs:  {750/850*(2.029-0.1)+0.1}, 0.425) {\textbf{e}};
\end{axis}

\begin{axis}[
width=0.4\textwidth, 
height=0.25\textwidth,
ymin=-0.1, ymax=0.55, 
xlabel = dictionary index,
xmin=0,xmax=1000,
yticklabel=\empty,
name=interior_gauss,
at=(soft_gauss.below south west),
anchor=above north west,
]
\addplot+[only marks, mark=x, mark size=5pt, very thick, red] file {dat_files/figure_4by2/gt_gau.dat};
\addlegendimage{no markers,very thick,blue}
\addplot[ycomb,ticks=none,very thick,blue] file {dat_files/figure_4by2/recon_cvx_gau.dat};
\addplot[black, thick, samples=2] coordinates {(0, 0)(1e3, 0)};
%\legend{Signal, Estimate}

%\node[right, inner sep=0mm,minimum size=5mm] at (axis cs:  {750/850*1000}, 0.425) {\textbf{f}};
\end{axis}
\end{tikzpicture}
\caption{The data in the top row are an additive superposition of two columns from the MRF (left) and the compressed-sensing (right) dictionaries depicted in Figure~\ref{fig:dictionary_correlations}. The correlation between the data and each dictionary column (second row) is much more informative for the incoherent dictionary (right) than for the coherent one (left). A fast first-order method~\cite{beck2009fast,becker2011templates} designed to solve problem~\eqref{eq:eq_const} (third row) achieves exact recovery of the true support for the compressed-sensing problem (right), but fails for MRF (left). In contrast, solving~\eqref{eq:eq_const} using a high-precision solver~\cite{grantcvx} achieves exact recovery in both cases (bottom row).}
\label{fig:thresholding}
\end{figure}

Coherence is not only a problem from a theoretical point of view. Most fast-estimation methods for sparse recovery exploit the fact that for incoherent dictionaries like the Gaussian dictionary in Figure~\ref{fig:dictionary_correlations} the Gram matrix $D^TD$ is close to the identity. As a result, multiplying the data by the transpose of the dictionary yields a noisy approximation to the sparse coefficients 
\begin{align}
D^Ty = D^TDc \approx c,
\end{align}
which can be \emph{cleaned up} using some form of thresholding. This concept is utilized by greedy approaches such as orthogonal matching pursuit~\cite{pati1993orthogonal} or iterative hard-thresholding methods~\cite{blumensath2009iterative}, as well as fast algorithms for $\ell_1$-norm minimization such as proximal methods~\cite{combettes2005signal} and coordinate descent~\cite{friedman2010regularization}, which use iterative soft-thresholding instead. When the dictionary is coherent, and $D^Ty$ is no longer approximately sparse, these techniques are extremely slow and often fail to converge to a sparse estimate. As an example Figure~\ref{fig:thresholding} shows the result of applying the fast iterative shrinkage-thresholding algorithm (FISTA) \cite{beck2009fast} to an MRF dictionary. The estimated coefficients are not sparse, which is consistent with numerical experiments reported in other works~\cite{McGivney2017}. 

Recently, theoretical guarantees for sparse-decomposition methods have been established for dictionaries arising in super-resolution~\cite{candes2014towards,fernandez2016super} and deconvolution problems~\cite{bernstein_deconvolution}. As in the case of MRF, these dictionaries have very high local correlations between columns. These papers show that although robust recovery of all sparse signals in such dictionaries is not possible due to their high coherence, $\ell_1$-norm minimization achieves exact recovery of a more restricted class of signals: sparse superpositions of dictionary columns that are not highly correlated. To be clear, there may be arbitrarily high local correlations in the dictionary, as long the columns that are actually present in the measured signal are relatively uncorrelated with each other. In the case of MRF, this corresponds to a multicompartment model where the fingerprints of the tissues present in each voxel are sufficiently different. Our numerical results indicate that solving the convex program
\begin{alignat}{2}
& \underset{\tilde{c} \, \in \, \R^{m}}{\text{minimize}} \quad && ||\tilde{c}||_1 \\
& \text{subject to} \quad    && D \tilde{c}  = x \\
& && \quad \tilde{c}  \geq 0.
\label{eq:eq_const}
\end{alignat}
makes it possible to successfully fit the sparse multicompartment model, as long as the parameters of each compartment are not too close (see Figure~\ref{fig:noiseless}). Crucially, we observe that it is necessary to use high-precision second-order methods~\cite{grantcvx} in order to achieve exact recovery. In contrast, fast $\ell_1$-norm minimization algorithms based on soft-thresholding or coordinate descent converge extremely slowly or fail to achieve exact recovery due to the coherence of the dictionary, as illustrated by the example in Figure~\ref{fig:thresholding}. The nonnegative constraint in~\eqref{eq:eq_const} stems from the fact that the coefficients in the multicompartment model correspond to the proton densities of each tissue and, therefore, cannot be negative. In practice, this constraint implies that all compartments have the same complex phase. 

\subsection{Robust sparse estimation via reweighted-$\ell_1$-norm minimization}
\label{sec:reweightedl1}
In the previous section, we discuss exact recovery of the sparse coefficients in a multicompartment model under the assumptions that (1) there is no noise and (2) the fingerprints corresponding to the true parameters are present in the dictionary. In practice, neither of these assumptions holds: noise is unavoidable, the tissue parameters may have any value in the continuous parameter space and the data $x$ used to fit the model is perturbed by aliasing artifacts (recall that it corresponds to a column in $X^{\brac{l}} - \frac{1}{\mu} \Lambda^{\brac{l}}$ at iteration $l$ of the alternating scheme described in Section~\ref{sec:admm}). A standard way to account for additive noise and model imprecisions in sparse-recovery problems is to relax the $\ell_1$-norm minimization problem~\eqref{eq:eq_const} to a regularized least-squares problem of the form
\begin{equation}
\begin{aligned}
& \underset{\tilde{c} \, \in \, \R^{m}}{\text{minimize}}
& & \frac{1}{2} ||D\tilde{c}-x||^2 + \lambda ||\tilde{c}||_1 \\
& \text{subject to}
& & \tilde{c} \geq 0
\end{aligned}
\label{eq:eq_lasso_nonneg}
\end{equation}
where $\lambda \geq 0$ is a regularization parameter that governs the trade-off between the least-squares data-consistency term and the $\ell_1$-norm regularization term. Solving~\eqref{eq:eq_lasso_nonneg} to perform sparse recovery is a popular sparse-regression method known as the lasso in statistics~\cite{tibshirani1996regression} and basis pursuit in signal processing~\cite{chen2001atomic}. 

\begin{figure}[tp]
	\centering
\begin{tikzpicture}
\begin{semilogxaxis}[
width={0.4\textwidth}, 
height={0.25\textwidth}, 
xticklabel=\empty,
xmin=1e-1,
xmax=1.3,
ymax=0.7,
title={$\lambda=10^{-5}$},
title style = {yshift = -0.8cm},
ylabel = Estimate $\hat{c}$,
name = lambdam5,
]
\addplot[ycomb, ticks=none, very thick,blue] file {dat_files/lambda_reconstruction/snr100/lambda_-5.dat};
\addplot+[only marks, mark=x, mark size=5pt, very thick, red] file {dat_files/lambda_reconstruction/gt.dat};
\addplot[black, thick, samples=2] coordinates {(1e-1, 0)(3, 0)};
\end{semilogxaxis}

\begin{semilogxaxis}[
width=0.4\textwidth, 
height=0.25\textwidth, 
xmin=1e-1,xmax=1.3,
ymax=0.7,
xticklabel=\empty,
yticklabel=\empty,
title = {$\lambda=10^{-3}$},
title style = {yshift = -0.8cm},
name = lambdam3,
at=(lambdam5.right of north east),
anchor = left of north west,
]
\addplot+[only marks, mark=x, mark size=5pt, very thick, red] file {dat_files/lambda_reconstruction/gt.dat};
\addlegendimage{no markers,very thick,blue}
\addplot[ycomb, ticks=none, very thick,blue] file {dat_files/lambda_reconstruction/snr100/lambda_-3.dat};
\addplot[black, thick, samples=2] coordinates {(1e-1, 0)(3, 0)};
%\legend{Signal, Estimate}
\end{semilogxaxis}

\begin{semilogxaxis}[
width=0.4\textwidth, 
height=0.25\textwidth, 
xlabel = $T_1$ (s),
xmin=1e-1,xmax=1.3,
ymax=0.7,
title = {$\lambda=10^{-2}$},
title style = {yshift = -0.8cm},
ylabel = Estimate $\hat{c}$,
name = lambdam2,
at=(lambdam5.below south west),
anchor = above north west,
]
\addplot[ycomb, ticks=none, very thick,blue] file {dat_files/lambda_reconstruction/snr100/lambda_-2.dat};
\addplot+[only marks, mark=x, mark size=5pt, very thick, red] file {dat_files/lambda_reconstruction/gt.dat};
\addplot[black, thick, samples=2] coordinates {(1e-1, 0)(3, 0)};
\end{semilogxaxis}

\begin{semilogxaxis}[
width=0.4\textwidth, 
height=0.25\textwidth, 
xlabel = $T_1$ (s),
xmin=1e-1,xmax=1.3,ymax=0.7,
yticklabel=\empty,
title = {$\lambda=1$},
title style = {yshift = -0.8cm},
name = lambda0,
at=(lambdam3.below south west),
anchor = above north west,
legend pos=south west,
legend style={fill=white, fill opacity=0.6, draw opacity=1, text opacity=1},
legend style = {yshift = 0.5cm, xshift = -0.1cm},
]
\addplot+[only marks, mark=x, mark size=5pt, very thick, red] file {dat_files/lambda_reconstruction/gt.dat};
\addlegendimage{no markers,very thick,blue}
\addplot[ycomb, ticks=none, very thick,blue] file {dat_files/lambda_reconstruction/snr100/lambda_0.dat};
\addplot[black, thick, samples=2] coordinates {(1e-1, 0)(3, 0)};
\addplot[black, thick, samples=2] coordinates {(1e-1, 0)(3, 0)};
\legend{Ground truth, Estimate}
\end{semilogxaxis}

\end{tikzpicture} 
\caption{Estimates for the components of an MRF signal obtained by solving problem~\eqref{eq:eq_lasso_nonneg} for different values of the parameter $\lambda$ using a high-precision solver~\cite{grantcvx}. The data are generated by adding i.i.d. Gaussian noise to the two-compartment MRF signal shown in Figure~\ref{fig:thresholding} (top left). The signal-to-noise ratio (defined as the ratio between the $\ell_2$ norm of the signal and the noise) is equal to 100.}
\label{fig:l1_noise}
\end{figure}
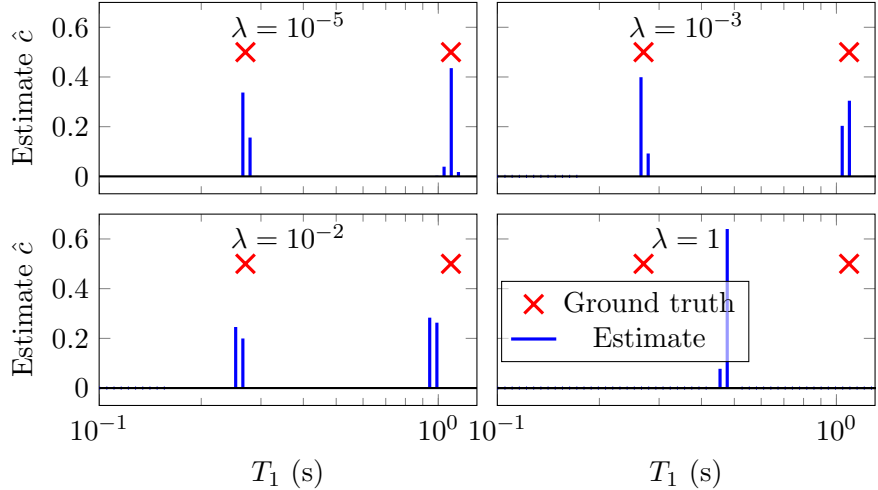

Figure~\ref{fig:l1_noise} shows the results of estimating the components of a noisy MRF signal by solving problem~\eqref{eq:eq_lasso_nonneg} for different values of the parameter $\lambda$. For large values of $\lambda$, the solution is sparse, but the fit to the data is not very accurate and the true support is not well approximated. For small values of $\lambda$, the solution to the convex program correctly locates the vicinity of the true coefficients, but it is contaminated by small spurious spikes. Reweighted-$\ell_1$-methods~\cite{candes2008enhancing,wipf2010iterative} are designed to enhance the performance of $\ell_1$-norm-regularized problems by promoting sparser solutions that are close to the initial estimate. In the case of the MRF dictionary, these methods are able to remove the spurious small spikes while retaining an accurate support estimate. This is achieved by solving a sequence of weighted $\ell_1$-norm regularized problems of the form 
\begin{equation}
\begin{aligned}
& \underset{\tilde{c} \, \in \, \R^{m}}{\text{minimize}}
& & \frac{1}{2} ||D\tilde{c}-x||^2 + \lambda \sum_{i=1}^{m} \omega_i^{\brac{k}} \tilde{c}_i \\
& \text{subject to}
& & \tilde{c} \geq 0
\end{aligned}
\label{eq:eq_lasso_nonneg_weighted}
\end{equation}
for $k=1,2,\ldots$. We denote the solution of this optimization problem by $\hat{c}^{(k)}$. The entries in the initial vector of weights $\omega^{\brac{1}}$ are initialized to one, so $\hat{c}^{(1)}$ is the solution to problem~\eqref{eq:eq_lasso_nonneg}. The weights are then updated depending on the subsequent solutions to problem~\eqref{eq:eq_lasso_nonneg_weighted}. 

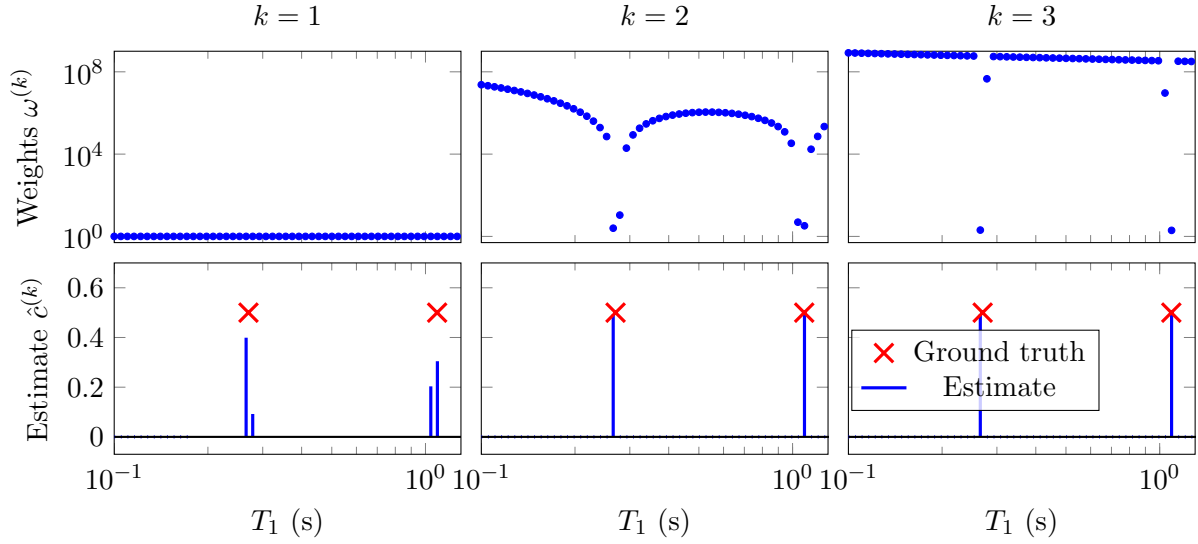
\begin{figure}[tp]
\begin{tikzpicture}
\begin{loglogaxis}[
width=0.375\textwidth, 
height=0.25\textwidth, 
xticklabel=\empty,
ylabel = {Weights $\omega^{\brac{k}}$},
xmin=1e-1,xmax=1.3,
ymax=1000000000,ymin=0.5,
name = k1_weights,
title = {$k = 1$},
]
\addplot[only marks, mark size=1.25pt, blue] file {dat_files/weight_map/w_init.dat};
\end{loglogaxis}

\begin{semilogxaxis}[
width=0.375\textwidth, 
height=0.25\textwidth, 
xlabel=$T_1$ (s),
ylabel = Estimate $\hat{c}^{\brac{k}}$,
xmin=1e-1,xmax=1.3,
ymax=0.7,
name = k1_est,
at=(k1_weights.below south west),
anchor=above north west,
]
\addplot[ycomb, ticks=none, very thick,blue] file {dat_files/weight_map/snr100/iter_candes_1.dat};
\addplot+[only marks, mark=x, mark size=5pt, very thick, red] file {dat_files/lambda_reconstruction/gt.dat};
\addplot[black, thick, samples=2] coordinates {(1e-1, 0)(3, 0)};
\end{semilogxaxis}

\begin{loglogaxis}[
width=0.375\textwidth, 
height=0.25\textwidth, 
xticklabel=\empty,
yticklabel=\empty,
xmin=1e-1,xmax=1.3,
ymax=1000000000,ymin=0.5,
name = k2_weights,
at=(k1_weights.right of north east),
anchor=left of north west,
title = {$k = 2$},
]
\addplot[only marks, mark size=1.25pt, blue] file {dat_files/weight_map/snr100/w_candes_1.dat};
% \addplot+[only marks, mark=x, mark size=5pt, very thick, red] file {dat_files/lambda_reconstruction/gt.dat};
% \addplot[black, thick, samples=2] coordinates {(1e-1, 0)(3, 0)};
\end{loglogaxis}

\begin{semilogxaxis}[
width=0.375\textwidth, 
height=0.25\textwidth, 
xlabel=$T_1$ (s),
yticklabel=\empty,
xmin=1e-1,xmax=1.3,
ymax=0.7,
name = k2_est,
at=(k2_weights.below south west),
anchor=above north west,
]
\addplot[ycomb, ticks=none, very thick,blue] file {dat_files/weight_map/snr100/iter_candes_2.dat};
\addplot+[only marks, mark=x, mark size=5pt, very thick, red] file {dat_files/lambda_reconstruction/gt.dat};
\addplot[black, thick, samples=2] coordinates {(1e-1, 0)(3, 0)};
\end{semilogxaxis}

\begin{loglogaxis}[
width=0.375\textwidth, 
height=0.25\textwidth, 
xticklabel=\empty,
yticklabel=\empty,
xmin=1e-1,xmax=1.3,
ymax=1000000000,ymin=0.5,
name = k3_weights,
at=(k2_weights.right of north east),
anchor=left of north west,
title = {$k = 3$},
]
\addplot[only marks,  mark size=1.25pt, blue] file {dat_files/weight_map/snr100/w_candes_2.dat};
% \addplot+[only marks, mark=x, mark size=5pt, very thick, red] file {dat_files/lambda_reconstruction/gt.dat};
% \addplot[black, thick, samples=2] coordinates {(1e-1, 0)(3, 0)};
\end{loglogaxis}

\begin{semilogxaxis}[
width=0.375\textwidth, 
height=0.25\textwidth, 
xlabel=$T_1$ (s),
yticklabel=\empty,
xmin=1e-1,xmax=1.3,
ymax=0.7,
name = k3_est,
at=(k3_weights.below south west),
anchor=above north west,
legend pos=south west,
legend style={fill=white, fill opacity=0.8, draw opacity=1, text opacity=1},
legend style = {yshift = 0.5cm, xshift = -0.1cm},
]
\addplot+[only marks, mark=x, mark size=5pt, very thick, red] file {dat_files/lambda_reconstruction/gt.dat};

\addlegendimage{no markers,very thick,blue}
\addplot[ycomb, ticks=none, very thick,blue] file {dat_files/weight_map/snr100/iter_candes_3.dat};

\addplot[black, thick, samples=2] coordinates {(1e-1, 0)(3, 0)};

\legend{Ground truth, Estimate}
\end{semilogxaxis}
\end{tikzpicture}
\caption{
Solutions to problem~\eqref{eq:eq_lasso_nonneg_weighted} (bottom) and corresponding weights (top) computed following the reweighting scheme~\eqref{eq:rwl1_candes}, with $\lambda:=10^{-3}$ and $\epsilon:=10^{-8}$. The data are the same as in Figure~\ref{fig:l1_noise}.
}
\label{fig:rwl1_candes}
\end{figure}

\begin{figure}[tp]
\begin{tikzpicture}
\begin{loglogaxis}[
width=0.375\textwidth, 
height=0.25\textwidth, 
xticklabel=\empty,
ylabel = {Weights $\omega^{\brac{k}}$},
xmin=1e-1,xmax=1.3,
ymax=4000,ymin=0.5,
name = k1_weights,
title = {$k = 1$},
]
\addplot[only marks, mark size=1.25pt, blue] file {dat_files/weight_map/w_init.dat};
% \addplot+[only marks, mark=x, mark size=5pt, very thick, red] file {dat_files/lambda_reconstruction/gt.dat};
% \addplot[black, thick, samples=2] coordinates {(1e-1, 0)(3, 0)};
\end{loglogaxis}

\begin{semilogxaxis}[
width=0.375\textwidth, 
height=0.25\textwidth, 
xlabel=$T_1$ (s),
ylabel = Estimate $\hat{c}^{\brac{k}}$,
xmin=1e-1,xmax=1.3,
ymax=0.7,
name = k1_est,
at=(k1_weights.below south west),
anchor=above north west,
]
\addplot[ycomb, ticks=none, very thick,blue] file {dat_files/weight_map/snr100/iter_sbl_1.dat};
\addplot+[only marks, mark=x, mark size=5pt, very thick, red] file {dat_files/lambda_reconstruction/gt.dat};
\addplot[black, thick, samples=2] coordinates {(1e-1, 0)(3, 0)};
\end{semilogxaxis}

\begin{loglogaxis}[
width=0.375\textwidth, 
height=0.25\textwidth, 
xticklabel=\empty,
yticklabel=\empty,
xmin=1e-1,xmax=1.3,
ymax=4000,ymin=0.5,
name = k2_weights,
at=(k1_weights.right of north east),
anchor=left of north west,
title = {$k = 2$},
]
\addplot[only marks, mark size=1.25pt, blue] file {dat_files/weight_map/snr100/w_sbl_1.dat};
% \addplot+[only marks, mark=x, mark size=5pt, very thick, red] file {dat_files/lambda_reconstruction/gt.dat};
% \addplot[black, thick, samples=2] coordinates {(1e-1, 0)(3, 0)};
\end{loglogaxis}

\begin{semilogxaxis}[
width=0.375\textwidth, 
height=0.25\textwidth, 
xlabel=$T_1$ (s),
yticklabel=\empty,
xmin=1e-1,xmax=1.3,
ymax=0.7,
name = k2_est,
at=(k2_weights.below south west),
anchor=above north west,
]
\addplot[ycomb, ticks=none, very thick,blue] file {dat_files/weight_map/snr100/iter_sbl_2.dat};
\addplot+[only marks, mark=x, mark size=5pt, very thick, red] file {dat_files/lambda_reconstruction/gt.dat};
\addplot[black, thick, samples=2] coordinates {(1e-1, 0)(3, 0)};
\end{semilogxaxis}

\begin{loglogaxis}[
width=0.375\textwidth, 
height=0.25\textwidth, 
xticklabel=\empty,
yticklabel=\empty,
xmin=1e-1,xmax=1.3,
ymax=4000,ymin=0.5,
name = k3_weights,
at=(k2_weights.right of north east),
anchor=left of north west,
title = {$k = 3$},
]
\addplot[only marks, mark size=1.25pt, blue] file {dat_files/weight_map/snr100/w_sbl_2.dat};
% \addplot+[only marks, mark=x, mark size=5pt, very thick, red] file {dat_files/lambda_reconstruction/gt.dat};
% \addplot[black, thick, samples=2] coordinates {(1e-1, 0)(3, 0)};
\end{loglogaxis}

\begin{semilogxaxis}[
width=0.375\textwidth, 
height=0.25\textwidth, 
xlabel=$T_1$ (s),
yticklabel=\empty,
xmin=1e-1,xmax=1.3,
ymax=0.7,
name = k3_est,
at=(k3_weights.below south west),
anchor=above north west,
legend pos=south west,
legend style={fill=white, fill opacity=0.8, draw opacity=1, text opacity=1},
legend style = {yshift = 0.5cm, xshift = -0.1cm},
]
\addplot+[only marks, mark=x, mark size=5pt, very thick, red] file {dat_files/lambda_reconstruction/gt.dat};

\addlegendimage{no markers,very thick,blue}
\addplot[ycomb, ticks=none, very thick,blue] file {dat_files/weight_map/snr100/iter_sbl_3.dat};

\addplot[black, thick, samples=2] coordinates {(1e-1, 0)(3, 0)};
\legend{Ground truth, Estimate}
\end{semilogxaxis}
\end{tikzpicture} 
%\begin{comment}
%\\
% 6 &
%\begin{tikzpicture}[scale=0.62]
%\begin{axis}[width=0.7\textwidth, height=0.4\textwidth, xlabel=Dictionary indices (indexed by $T_1$ values),xmin=1e-1,xmax=1.3,ymax=0.7]%, xlabel shift = 5 pt]
%\addplot[ycomb, ticks=none, very thick,blue] file {dat_files/weight_map/iter_sbl_6.dat};
%\addplot+[only marks, mark=x, mark size=5pt, very thick, red] file {dat_files/lambda_reconstruction/gt.dat};
%\addplot[black, thick, samples=2] coordinates {(1e-1, 0)(3, 0)};
%\end{axis}
%\end{tikzpicture} 
%&
%\begin{tikzpicture}[scale=0.62]
%\begin{semilogyaxis}[width=0.7\textwidth, height=0.4\textwidth, xlabel=Dictionary indices (indexed by $T_1$ values),xmin=1e-1,xmax=1.3,ymax=10000]%, xlabel shift = 5 pt]
%\addplot[only marks, thick, blue] file {dat_files/weight_map/w_sbl_5.dat};
%% \addplot+[only marks, mark=x, mark size=5pt, very thick, red] file {dat_files/lambda_reconstruction/gt.dat};
%% \addplot[black, thick, samples=2] coordinates {(1e-1, 0)(3, 0)};
%\end{semilogyaxis}
%\end{tikzpicture}
%\end{comment}
\caption{
Solutions to problem~\eqref{eq:eq_lasso_nonneg_weighted} (bottom) and corresponding weights (top) computed following the reweighting scheme~\eqref{eq:rwl1_wipf}, with $\lambda:=10^{-3}$ and $\epsilon:=10^{-8}$. The data are the same as in Figure~\ref{fig:l1_noise}.
}
\label{fig:rwl1_wipf}
\end{figure}
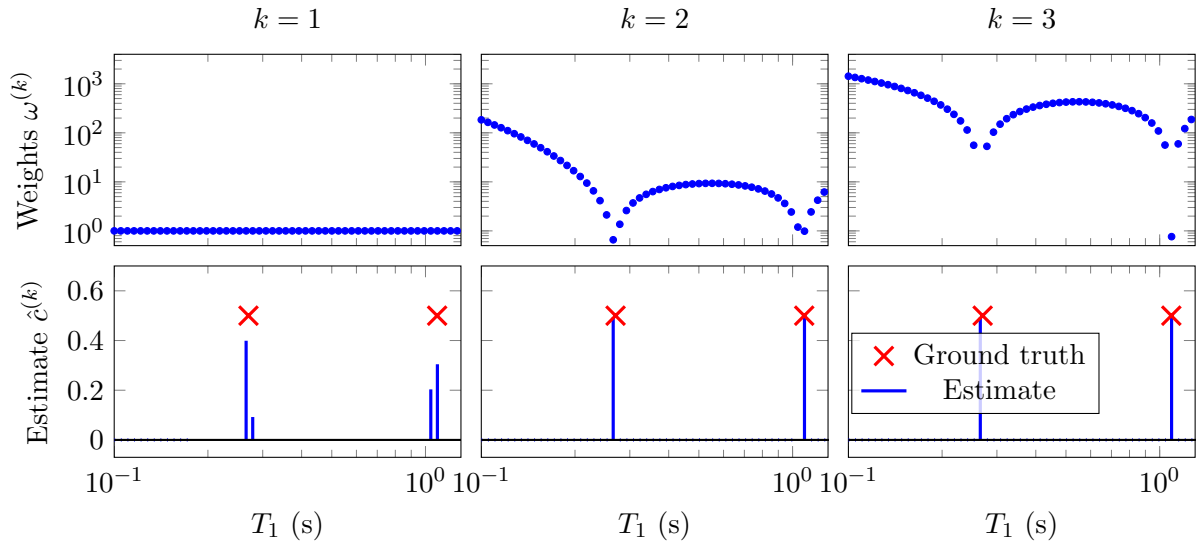

We consider two different reweighting schemes. The first, proposed in~\cite{candes2008enhancing}, simply sets the weights to be inversely proportional to the previous estimate $\hat{c}^{(k-1)}$
\begin{align}
\omega_i^{\brac{k}} := \frac{1}{\epsilon + \abs{\hat{c}^{(k-1)}}}, \quad 1\leq i \leq m,
\label{eq:rwl1_candes}
\end{align}
where $\epsilon$ is a parameter typically fixed to a very small value with respect to the expected magnitude of the nonzero coefficients (e.g. $10^{-8}$) in order to preclude division by zero. Intuitively, the weights penalize the coefficients that do not contribute to the fit in previous iterations, promoting increasingly sparse solutions. Mathematically, the algorithm can be interpreted as a majorization-minimization method for a non-convex sparsity-inducing penalty function~\cite{candes2008enhancing}. The second updating scheme, proposed in~\cite{wipf2010iterative} and inspired by sparse Bayesian learning methods~\cite{wipf2004sparse}, is of the form
\begin{align}
 \omega_i^{\brac{k}} & := \brac{ D_i^T \left( \epsilon I + DC^{-1}WD^T \right)^{-1}D_i }^{\frac{1}{2}}, \label{eq:rwl1_wipf} \\
 C & := \op{diag}\brac{\hat{c}^{(k-1)}}, \\
 W & := \op{diag}\brac{\omega^{(k-1)}},
% \label{eq:sbl}
\end{align}
where $\epsilon$ is a parameter which precludes the matrix $\epsilon I + DX^{-1}WD^T $ from being singular. The reweighting scheme sets $\omega_i^{\brac{k}}$ to be small if there is any large entry $\hat{c}^{(k-1)}_j$ such that the corresponding columns of the dictionary $D_i$ and $D_j$ are correlated. This produces smoother weights than~\eqref{eq:rwl1_candes} providing more robustness to initial errors in the support estimate.  

Figures~\ref{fig:rwl1_candes} and~\ref{fig:rwl1_wipf} show the results obtained by applying reweighted-$\ell_1$-norm regularization to the same problem as in Figure~\ref{fig:l1_noise}. As expected, the update~\eqref{eq:rwl1_wipf} yields smoother weights. Convergence to a sparse solution is achieved in just two iterations. The estimate is an accurate estimate of the parameters despite the presence of noise and the gridding error (the true parameters do not lie on the grid used to construct the dictionary). 

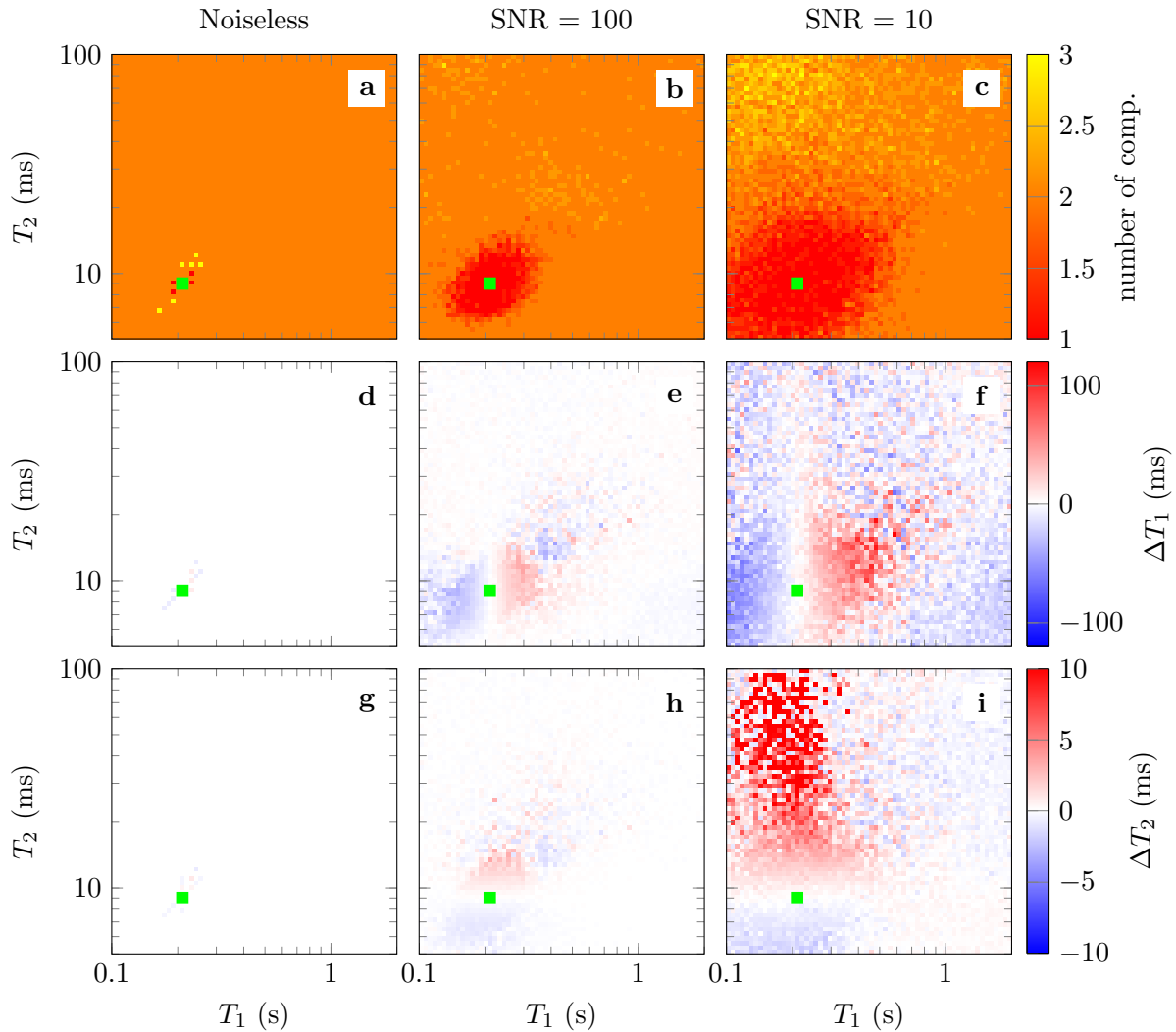
\begin{figure}
\def\s{110}

\centering
\begin{tikzpicture}
    \begin{loglogaxis}[width=\s,
                height=\s,
                axis on top,
                scale only axis,
                log ticks with fixed point,
                xmin=0.1,
                xmax=2,
                ymin=5,
                ymax=100,
                xticklabels = {},
                ylabel = {$T_2$~(ms)},
                name = natnoiseless,
                title={Noiseless},
              ]
        \addplot graphics[xmin=0.1,xmax=2,ymin=5,ymax=100] {asy_pictures/cluster/noiseless/nat.pdf};
        \addplot[mark=square*, only marks, mark options={scale=1pt, fill=green, draw=green, thick}] coordinates {(0.21, 9)};
        
        \node[right, inner sep=0mm, minimum size=5mm, fill=white, text=black] at (axis cs:1.2,70) {\textbf{a}};
    \end{loglogaxis}
    
    \begin{loglogaxis}[width=\s,
                height=\s,
                axis on top,
                scale only axis,
                log ticks with fixed point,
                xmin=0.1,
                xmax=2,
                ymin=5,
                ymax=100,
                xticklabels = {},
                yticklabels = {},
                name = nat100,
                at = (natnoiseless.south east),
                anchor = south west,
                xshift = 0.3cm,
                title={SNR = 100},
              ]
        \addplot graphics[xmin=0.1,xmax=2,ymin=5,ymax=100] {asy_pictures/cluster/snr100/nat.pdf};
        \addplot[mark=square*, only marks, mark options={scale=1pt, fill=green, draw=green, thick}] coordinates {(0.21, 9)};
        
        \node[right, inner sep=0mm, minimum size=5mm, fill=white, text=black] at (axis cs:1.2,70) {\textbf{b}};
    \end{loglogaxis}
    
    \begin{loglogaxis}[width=\s,
                height=\s,
                axis on top,
                scale only axis,
                log ticks with fixed point,
                xmin=0.1,
                xmax=2,
                ymin=5,
                ymax=100,
                xticklabels = {},
                yticklabels = {},
                title={SNR = 10},
                point meta min=1.,
                point meta max= 3.,
                name = nat10,
                at = (nat100.south east),
                anchor = south west,
                xshift = 0.3cm,
                colorbar,
                colormap/redyellow,
%                colorbar sampled={surf,shader=interp},
                colorbar style={ylabel= {number of comp.}, width=0.3cm, xshift=-0.1cm},
              ]
        \addplot graphics[xmin=0.1,xmax=2,ymin=5,ymax=100] {asy_pictures/cluster/snr10/nat.pdf};
       \addplot[mark=square*, only marks, mark options={scale=1pt, fill=green, draw=green, thick}] coordinates {(0.21, 9)};
        
        \node[right, inner sep=0mm, minimum size=5mm, fill=white, text=black] at (axis cs:1.2,70) {\textbf{c}};
    \end{loglogaxis}
    
    \begin{loglogaxis}[width=\s,
                height=\s,
                axis on top,
                scale only axis,
                log ticks with fixed point,
                xmin=0.1,
                xmax=2,
                ymin=5,
                ymax=100,
                ylabel = {$T_2$~(ms)},
                xticklabels = {},
                name = t1noiseless,
                at = (natnoiseless.south west),
                anchor = north west,
                yshift = -0.3cm,
              ]
        \addplot graphics[xmin=0.1,xmax=2,ymin=5,ymax=100] {asy_pictures/cluster/noiseless/t1.pdf};
       \addplot[mark=square*, only marks, mark options={scale=1pt, fill=green, draw=green, thick}] coordinates {(0.21, 9)};
        
        \node[right, inner sep=0mm, minimum size=5mm, fill=white, text=black] at (axis cs:1.2,70) {\textbf{d}};
    \end{loglogaxis}
    
    \begin{loglogaxis}[width=\s,
                height=\s,
                axis on top,
                scale only axis,
                log ticks with fixed point,
                xmin=0.1,
                xmax=2,
                ymin=5,
                ymax=100,
                xticklabels = {},
                yticklabels = {},
                name = t1100,
                at = (t1noiseless.south east),
                anchor = south west,
                xshift = 0.3cm,
              ]
        \addplot graphics[xmin=0.1,xmax=2,ymin=5,ymax=100] {asy_pictures/cluster/snr100/t1.pdf};
       \addplot[mark=square*, only marks, mark options={scale=1pt, fill=green, draw=green, thick}] coordinates {(0.21, 9)};
        
        \node[right, inner sep=0mm, minimum size=5mm, fill=white, text=black] at (axis cs:1.2,70) {\textbf{e}};
    \end{loglogaxis}
    
    \begin{loglogaxis}[width=\s,
                height=\s,
                axis on top,
                scale only axis,
                log ticks with fixed point,
                xmin=0.1,
                xmax=2,
                ymin=5,
                ymax=100,
                xticklabels = {},
                yticklabels = {},
                name = t110,
                at = (t1100.south east),
                anchor = south west,
                xshift = 0.3cm,
                colorbar,
                colorbar sampled={surf,shader=interp},
                point meta min=-120,
                point meta max= 120,
                colormap name = bwr,
                colorbar style={ylabel=$\Delta T_1$~(ms), width=0.3cm, xshift=-0.1cm},
              ]
        \addplot graphics[xmin=0.1,xmax=2,ymin=5,ymax=100] {asy_pictures/cluster/snr10/t1.pdf};
       \addplot[mark=square*, only marks, mark options={scale=1pt, fill=green, draw=green, thick}] coordinates {(0.21, 9)};
        
        \node[right, inner sep=0mm, minimum size=5mm, fill=white, text=black] at (axis cs:1.2,70) {\textbf{f}};
    \end{loglogaxis}
    
    \begin{loglogaxis}[width=\s,
                height=\s,
                axis on top,
                scale only axis,
                log ticks with fixed point,
                xmin=0.1,
                xmax=2,
                ymin=5,
                ymax=100,
                xlabel = {$T_1$~(s)},
                ylabel = {$T_2$~(ms)},
                name = t2noiseless,
                at = (t1noiseless.south west),
                anchor = north west,
                yshift = -0.3cm,
              ]
        \addplot graphics[xmin=0.1,xmax=2,ymin=5,ymax=100] {asy_pictures/cluster/noiseless/t2.pdf};
      \addplot[mark=square*, only marks, mark options={scale=1pt, fill=green, draw=green, thick}] coordinates {(0.21, 9)};
        
        \node[right, inner sep=0mm, minimum size=5mm, fill=white, text=black] at (axis cs:1.2,70) {\textbf{g}};
    \end{loglogaxis}
    
    \begin{loglogaxis}[width=\s,
                height=\s,
                axis on top,
                scale only axis,
                log ticks with fixed point,
                xmin=0.1,
                xmax=2,
                ymin=5,
                ymax=100,
                xlabel = {$T_1$~(s)},
                yticklabels = {},
                name = t2100,
                at = (t2noiseless.south east),
                anchor = south west,
                xshift = 0.3cm,
              ]
        \addplot graphics[xmin=0.1,xmax=2,ymin=5,ymax=100] {asy_pictures/cluster/snr100/t2.pdf};
      \addplot[mark=square*, only marks, mark options={scale=1pt, fill=green, draw=green, thick}] coordinates {(0.21, 9)};
        
        \node[right, inner sep=0mm, minimum size=5mm, fill=white, text=black] at (axis cs:1.2,70) {\textbf{h}};
    \end{loglogaxis}
    
    \begin{loglogaxis}[width=\s,
                height=\s,
                axis on top,
                scale only axis,
                log ticks with fixed point,
                xmin=0.1,
                xmax=2,
                ymin=5,
                ymax=100,
                xlabel = {$T_1$~(s)},
                yticklabels = {},
                name = t2_10,
                colorbar,
                colormap name = bwr,
                colorbar sampled={surf,shader=interp},
                point meta min=-10,
                point meta max= 10,
                name = t210,
                at = (t2100.south east),
                anchor = south west,
                xshift = 0.3cm,
                colorbar style={ylabel=$\Delta T_2$~(ms), width=0.3cm, xshift=-0.1cm},
              ]
              
        \addplot graphics[xmin=0.1,xmax=2,ymin=5,ymax=100] {asy_pictures/cluster/snr10/t2.pdf};
        \addplot[mark=square*, only marks, mark options={scale=1pt, fill=green, draw=green, thick}] coordinates {(0.21, 9)};
        
        \node[right, inner sep=0mm, minimum size=5mm, fill=white, text=black] at (axis cs:1.2,70) {\textbf{i}};
    \end{loglogaxis}
\end{tikzpicture}
\caption{This figure depicts the results of reconstructing one compartment with $T_1 = 0.21$~s and $T_2 = 9$~ms (marked by the green square), in the presence of a second compartment with $T_1$ values between $0.1$~s and $2.0$~s and $T_2$ values between $5$~ms and $100$~ms (the proton density of each tissue is 0.5). The fingerprints were generated for a single-voxel without k-space undersampling and following Refs.~\cite{asslander2016pseudo,Asslander2017arxiv}. I.i.d. Gaussian noise is added to the simulated data. The reconstruction was performed with the reweighting scheme defined in Eq.~\eqref{eq:rwl1_wipf}, where the weighted $\ell_1$-norm minimization problems are solved using the method described in Section~\ref{sec:interior_point}. The first row depicts the number of compartments resulting from the reconstruction (the ground truth is two). The second and third row show the $T_1$ and $T_2$ errors, respectively. These errors correspond to the signed distance between the relaxation times of the fixed compartment ($T_1 = 0.21$~s and $T_2 = 9$) and the closest relaxation time of the reconstructed atoms. The results are averaged over 5 repetitions with different noise realizations.}
\label{fig:noiseless}
\end{figure}

In order to evaluate the performance of the sparse recovery algorithm, we carried out an experiment with data simulated from a voxel containing a 50\%-50\% mix of two compartments. The parameters of the first compartment are fixed ($T_1 = 0.21$~s, $T_2 = 9$~ms), whereas the parameters of the second compartment vary on a fine grid ($T_1$ between $0.1$~s and $2.0$~s and $T_2$ between $5$~ms and $100$~ms). The data were generated by adding the fingerprints from the two compartments and perturbing the result with i.i.d. Gaussian noise. Figure~\ref{fig:noiseless} shows the result of fitting the multicompartment model using the reweighting scheme defined in Eq.~\eqref{eq:rwl1_wipf}, as the parameters of the second compartment vary. In the noiseless case, exact recovery occurs (even without reweighting) except for very close values. For the noisy data, the algorithm tends to detect the correct number of components as long as the parameters of the two components are not too similar, i.e. they are sufficiently separated in parameter space (b, c). The accuracy of the parameter estimate then decreases as the noise level increases (e, f, h, i), resulting in some cases in the appearance of a spurious third compartment (b, c). When the parameters of both compartments are very close in parameter space, the algorithm tends to estimate a single compartment. This occurs in the red region around the green square in subfigure b. Increasing the noise increases the size of the region (c). Within it, one can observe that the reconstructed compartment lies between the two original compartments, as indicated by the slope in the error (e, f, h, i). Areas with $T_1 \approx T_2$ seem to be subject to a systematic overestimation of the number of compartments, as well as an overestimation of $T_2$ (c, i), while the error in the relaxation times is less structured in the rest of the parameter space (e, f, h, i). 

%Figure~\ref{fig:noiseless} shows the results of a more exhaustive computational experiment evaluating the performance of the method. We fix $x$ to consist of two fingerprints, one fixed to a particular value (represented by a cross on the images) and the other varying over different values of $T_1$ and $T_2$. For each of these values, the figure shows the number of compartments recovered by our method, as well as the relative error in the $T_1$ and $T_2$ estimates. When no noise is added to the data, recovery is very accurate except when the two fingerprints are very close to each other in $T_1$-$T_2$ space. As the noise level is increased, the error degrades gracefully, deteriorating more when the two fingerprints are closer in $T_1$-$T_2$ space and hence more similar. 

\subsection{Fast interior-point solver for $\ell_1$-norm regularization in coherent dictionaries}
\label{sec:interior_point}
In this section we describe an interior-point solver for the $\ell_1$-norm regularized problem
\begin{equation}
\begin{aligned}
& \underset{\tilde{c} \, \in \, \R^{m}}{\text{minimize}}
& & \frac{1}{2} ||D\tilde{c}-x||^2 + \lambda \sum_{i=1}^{m} \tilde{c}_i \\
& \text{subject to}
& & \tilde{c} \geq 0,
\end{aligned}
\label{eq:IPL1}
\end{equation}
which allows us to apply the reweighting schemes described in Section~\ref{sec:reweightedl1} efficiently. Interior-point methods enforce inequality constraints by using a barrier function~\cite{boyd2004convex,nocedal2006numerical,wright1997primal}. In this case we use a logarithmic function that forces the coefficients to be nonnegative. Parametrizing the modified optimization problem with $t>0$, we obtain a sequence of cost functions of the form
\begin{equation}
\begin{aligned}
%& \underset{\tilde{c} \, \in \, \R^{m}}{\text{minimize}}
%& & 
\phi_t(\tilde{c}) := \frac{1}{2} t ||D\tilde{c}-x||^2 + t\lambda \sum_{i=1}^n{\tilde{c}_i} -\sum_{i=1}^n \log \tilde{c}_i.
\end{aligned}
\label{eq:eq_logbarrier_objfun}
\end{equation}
The \emph{central path} of the interior-point scheme consists of the minimizers of $\phi_t$ as $t$ varies from $0$ to $\infty$. To find a solution for problem~\ref{eq:IPL1}, we find a sequence of points $c^{(1)}, c^{(2)}, \ldots$ in the central path by iteratively solving the Newton system 
\begin{align}\label{eq:Newton}
%\begin{bmatrix} 
    \brac{Z+2tD^TD} \brac{ c^{(k)} - c^{(k-1)} } & = - \nabla \phi_t \brac{c^{(k-1)} } 
\end{align}
where   
\begin{align}
Z & := \text{diag}\left(\frac{1}{\tilde{c}_1^2},...,\frac{1}{\tilde{c}_m^2}\right). %, \\
% g & := - \nabla_{\tilde{c}}\phi_t(\tilde{c}).
\end{align}
This is essentially the approach taken in~\cite{candes2005l1,kim2007interior} to tackle $\ell_1$-norm regularized least squares. Solving~\eqref{eq:Newton} is the computational bottleneck of this method. In the case of the MRF dictionary, $D^TD$ is a matrix of rank $k$, where $k$ is the order of the low-rank approximation to the dictionary described in Section~\ref{sec:compression}. As a result, solving the Newton system~\eqref{eq:Newton} directly is extremely slow for large values of $t$. Using a preconditioner of the form 
\begin{align}
P :=     2t\op{diag}(D^TD)+Z \label{eq:precond}
\end{align}
combined with an iterative conjugate-gradient method to invert the system, as suggested in~\cite{kim2007interior}, does not alleviate this problem due to the ill-conditioning of the matrix as $t$ grows large. 

\begin{figure}[tp]
\begin{center}
\begin{tikzpicture}
\begin{semilogyaxis}[width=0.7\textwidth, height=0.4\textwidth,xlabel=Number of fingerprints,ylabel=Computation time (s),legend pos=north west,xmode=log]
\addplot[mark=*,thick,violet,dashed,mark options=solid] file {dat_files/time_computation/cvx2.dat};
\addplot[mark=*,thick,red,dashed,mark options=solid] file {dat_files/time_computation/boyd2.dat};
\addplot[mark=*,thick,blue,dashed,mark options=solid] file {dat_files/time_computation/l1ls2.dat};
\legend{General purpose,Preconditioned CG,Woodbury inversion}
\end{semilogyaxis}
\end{tikzpicture}
\end{center}
\caption{Computation times of the proposed Woodbury-inversion method compared to a general-purpose solver~\cite{grantcvx} and an approach based on preconditioned conjugate-gradients (CG)~\cite{kim2007interior}. MRF dictionaries containing different numbers of fingerprints in the same range ($T_1$ values from $0.1$ s to $2$ s, $T_2$ values from $0.005$ s to $0.1$ s) were generated using the approach described in Refs.~\cite{asslander2016pseudo,Asslander2017arxiv}. Each method was applied to solve 10 instances of problem~\eqref{eq:IPL1} for each dictionary size on a computer cluster (Four AMD Opteron 6136, each with 32 cores, 2.4 GHz, 128 GB of RAM).}
% to solve \ref{eq:IPL1} with three different algorithms, using different size for the dictionary. The signal used here is a mix of Water ($T_1 = 0.21$s, $T_2 = 0.009$s) and Grey Matter ($T_1 = 0.91$s, $T_2=0.062$s) on top of which a gaussian noise of SNR $100$ is added. Each dictionary has a range of $T_1$ value going from $0.1$s to $2$s and $T_2$ value from $0.005$s to $0.1$s. The distance between two successive elements in these lists is log-constant for each dictionary, the multiplication factor being from $1.03$ to $1.2$ in a range of $18$ values linearly equally spaced. The number of atoms therefore ranges from 324 to 10609. The three algorithms are \textit{cvx}, the Preconditioned algorithm uses \cite{kim2007interior} and the Woodburry inversion method is described in section \ref{sec:interior_point}. This experiment has been run $10$ times successivly on a cluster whose specs are Four AMD Opteron 6136 (2.4 GHz) (32 cores) 128 GB of Memory}
\label{fig:computation_times}
\end{figure}
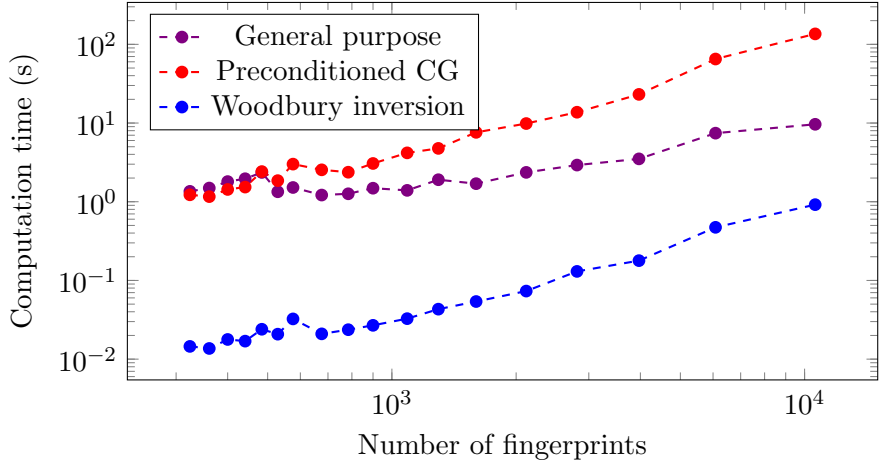

In order to solve the Newton system~\eqref{eq:Newton} efficiently, we take a different route: exploiting the low-rank structure of the Hessian by applying the Woodbury inversion lemma, also known as Sherman-Morrison-Woodbury formula.

\begin{lemma}[Woodbury inversion lemma] 
Assume $A \in \R^{n\times n}$, $B, G \in \R^{n\times r}$ and  $L \in \R^{r\times r}$. Then, 
\begin{equation}
(A+BLG^T)^{-1} = A^{-1} -A^{-1}B(L^{-1}+GA^{-1}B)^{-1}GA^{-1}
\end{equation}
\end{lemma}
Setting $A:=Z$, $B:=D$, $G:=D^T$ and $L := I \in \R^{k}$ in the lemma yields the following expression for the inverse of the system matrix 
\begin{equation}
(Z+2tD^TD)^{-1} = Z^{-1} -2tZ^{-1}D^T(I^{-1}+2tDZ^{-1}D^T)^{-1}DZ^{-1}.
\end{equation}
Since in practice $k$ is very small (typically between 10 and 15) due to the dictionary-compression scheme described in Section~\ref{sec:compression} and $Z$ is diagonal, using this formula accelerates the inversion of the Newton system dramatically.
%the complexity of solving system is reduced to $\ml{O}(m)$ as opposed to $\ml{O}(m^2)$. 

Figure~\ref{fig:computation_times} compares the computational cost of our method based on Woodbury-inversion with a general-purpose convex-programming solver based on an interior-point method~\cite{grantcvx} and an interior-point method tailored to large-scale problems, which applies the preconditioner in Eq.~\eqref{eq:precond} and uses an iterative method to solve the Newton system~\cite{kim2007interior}. Our method is almost an order of magnitude faster for dictionaries containing up to $10^4$ columns.  

To conclude the section, we note that this algorithm can be directly applied to weighted $\ell_1$-norm-regularization least squares problems of the form 
\begin{equation}
\begin{aligned}
& \underset{\tilde{c} \, \in \, \R^{m}}{\text{minimize}}
& & \frac{1}{2} ||D\tilde{c}-x||^2 + \lambda \sum_{i=1}^{m} {W\tilde{c}_i} \\
& \text{subject to}
& & \tilde{c} \geq 0,
\end{aligned}
\end{equation}
where $W$ is a diagonal matrix containing the weights. We just need to apply the change of variable $\tilde{c}'=W\tilde{c}$ and $D' = DW^{-1}$. Since the weights are all positive, the constraint $\tilde{c} \geq 0$ is equivalent to $\tilde{c}' \geq 0$. 

%\begin{table}
%\label{PCGconv}
%\begin{center}
%\begin{tabular}{ | c | c | c |}
%\hline
%Interior-Point Iteration No. & Condition Number & PCG Iteration to Converge \\ \hline
%%1 & 2.224081e-08 & 4 \\ \hline
%%2 & 1.108188e-08  & 17 \\ \hline
%%3 & 2.222386e-08  & 16 \\ \hline
%4 & 4.444851e-08  & 16 \\ \hline
%%5 & 8.889299e-08 & 16 \\ \hline
%%6 & piter = 16, rcond = 1.777493e-07
%7 & 3.544270e-07 & 16 \\ \hline
%%8 & piter = 19, rcond = 5.195380e-07
%%9 & piter = 18, rcond = 1.508823e-06
%10 & 3.360125e-06 & 17 \\ \hline
%%11 & piter = 20, rcond = 4.727376e-06
%%12 & piter = 21, rcond = 4.552198e-06
%13 & 2.884776e-06 & 30  \\ \hline
%%14 & piter = 53, rcond = 1.120963e-06
%% 15 & 4.114272e-07& 103  \\ \hline
%16 & 8.423252e-08 & 234 \\ \hline
%%17 & piter = 574, rcond = 1.322897e-08
%%18 & piter = 1698, rcond = 1.266721e-09
%19 & 1.104555e-10 & 4959 \\ \hline
%%20 & 3.435080e-11& 4959  \\ \hline
%%21 & piter = 4800, rcond = 1.288498e-11
%22 & 1.023070e-11 & 4167 \\ \hline
%%23 & piter = 4274, rcond = 5.569065e-12
%%24 & piter = 4954, rcond = 5.834858e-12
%25 & 4.036563e-12 & 4269  \\ 
%%26 & piter = 4359, rcond = 3.599937e-12
%
%    \hline
%\end{tabular}
%\caption{The Condition number and PCG iterations required to converge to suboptimal duality gap as the interior point iteration proceeds. Hessian Matrix $H \in \mathcal{R}^{5980 \times 5980}$.}
%\end{center}
%\label{eq:table}
%\end{table}

\section{Results}
\label{sec:numerical}

\subsection{Simulations}
\label{sec:simulated_phantom}
\begin{figure}[tp]
\begin{center}
{ \footnotesize
  \begin{tabular}{ |c | c | c | c | c | }
       \hline tissue & A & B & C & D \\ \hline
%    $T_1 (s)$ & 0.2133 & 0.6579 & 0.9066  & 2.3163 \\ \hline
%    $T_2 (s)$ & $0.0090$ &  $0.0421$ & $0.0623$  &0.4157 \\ \hline
    $T_1 (s)$ & $0.21$ & $0.66$ & $0.91$  & $2.32$ \\ \hline
    $T_2 (s)$ & $0.009$ &  $0.042$ & $0.062$  & $0.416$ \\ \hline
  \end{tabular} \vspace{0.5cm}\\
    \begin{tabular}{ |c | c | c | c | c | }
       \hline&Column 1 & Column 2 & Column 3 & Column 4 \\ \hline
    Row 1 & $100\%$ A & $50 \%$ A + $50 \%$ B & $50 \%$ A + $50 \%$ C  & $50 \%$ A + $50 \%$ D \\ \hline
    Row 2 & $50 \%$ B + $50 \%$ A &  $100\%$ B & $50 \%$ B + $50 \%$ C  & $50 \%$ B + $50 \%$ D \\ \hline
    Row 3 & $50 \%$ C + $50 \%$ A  & $50 \%$ C + $50 \%$ B  & $100\%$ C & $50 \%$ C + $50 \%$ D \\ \hline
    Row 4 & $50 \%$ D + $50 \%$ A & $50 \%$ D + $50 \%$ B & $50 \%$ D + $50 \%$ C  & $100\%$ D \\
    \hline
  \end{tabular}
}
\end{center}
\centering
\begin{tabular}{ >{\centering\arraybackslash}m{0.17\linewidth} >{\centering\arraybackslash}m{0.17\linewidth}  >{\centering\arraybackslash}m{0.08\linewidth} >{\centering\arraybackslash}m{0.01\linewidth}}
1\textsuperscript{st} compart.
&
2\textsuperscript{nd} compart.
\\
\includegraphics[width = \heatSubPlotWidthSmall cm]{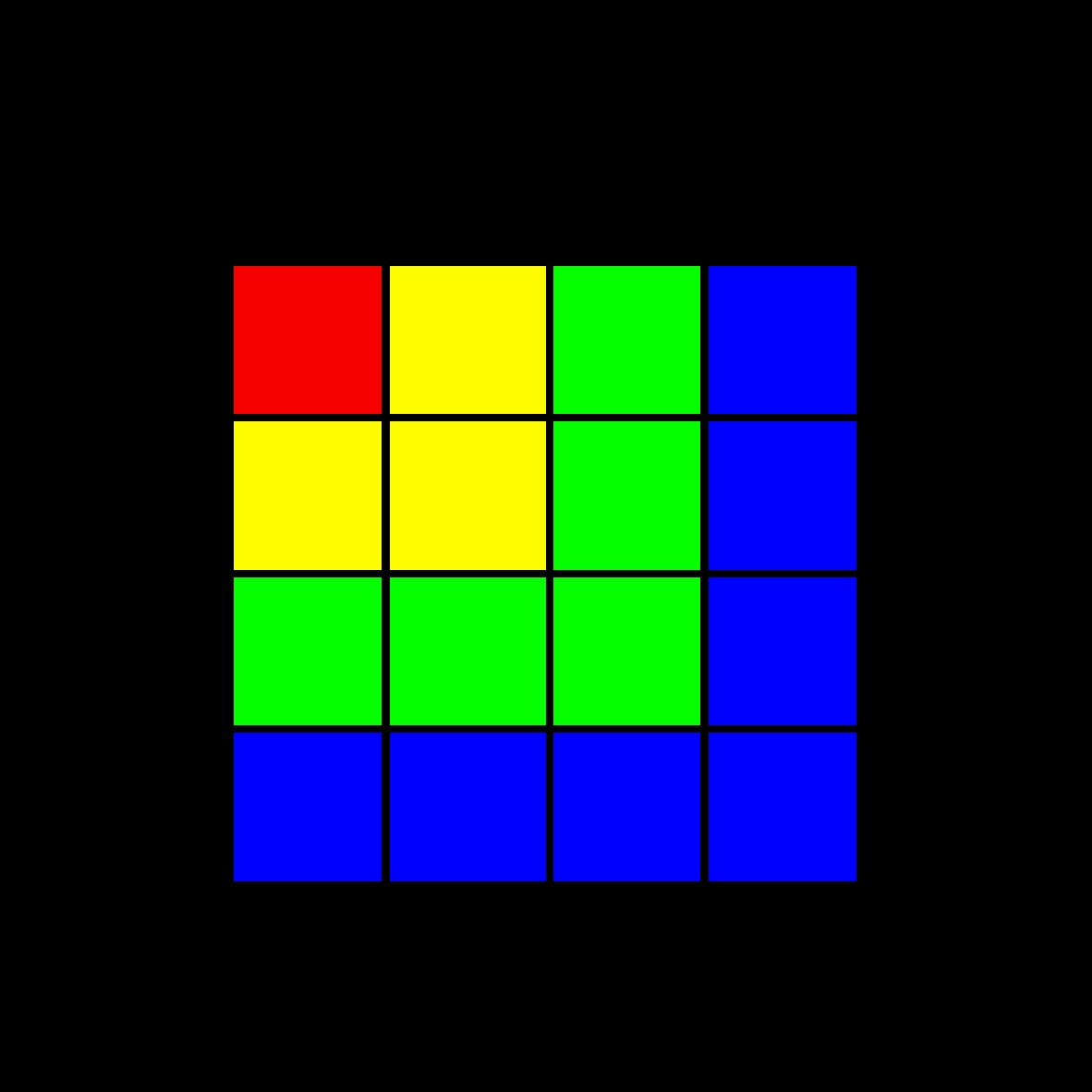}
&
\includegraphics[width = \heatSubPlotWidthSmall cm]{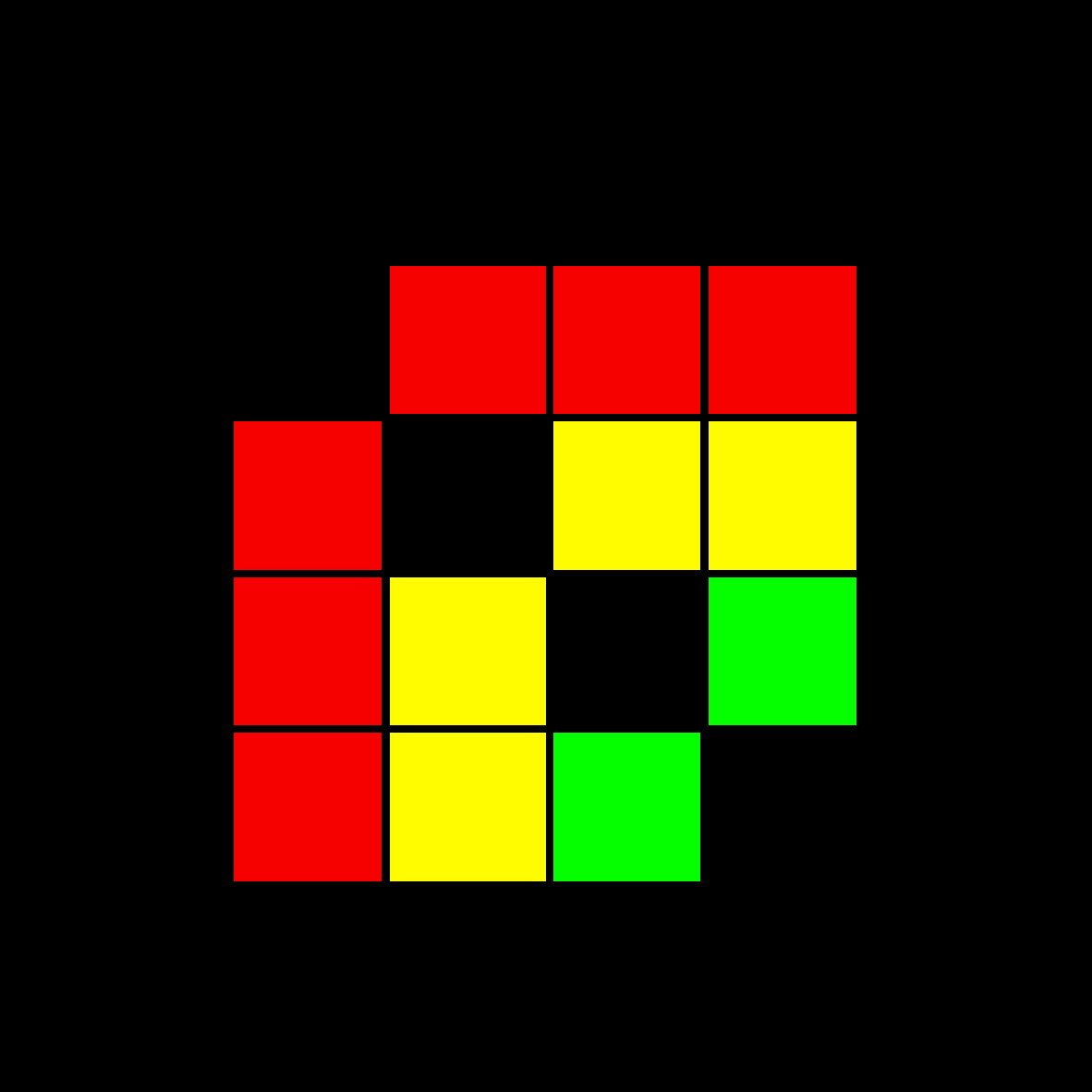}
&
\includegraphics[width = 1.5 cm]{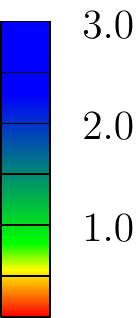}
&
\rotatebox[origin=c]{90}{$T_1$~(s)}
\\
\includegraphics[width = \heatSubPlotWidthSmall cm]{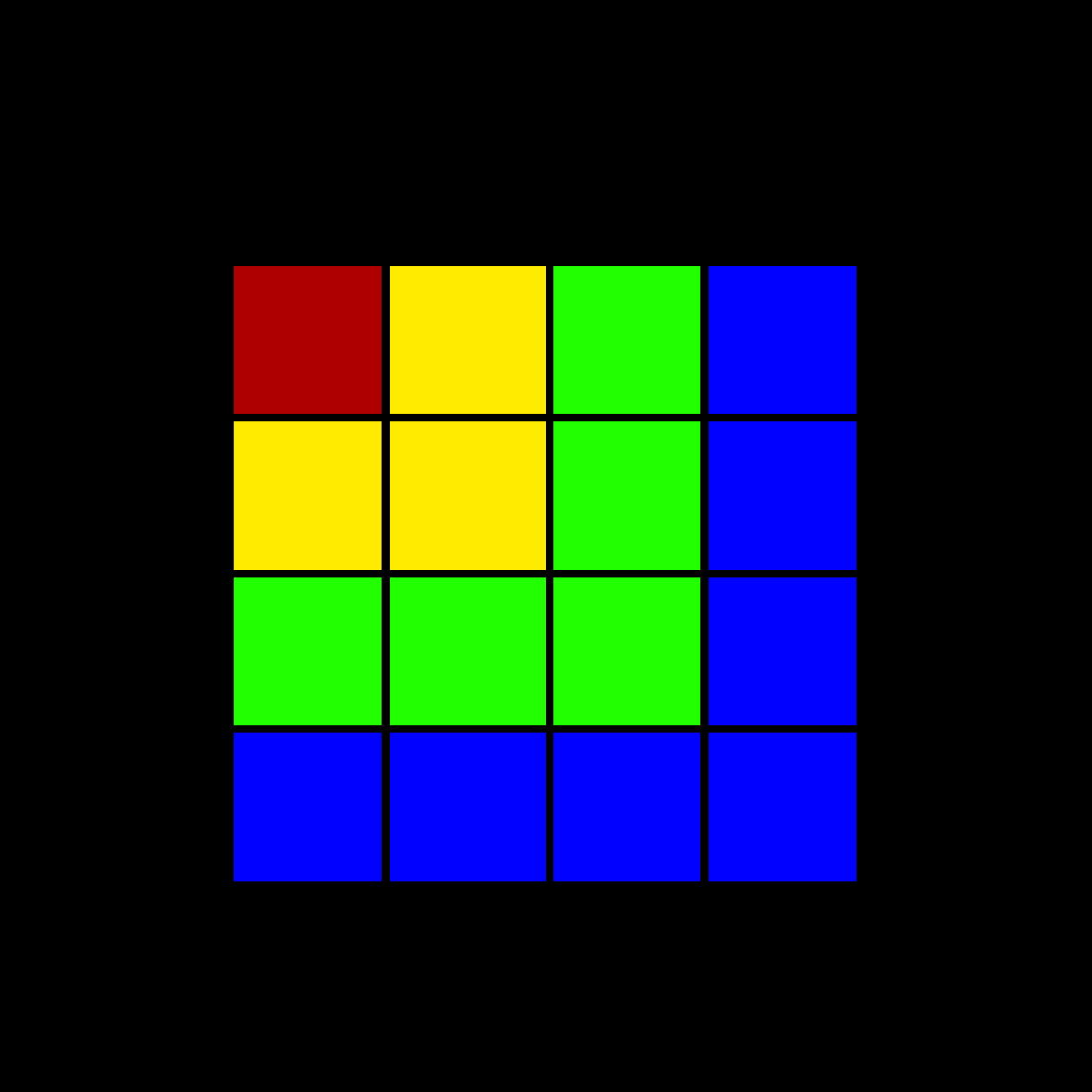}
&
\includegraphics[width = \heatSubPlotWidthSmall cm]{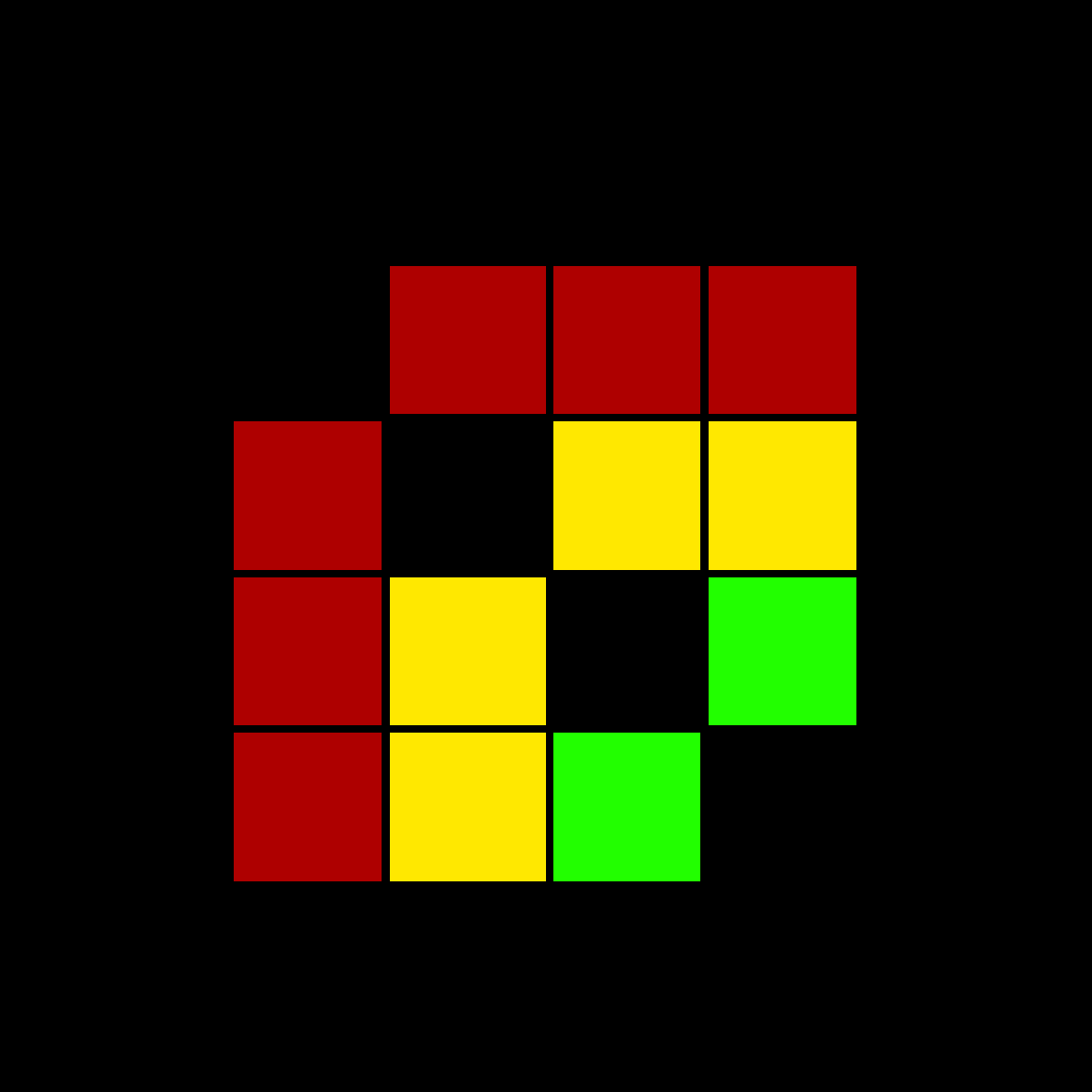}
&
\includegraphics[width = 1.5 cm]{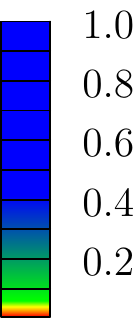}
&
\rotatebox[origin=c]{90}{$T_2$~(s)}
\\
\includegraphics[width = \heatSubPlotWidthSmall cm]{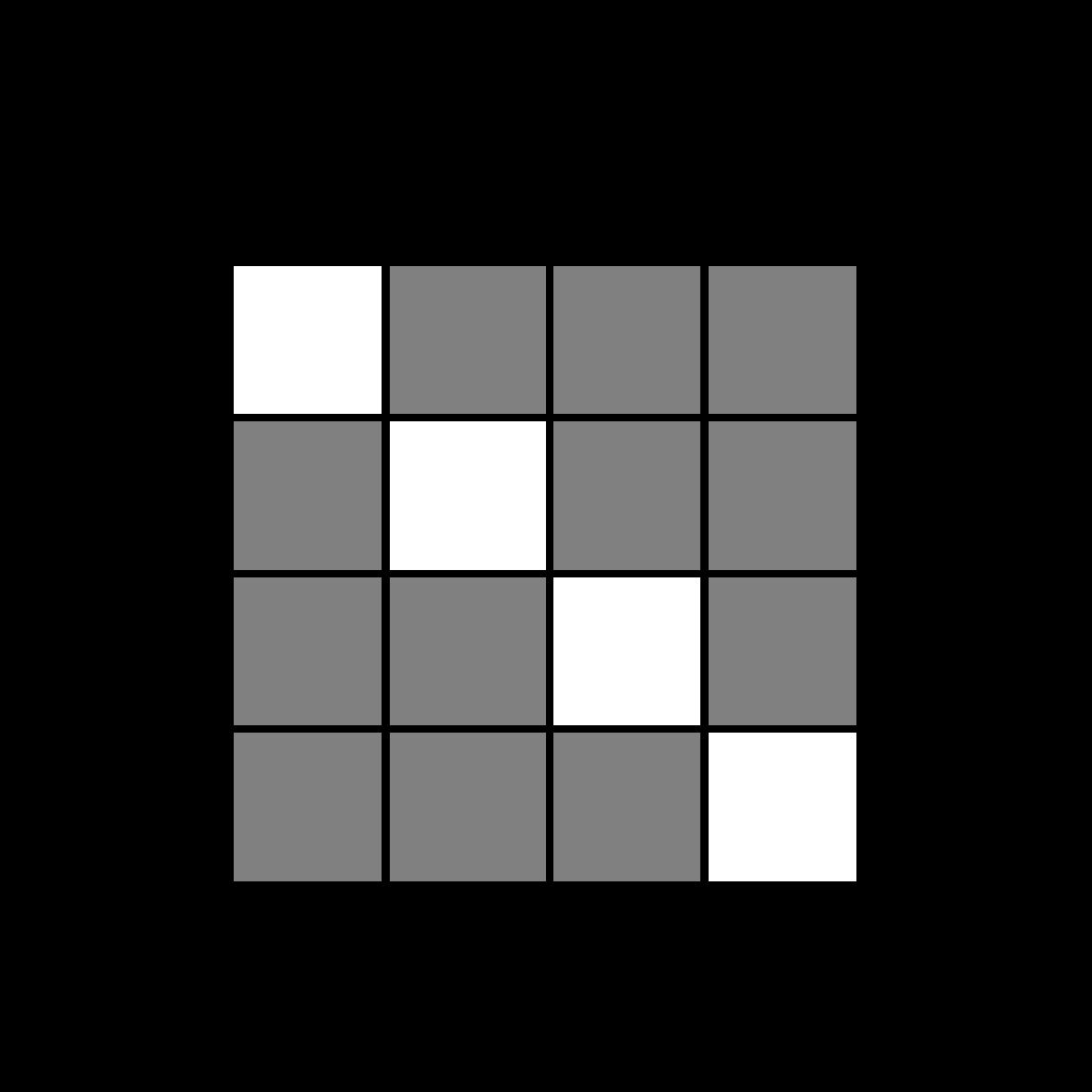}
&
\includegraphics[width = \heatSubPlotWidthSmall cm]{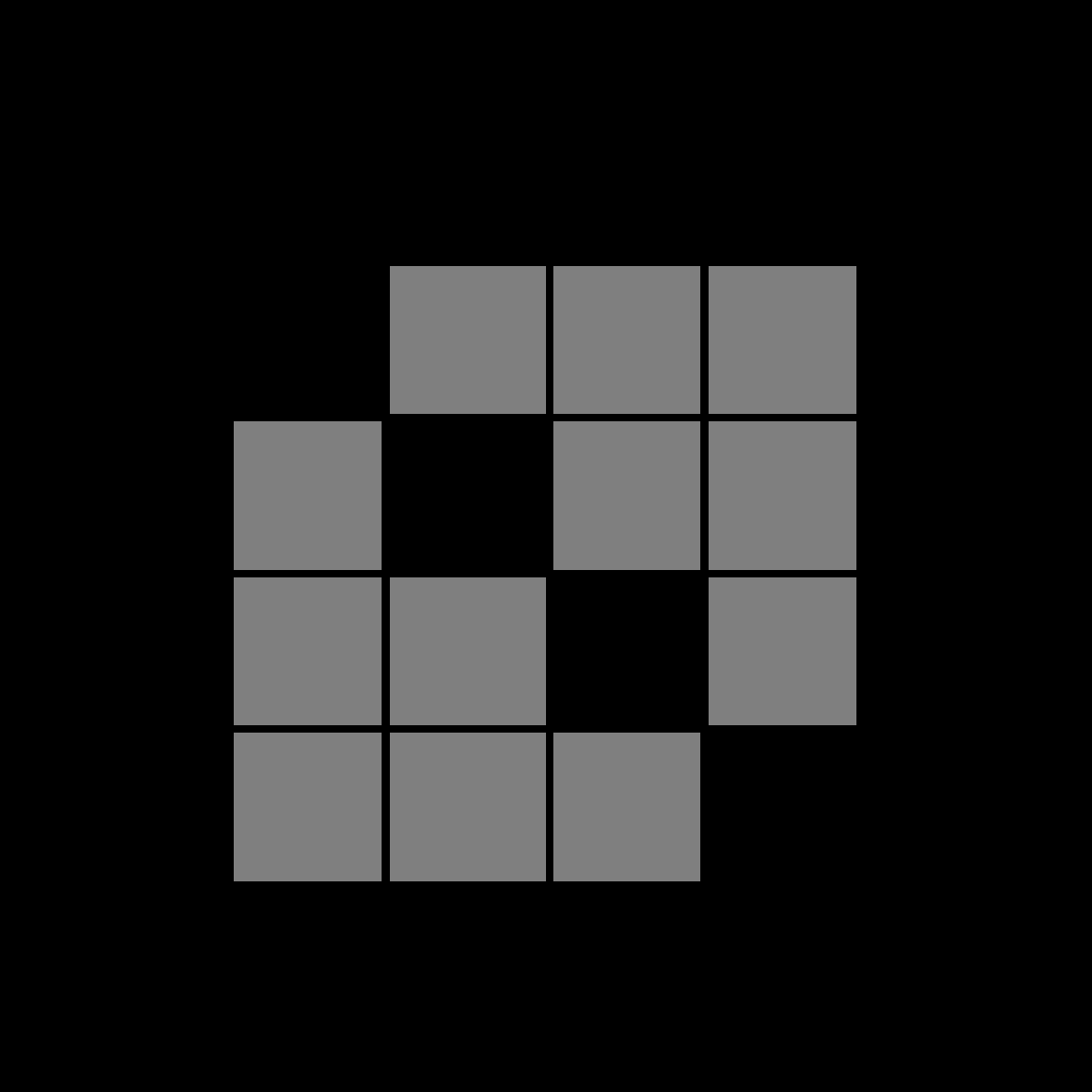}
&
\includegraphics[width = 1.5 cm]{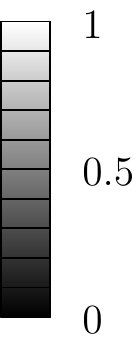}
&
\rotatebox[origin=c]{90}{$PD$~(a.u.)}
\end{tabular}
   \caption{The numerical phantom consists of four different synthetic tissues with relaxation times in the range commonly found in biological brain tissue. In particular, the relaxation times of tissues A-D are in the range found in myelin water, gray matter, white matter, and cerebrospinal fluid. The phantom is a sum of rows and columns of the corresponding types of tissue in increasing order of the $T_1$ relaxation time. When sorting the compartments of each voxel by the $T_1$ relaxation time, the numerical phantom has the depicted diagonal structure.}
   \label{tab:grid_description}
\end{figure}

In this section, we evaluate our methods on a numerical phantom that mimics an MRF experiment with radial $k$-space sampling. The numerical phantom has 16 different 19x19 voxel regions. The voxels in each region consist of either one or two compartments with relaxation times in the same range as biological tissues such as myelin water (A), gray matter (B), white matter (C), and cerebrospinal fluid (C). Figure~\ref{tab:grid_description} depicts the structure of the phantom along with the corresponding relaxation times. For each compartment, a fingerprint was calculated with Bloch simulations utilizing the flip angle pattern described in Fig.~3k in~\cite{Asslander2017arxiv}. These fingerprints are then combined additively to yield the multicompartment fingerprints of each voxel. The data correspond to undersampled k-space data of each time frame, calculated with a non-uniform fast Fourier transform \cite{Fessler2003,asslander2017low} with 16 radial k-space spokes per time frame. The fingerprints consist of 850 time frames, and the trajectory of successive time frames are rotated by 16 times the golden angle with respect to each other~\cite{Winkelmann2007b}. Noise with an i.i.d Gaussian distribution was added to the simulated $k$-space data to achieve different signal-to-noise ratios. 

\begin{figure}[tp]
%\begin{tabular}{ |>{\centering\arraybackslash}m{0.05\linewidth} |>{\centering\arraybackslash}m{0.4\linewidth}  |>{\centering\arraybackslash}m{0.4\linewidth}| >{\centering\arraybackslash}m{0.05\linewidth}|}
\begin{tabular}{ >{\centering\arraybackslash}m{0.02\linewidth} >{\centering\arraybackslash}m{0.4\linewidth}  >{\centering\arraybackslash}m{0.4\linewidth} >{\centering\arraybackslash}m{0.05\linewidth}}
&&& Corr. Coef.
\\
\rotatebox[origin=c]{90}{$T_2$~(s)}
&
\includegraphics[width = 1.02\linewidth]{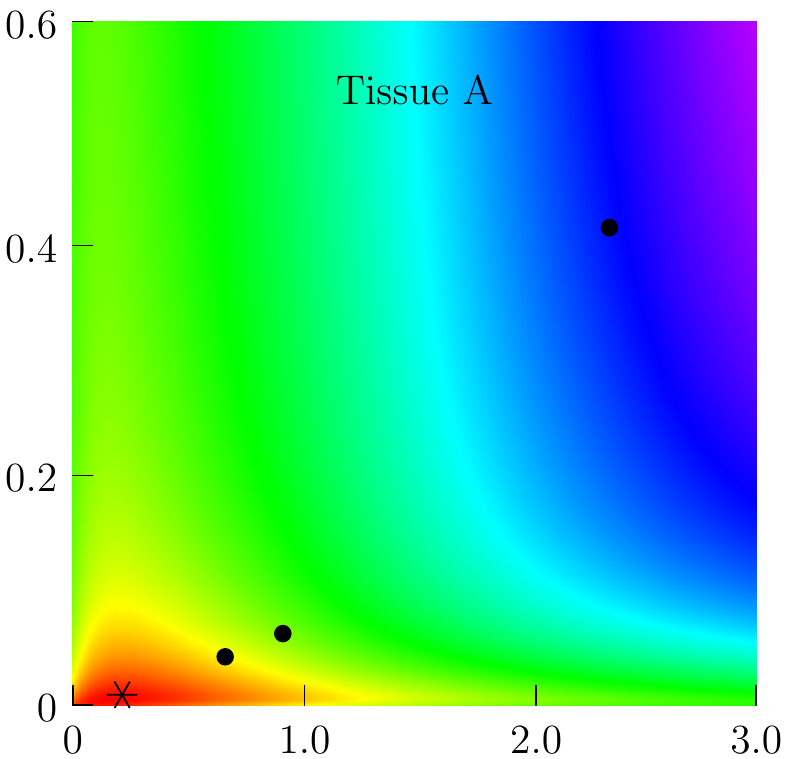}
&
\includegraphics[width = 1.0\linewidth]{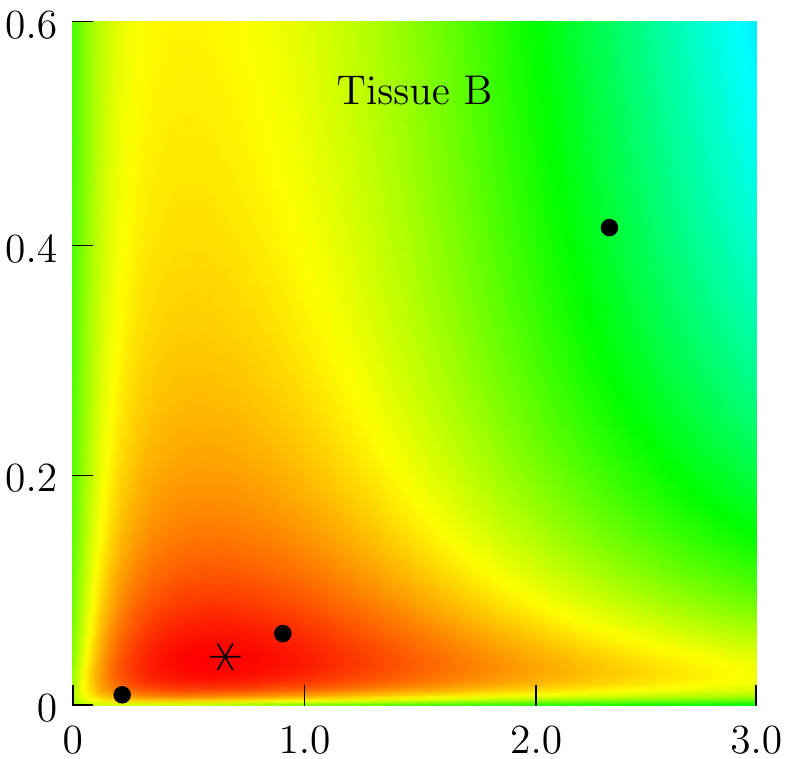}
&
\includegraphics[width = 1.3cm]{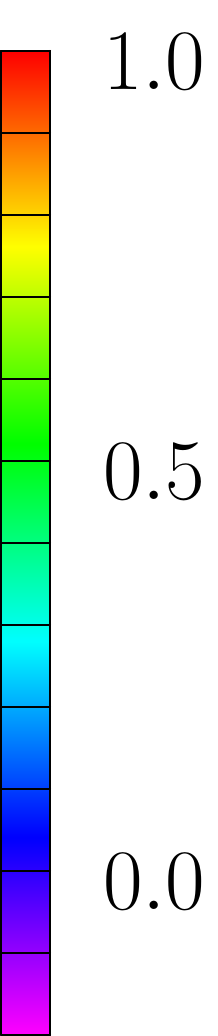}
\\
\rotatebox[origin=c]{90}{$T_2$~(s)}
&
\includegraphics[width = 1.0\linewidth]{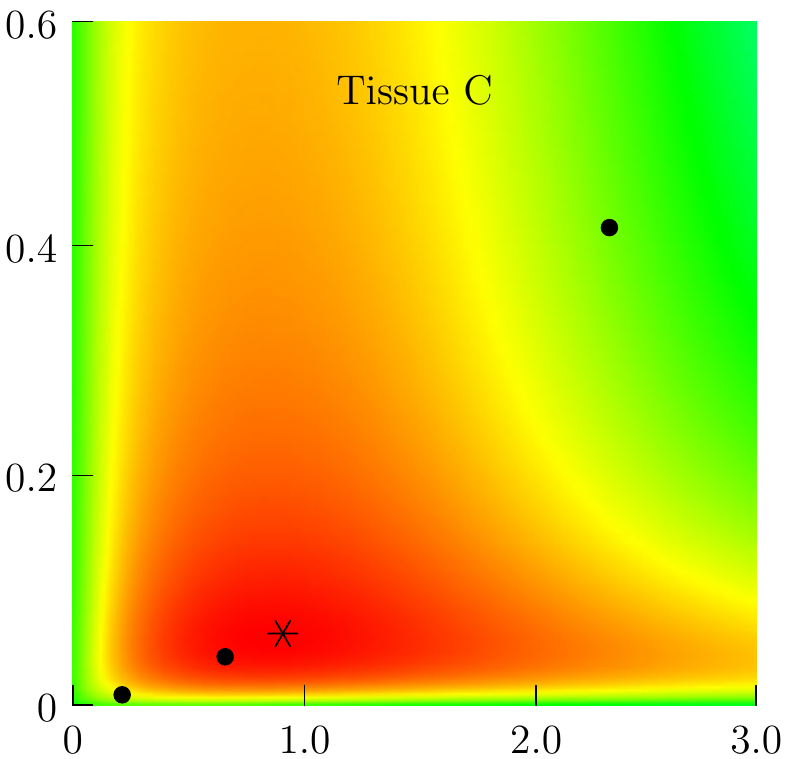}
&
\includegraphics[width = 1.0\linewidth]{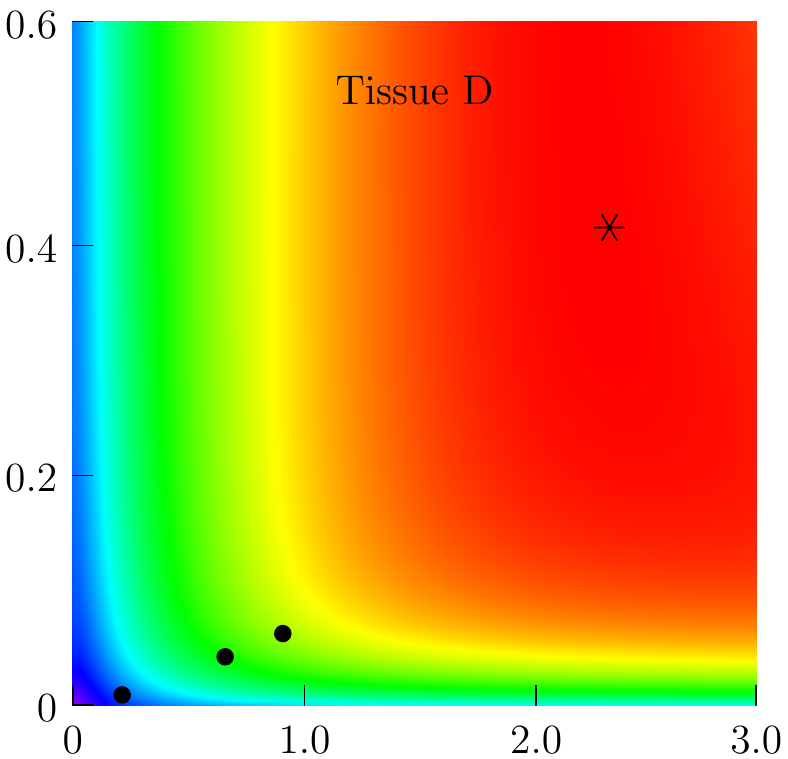}
&
\includegraphics[width = 1.3cm]{corr2d_cbar}
\\
& $T_1$(s)& $T_1$(s)&
\end{tabular}
\caption{Correlation between the fingerprints corresponding to the four tissues with other fingerprints in the dictionary, indexed by their $T_1$ and $T_2$ values. The tissue corresponding to the fixed fingerprint is marked with a star, the position of the other three tissues also present in the phantom are marked with dots.}
\label{fig:corr_2d}
\end{figure}

\begin{figure}
	\centering
\begin{tikzpicture}
\begin{scope}[spy using outlines={circle, magnification=4, connect spies}]
\begin{axis}[
width=0.75\textwidth,
xmin=0,xmax=850,
height=0.4\textwidth, 
xlabel = {time (s)},
ylabel = {signal (a.u.)}, 
y dir=reverse, 
scaled y ticks={real:-1}, 
ytick scale label code/.code={}, 
scaled x ticks={real:222.222}, 
xtick scale label code/.code={}, 
xtick distance=222.222,
legend pos = south east,
]
% \addlegendimage{empty legend}
\addplot[orange] file {dat_files/wm_gm/sig_wm.dat};
\addplot[cyan] file {dat_files/wm_gm/sig_gm.dat};
\addplot[blue] file {dat_files/wm_gm/sig_data.dat};
\addplot[magenta] file {dat_files/wm_gm/sig_sc.dat};
\legend{tissue B, tissue C, 50\%-50\% Comb., SC reconstruction}

% Spy
\coordinate (spypoint) at (axis cs:40,0.175);
\coordinate (spyviewer) at (axis cs:300,0.2);
\spy[black,size=3cm] on (spypoint) in node (myspy) [fill=white] at (spyviewer);
\end{axis}
\end{scope}
\end{tikzpicture}
\caption{Simulated fingerprints of tissue B and C, and a 50\%-50\% combination of those tissues are shown, along with a single-compartment (SC) reconstruction, i.e. the fingerprint in the dictionary that is closest to the 50\%-50\% combination. In this particular example, the fingerprints of those tissues are highly correlated, and the single-compartment reconstruction represents the data well (the relative error in the $\ell_2$-norm is 1.42\%). As a result, the multicompartment reconstruction cannot distinguish the two compartments if the data contain even a small level of noise. 
}
\label{fig:wm_gm}
\end{figure}
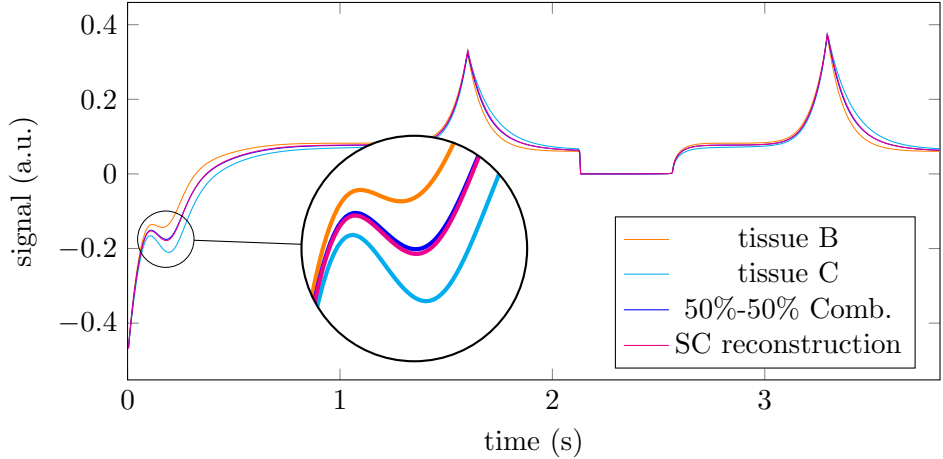

When two different tissues are present in a voxel, two conditions are necessary so that the problem of distinguishing their contributions to the signal is well posed. First, their corresponding fingerprints should be sufficiently distinct, i.e. have a low correlation coefficient (c.f. Section~\ref{sec:sparse_estimation_l1}). Second, the weighted sum of their fingerprints should be different from any other fingerprint in the dictionary. Otherwise, a single-compartment model will fit the data well. Figure~\ref{fig:corr_2d} shows the correlations between the fingerprints corresponding to each of the tissues present in the numerical phantom and the rest of the fingerprints in the dictionary. For tissues that have similar $T_1$ and $T_2$ values, the correlation is very high. In particular, the fingerprints corresponding to tissue B and C are extremely similar. Furthermore, their additive combination is almost indistinguishable from another fingerprint also present in the dictionary, as illustrated in Figure~\ref{fig:wm_gm}. It will consequently be hardly possible to distinguish a voxel containing a single tissue with that particular fingerprint and a two-compartment voxel containing tissue B and C if there is even a very low level of noise in the data. For the other combinations of the four examined relaxation times, this is not the case.

To estimate the relaxation times $T_1$ and $T_2$ from the simulated data we apply the method described in Section~\ref{sec:admm} while fixing the number of ADMM iterations to 10. We compress the dictionary by truncating its SVD to obtain a rank-10 approximation, as described in Section~\ref{sec:compression}. The magnetization vector $\tilde{X}$ is updated by applying 10 conjugate-gradient iterations. The coefficient variable $\tilde{C}$ is updated by applying the reweighted-$\ell_1$ method described in Section~\ref{sec:reweightedl1} over 5 iterations with the reweighting scheme in equation~\eqref{eq:rwl1_candes} (the reweighting scheme~\eqref{eq:rwl1_wipf} yields similar results). 

\begin{figure}[tp]
\begin{center}
%\begin{tabular}{cccc}
\begin{tabular}{ >{\centering\arraybackslash}m{0.225\linewidth}  >{\centering\arraybackslash}m{0.225\linewidth} >{\centering\arraybackslash}m{0.225\linewidth} >{\centering\arraybackslash}m{0.05\linewidth} >{\centering\arraybackslash}m{0.08\linewidth}}
	1\textsuperscript{st} Compartment&
	2\textsuperscript{nd} Compartment&
	3\textsuperscript{rd} Compartment&
\\
\includegraphics[width = \heatSubPlotWidth cm]{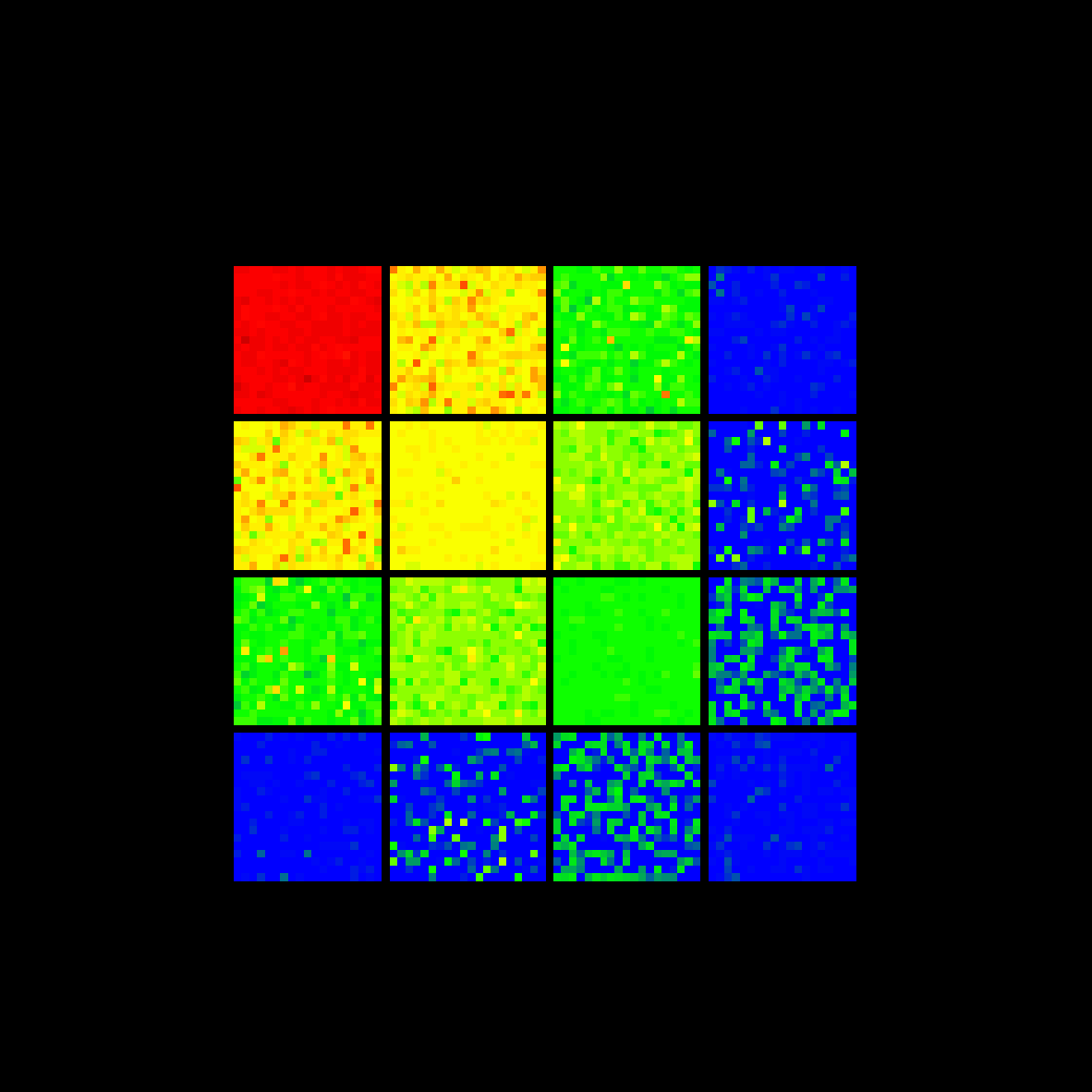}
&
\includegraphics[width = \heatSubPlotWidth cm]{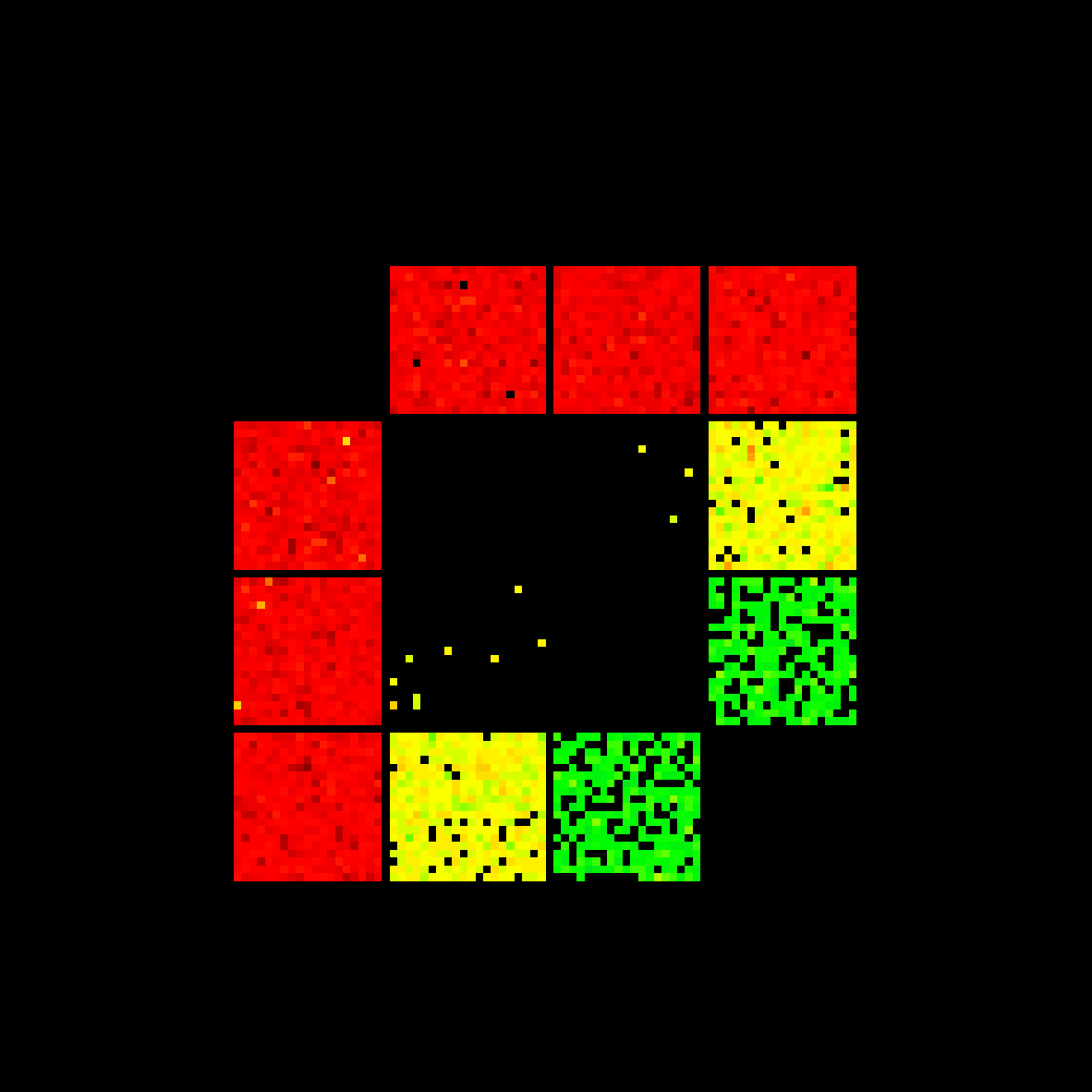}
&
\includegraphics[width = \heatSubPlotWidth cm]{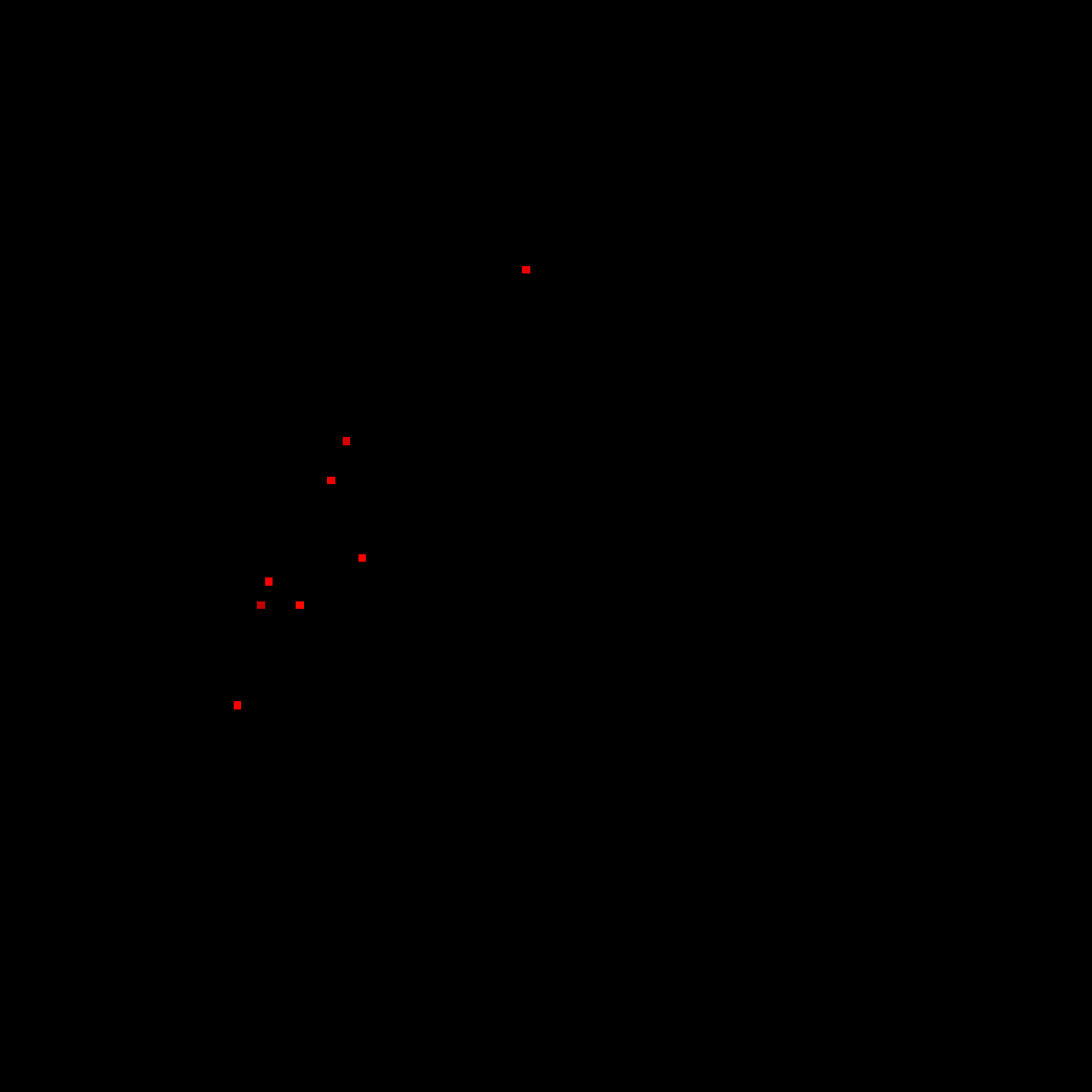}
&
\includegraphics[width = 1.5 cm]{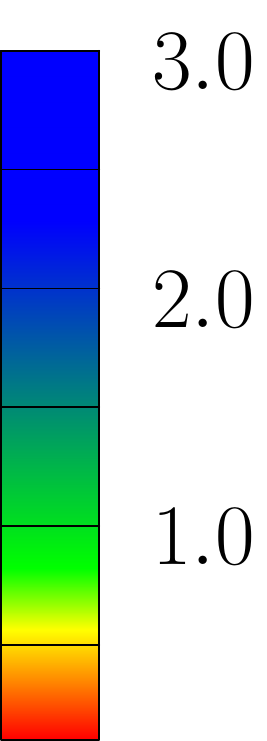}
&
\rotatebox[origin=c]{90}{$T_1$~(s)}
\\
\includegraphics[width = \heatSubPlotWidth cm]{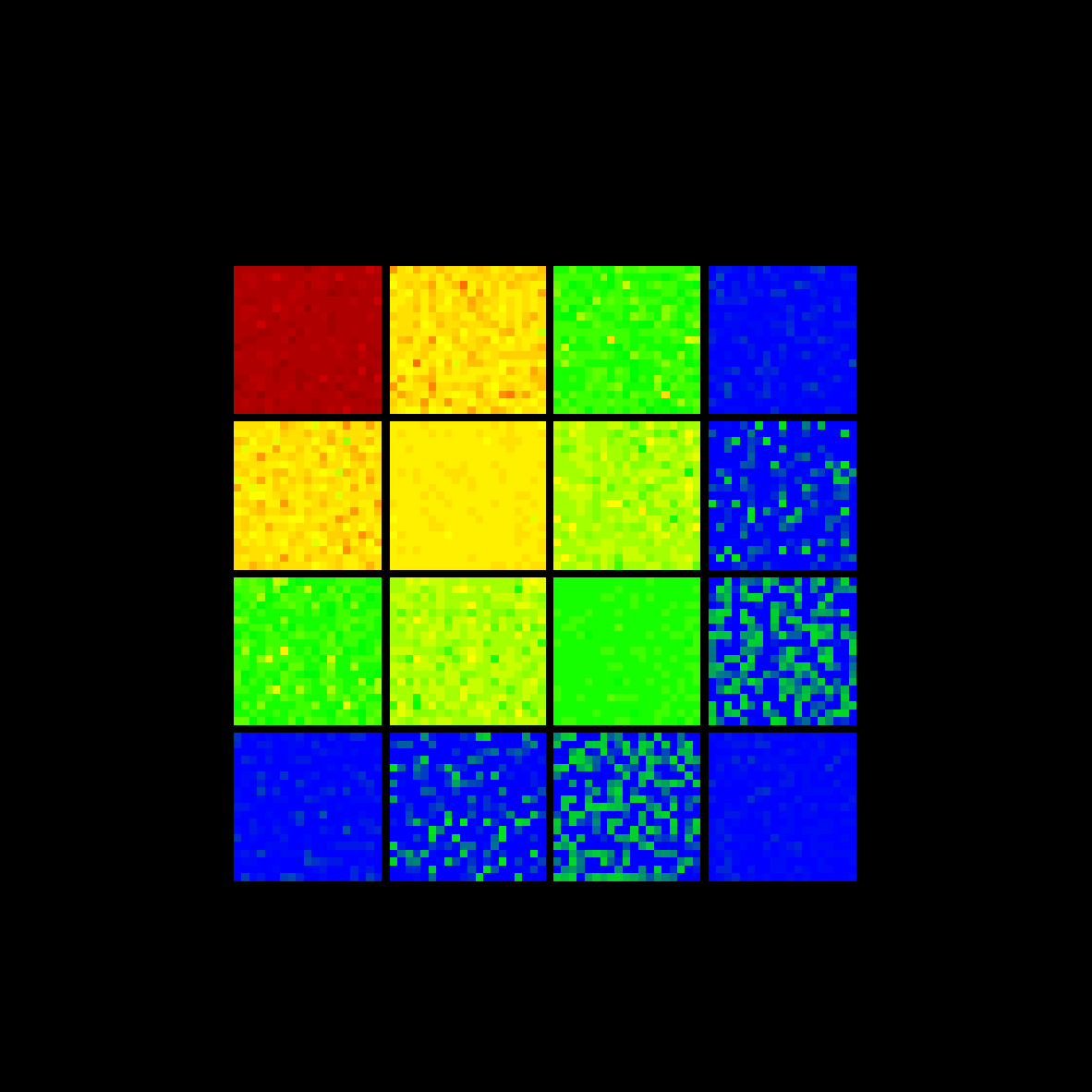}
&
\includegraphics[width = \heatSubPlotWidth cm]{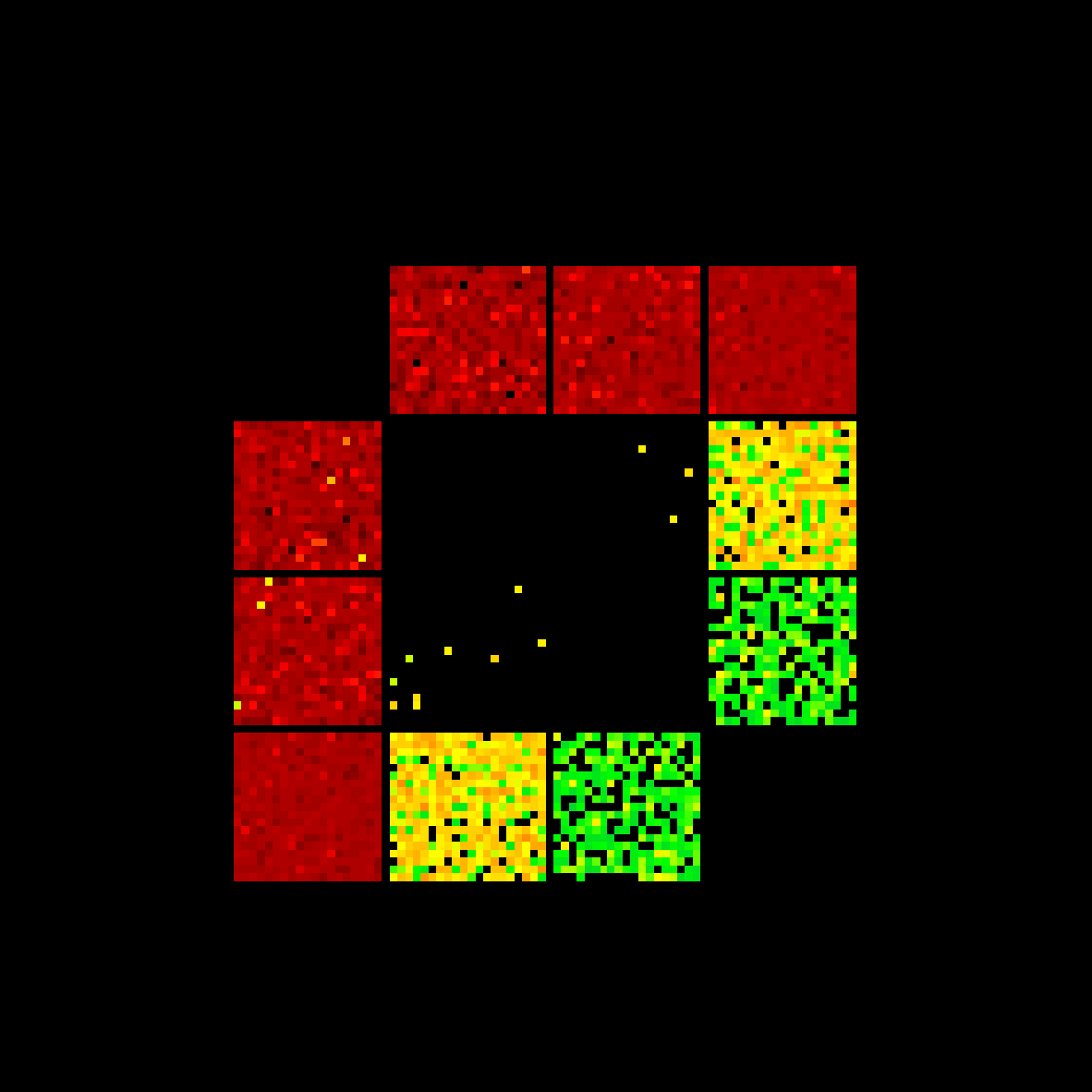}
&
\includegraphics[width = \heatSubPlotWidth cm]{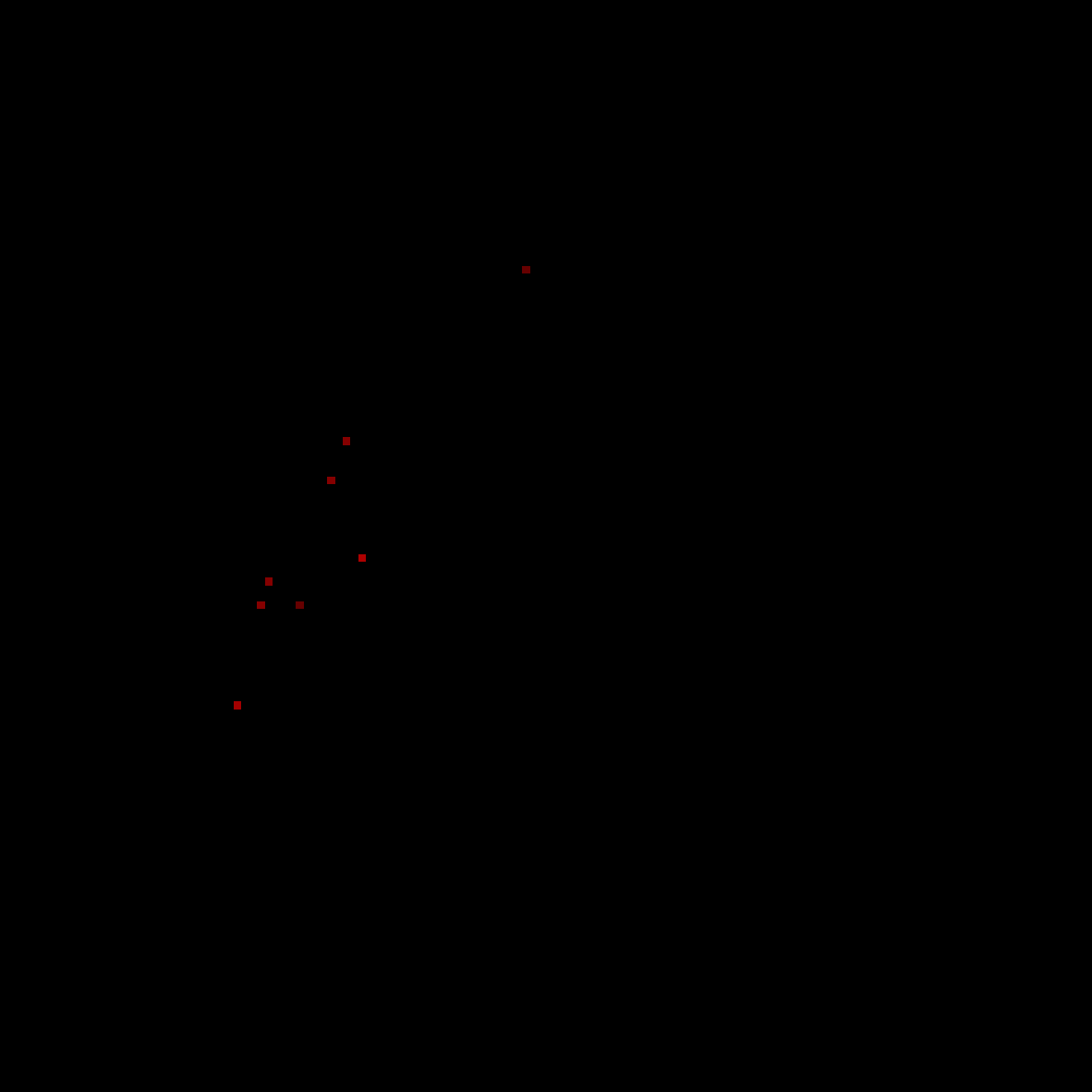}
&
\includegraphics[width = 1.5 cm]{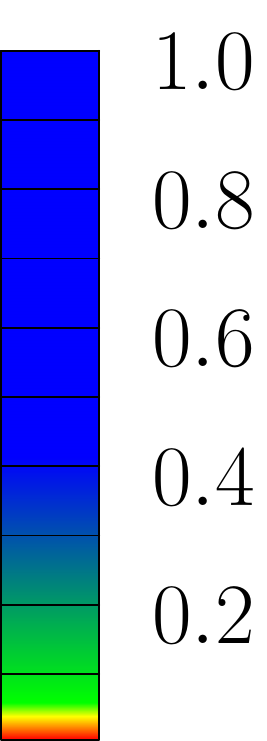}
&
\rotatebox[origin=c]{90}{$T_2$~(s)}
\\
\includegraphics[width = \heatSubPlotWidth cm]{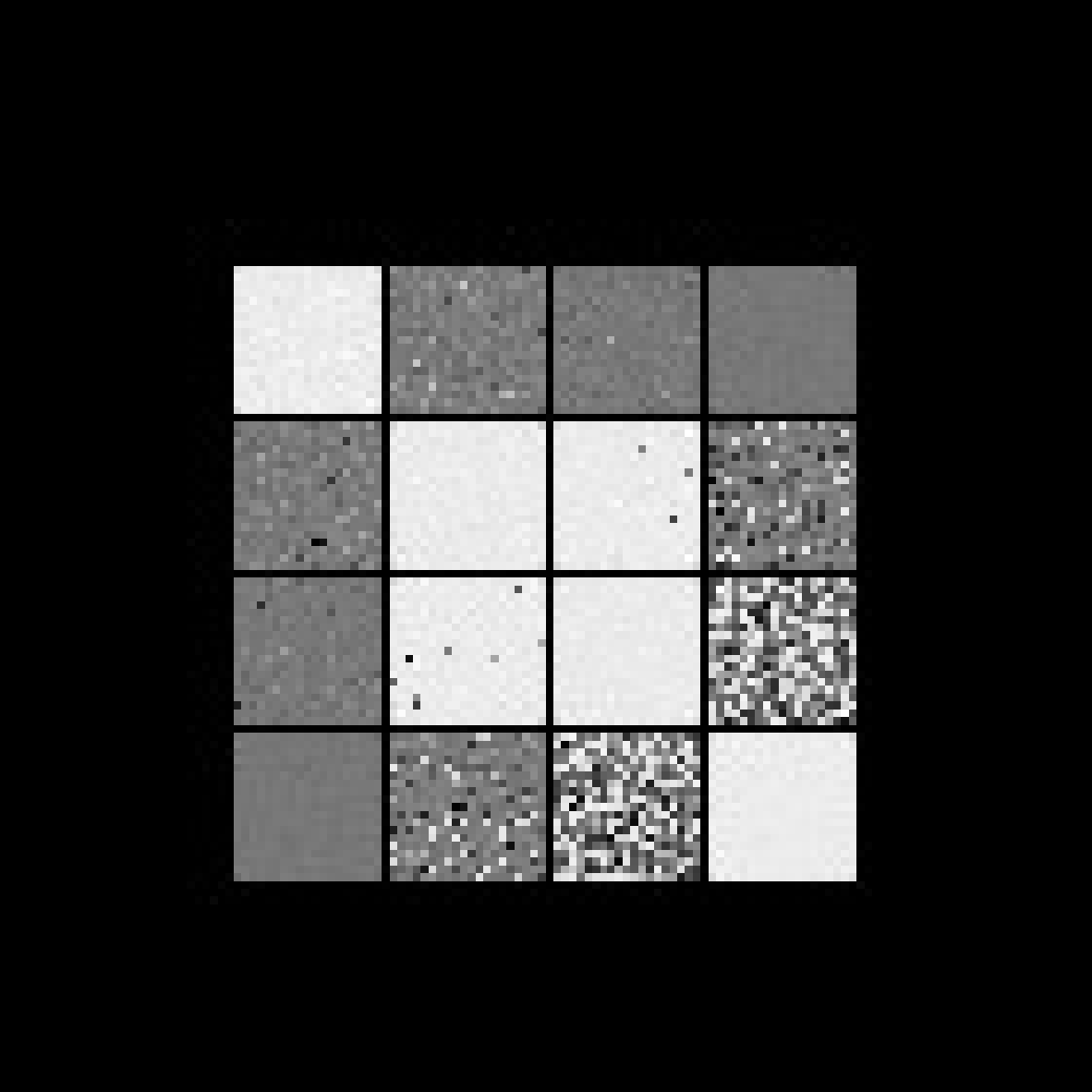}
&
\includegraphics[width = \heatSubPlotWidth cm]{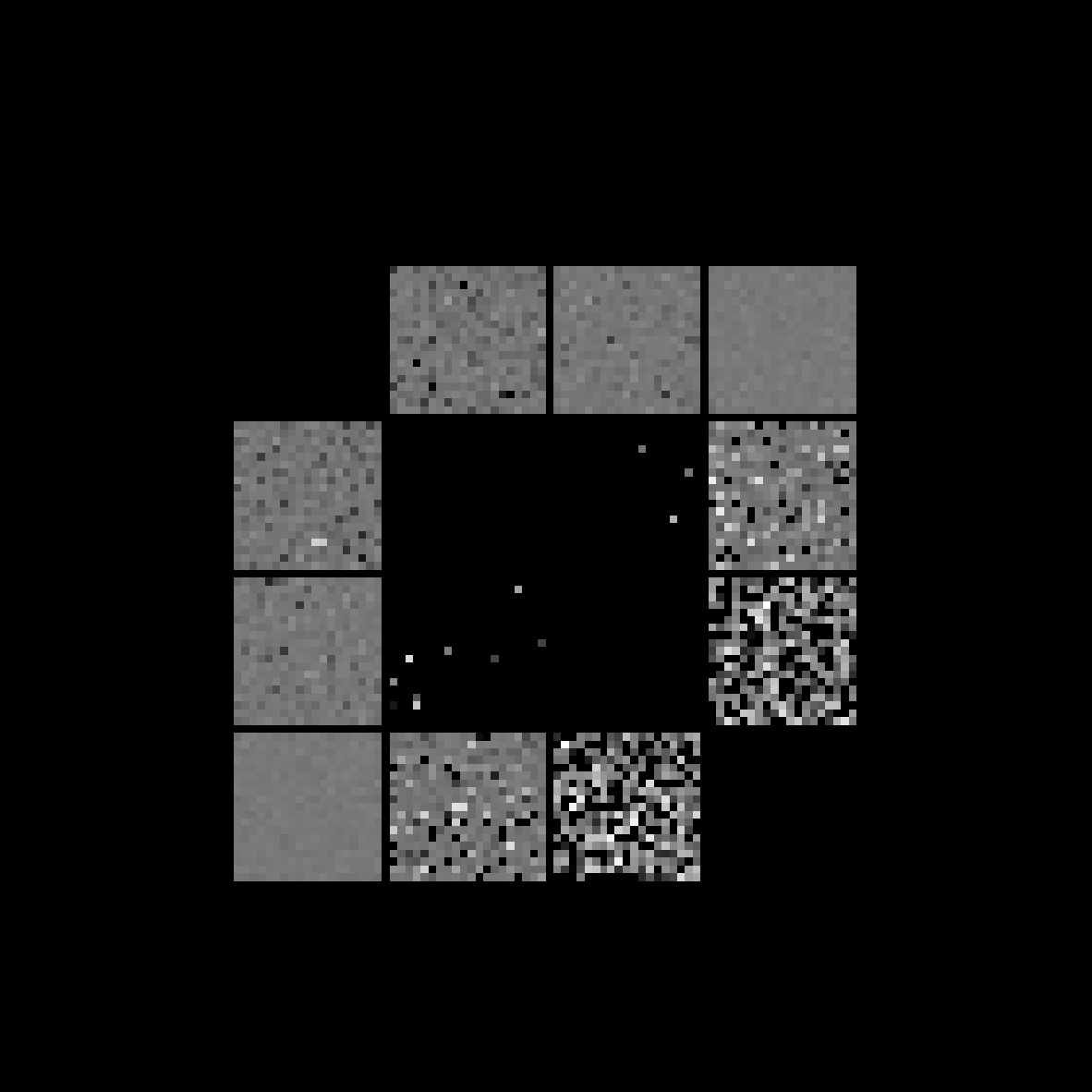}
&
\includegraphics[width = \heatSubPlotWidth cm]{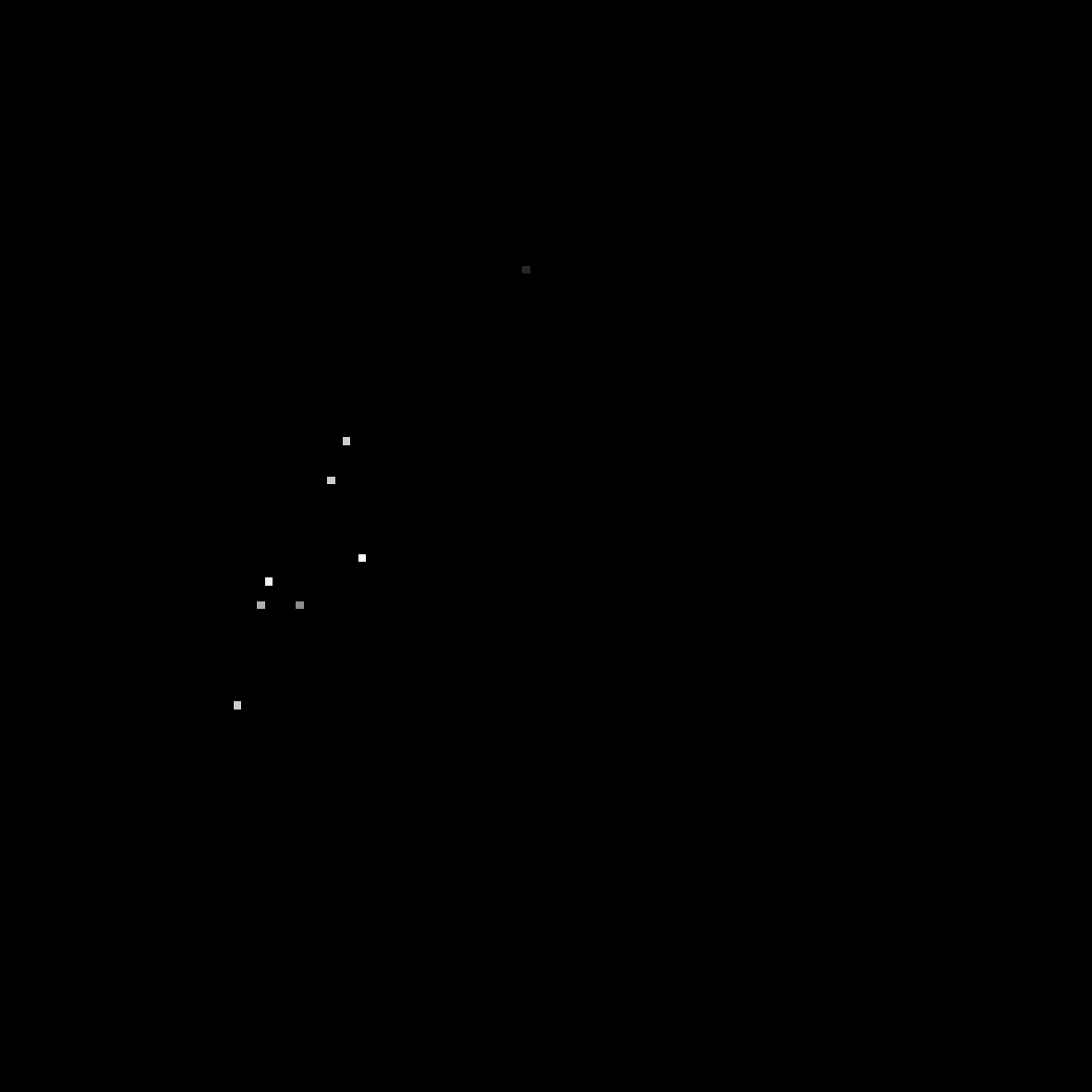}
&
\includegraphics[width = 1.5 cm]{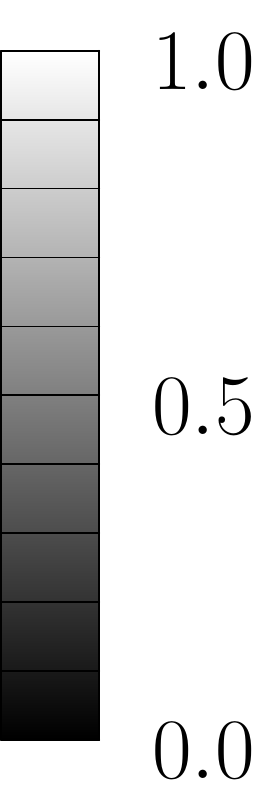}
&
\rotatebox[origin=c]{90}{$PD$~(a.u.)}
\\
\end{tabular}
\caption{The depicted parameter maps were reconstructed from data simulated with an SNR = $10^3$ and using 16 radial k-space spokes. The proton density (PD, third row) is here equivalent to the fraction of the recovered compartments. The compartments in each voxel are sorted according to their $\ell_2$-norm distance to the origin in $T_1$-$T_2$ space. Compare to Figure~\ref{tab:grid_description}, which shows the ground truth.}
\label{fig:hms16n3}
\end{center}
\end{figure}

\begin{figure}[htbp]
\begin{center}
%\begin{tabular}{cccc}
\begin{tabular}{ >{\centering\arraybackslash}m{0.225\linewidth}  >{\centering\arraybackslash}m{0.225\linewidth} >{\centering\arraybackslash}m{0.225\linewidth} >{\centering\arraybackslash}m{0.05\linewidth} >{\centering\arraybackslash}m{0.08\linewidth}}
1\textsuperscript{st} Compartment&
2\textsuperscript{nd} Compartment&
3\textsuperscript{rd} Compartment&
\\
\includegraphics[width = \heatSubPlotWidth cm]{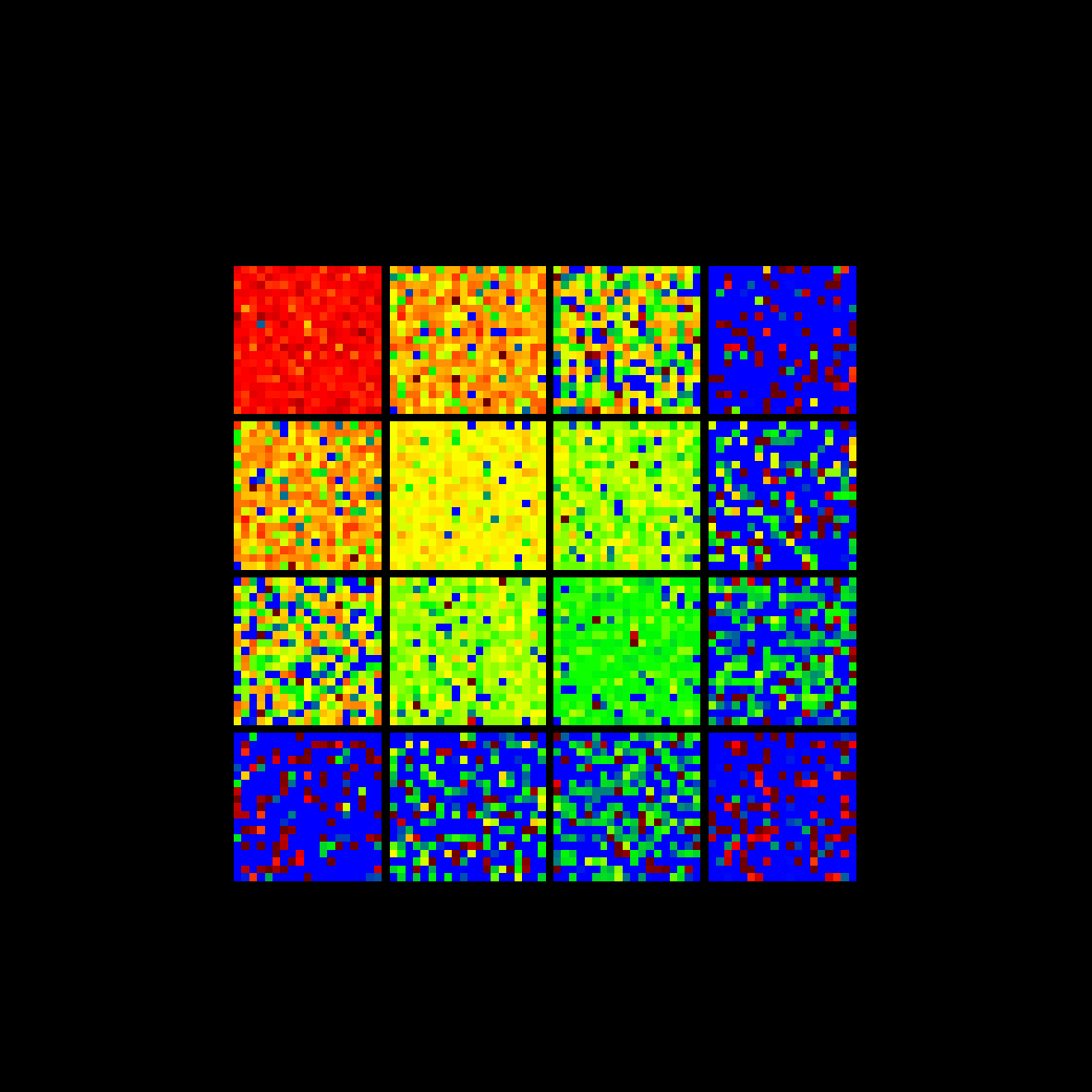}
&
\includegraphics[width = \heatSubPlotWidth cm]{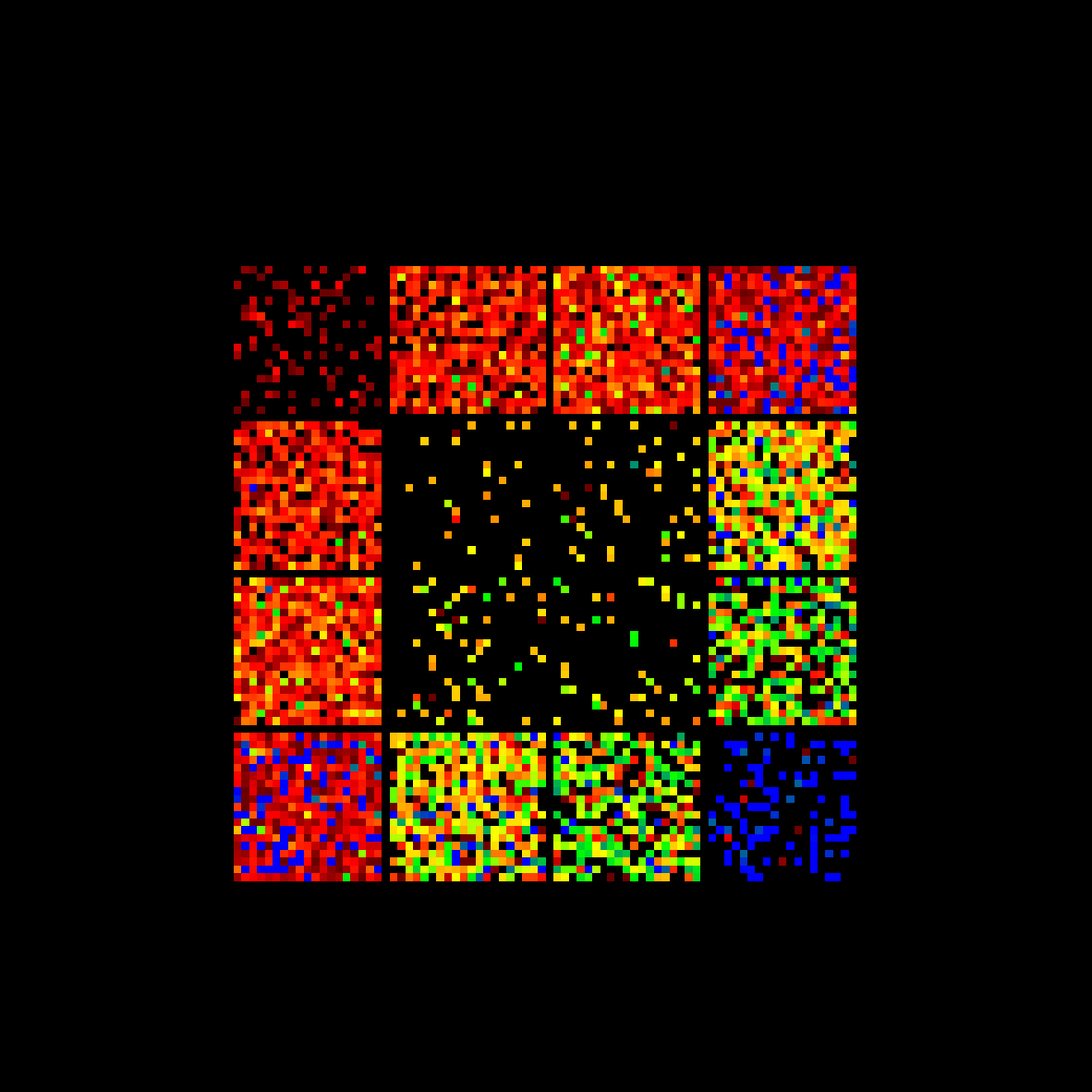}
&
\includegraphics[width = \heatSubPlotWidth cm]{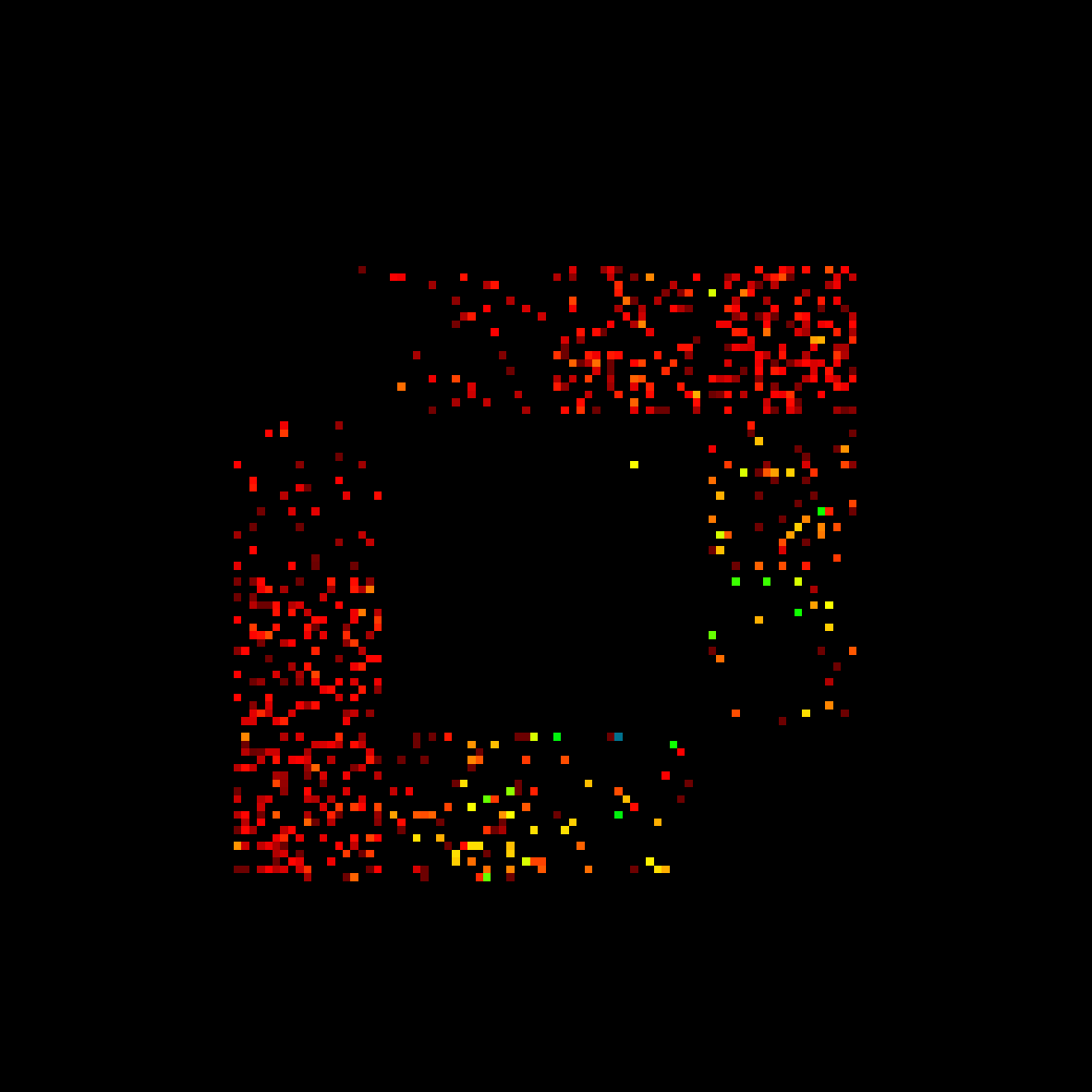}
&
\includegraphics[width = 1.5 cm]{heatmap_simulateds4n3T1cbar_noblack}
&
\rotatebox[origin=c]{90}{$T_1$~(s)}
\\
\includegraphics[width = \heatSubPlotWidth cm]{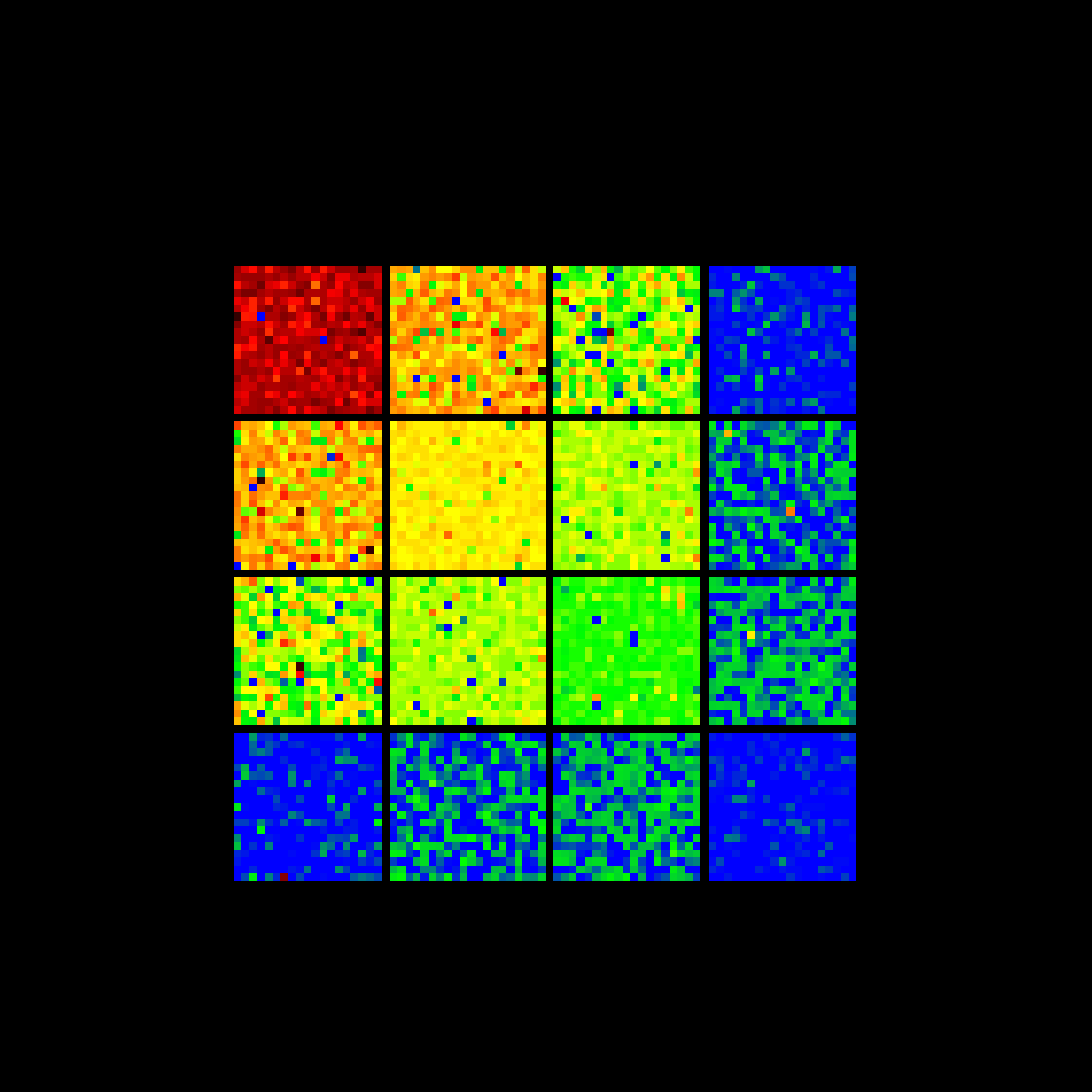}
&
\includegraphics[width = \heatSubPlotWidth cm]{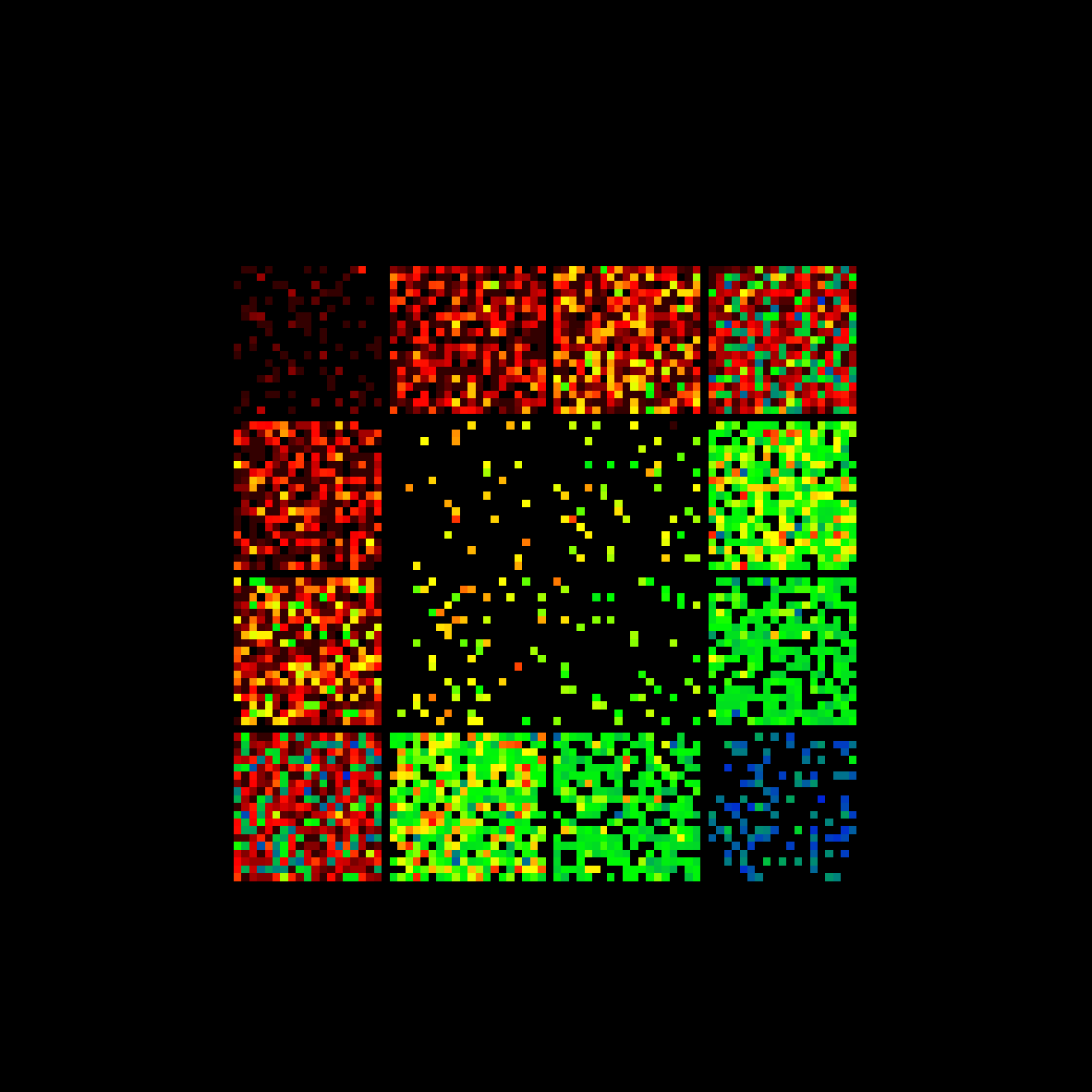}
&
\includegraphics[width = \heatSubPlotWidth cm]{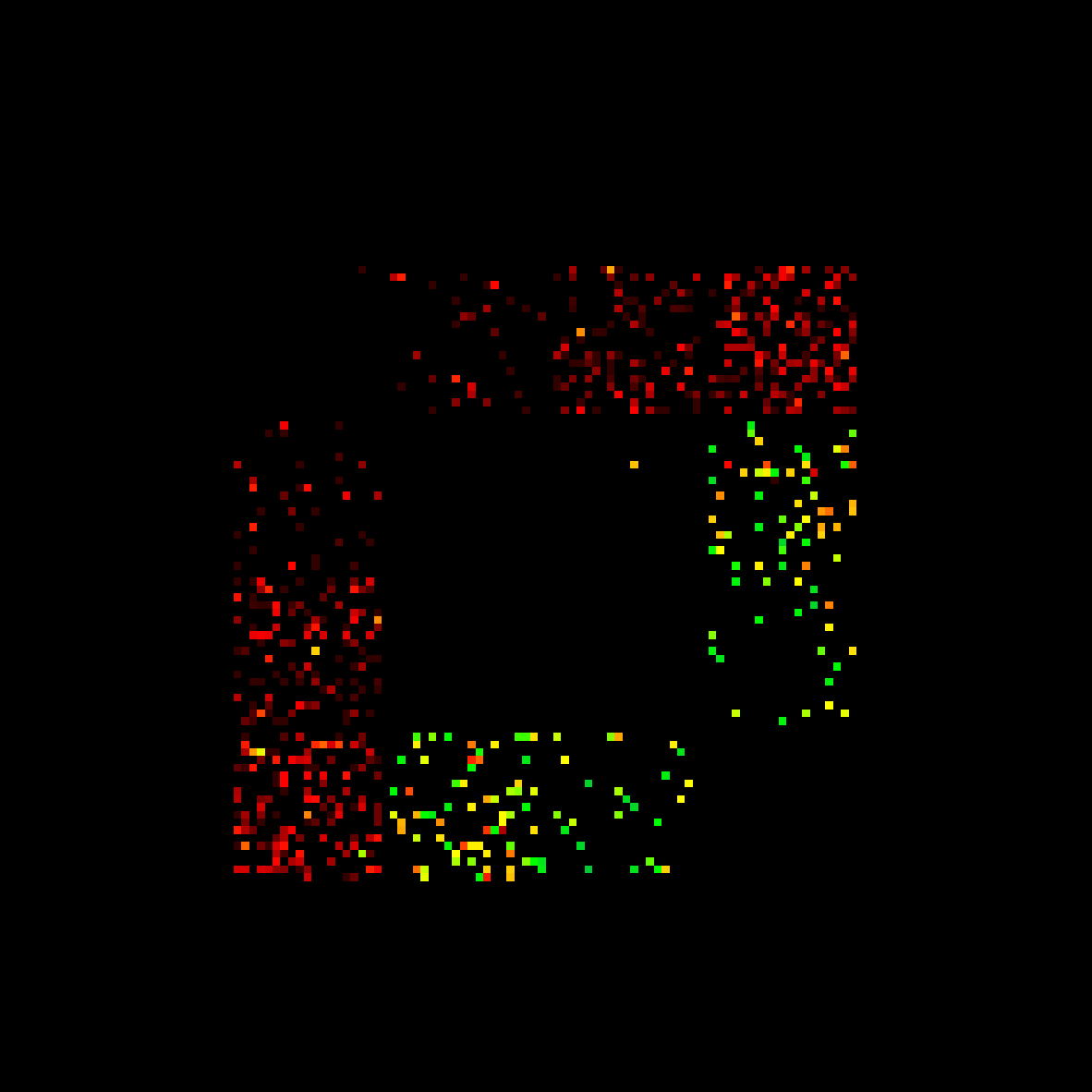}
&
\includegraphics[width = 1.5 cm]{heatmap_simulateds4n3T2cbar_noblack}
&
\rotatebox[origin=c]{90}{$T_2$~(s)}
\\
\includegraphics[width = \heatSubPlotWidth cm]{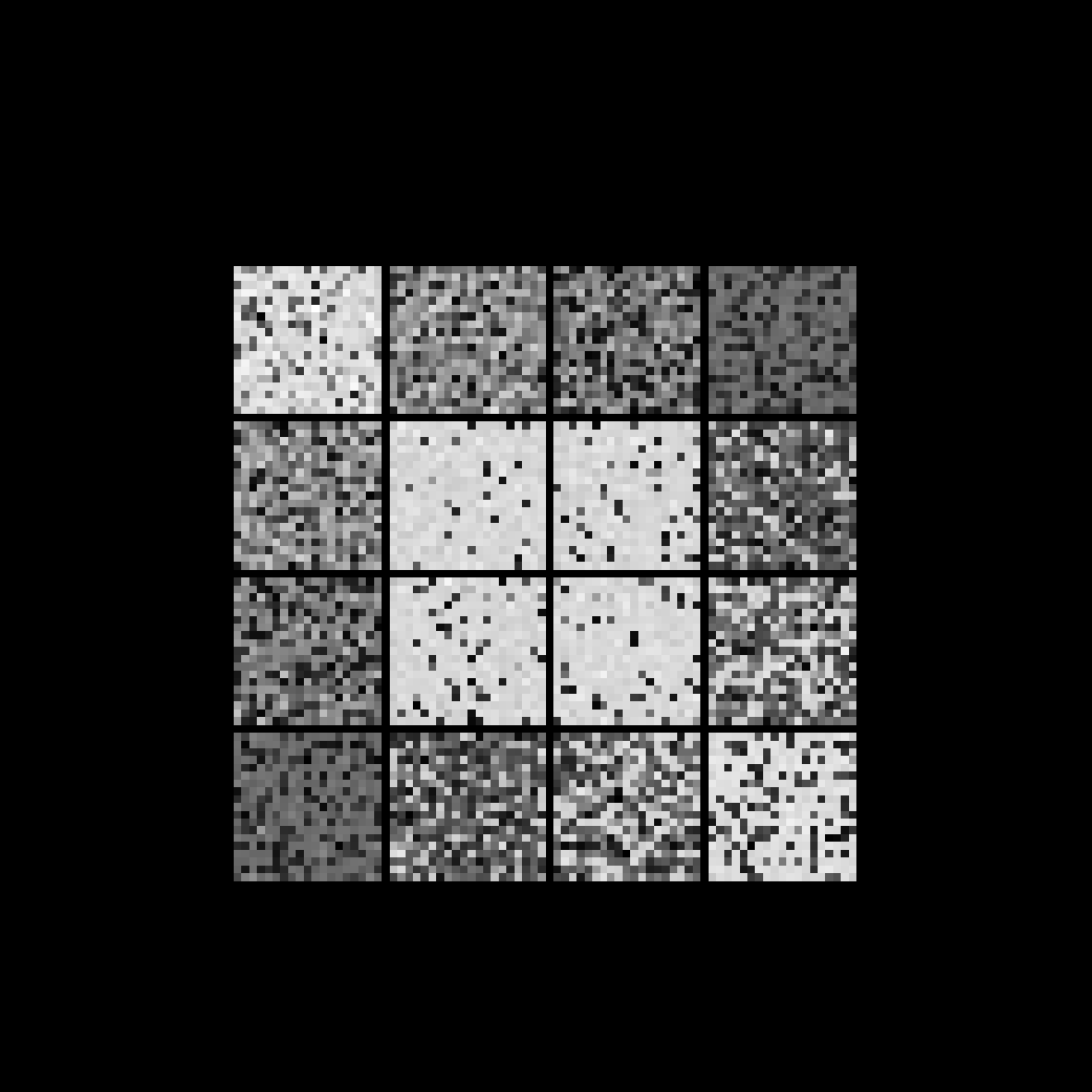}
&
\includegraphics[width = \heatSubPlotWidth cm]{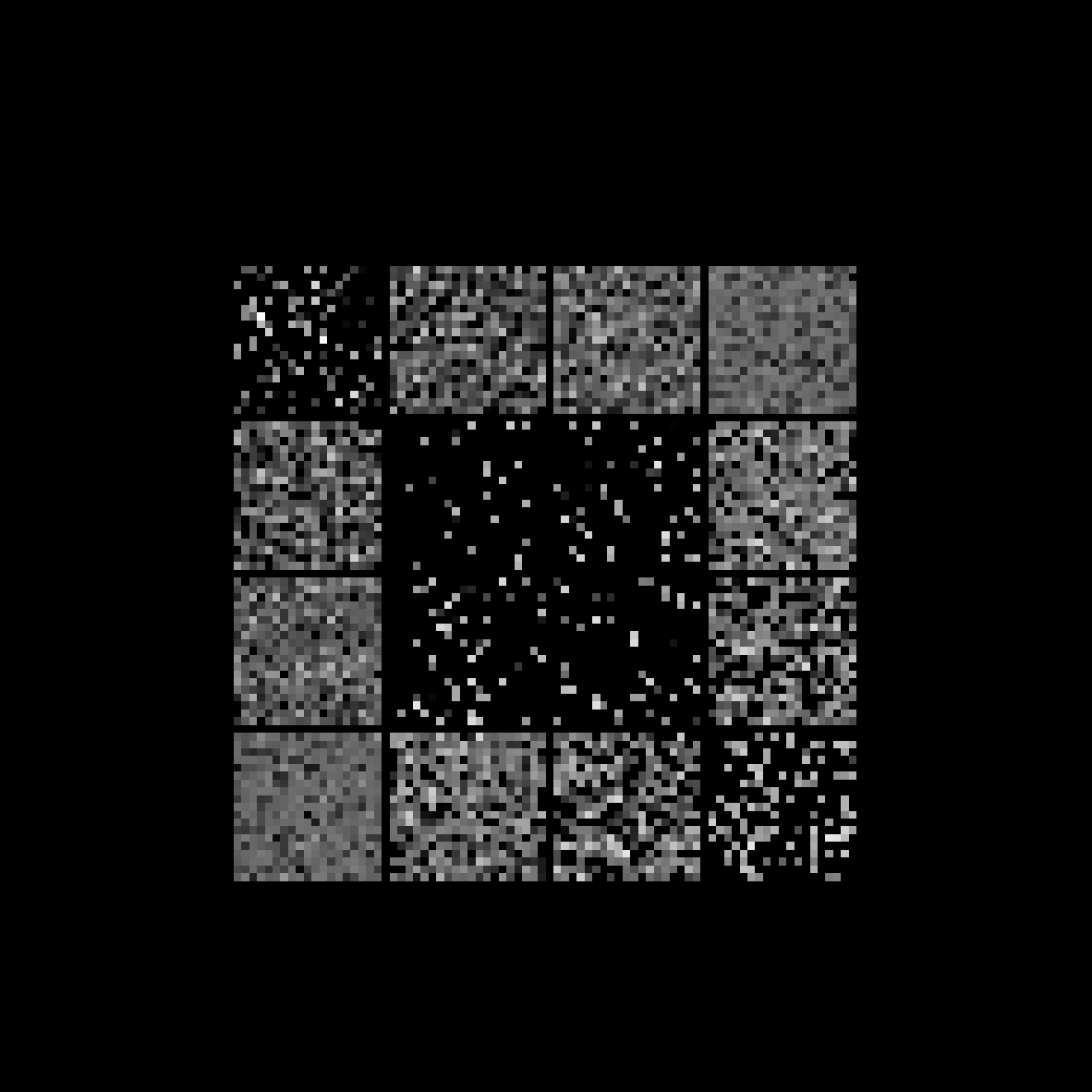}
&
\includegraphics[width = \heatSubPlotWidth cm]{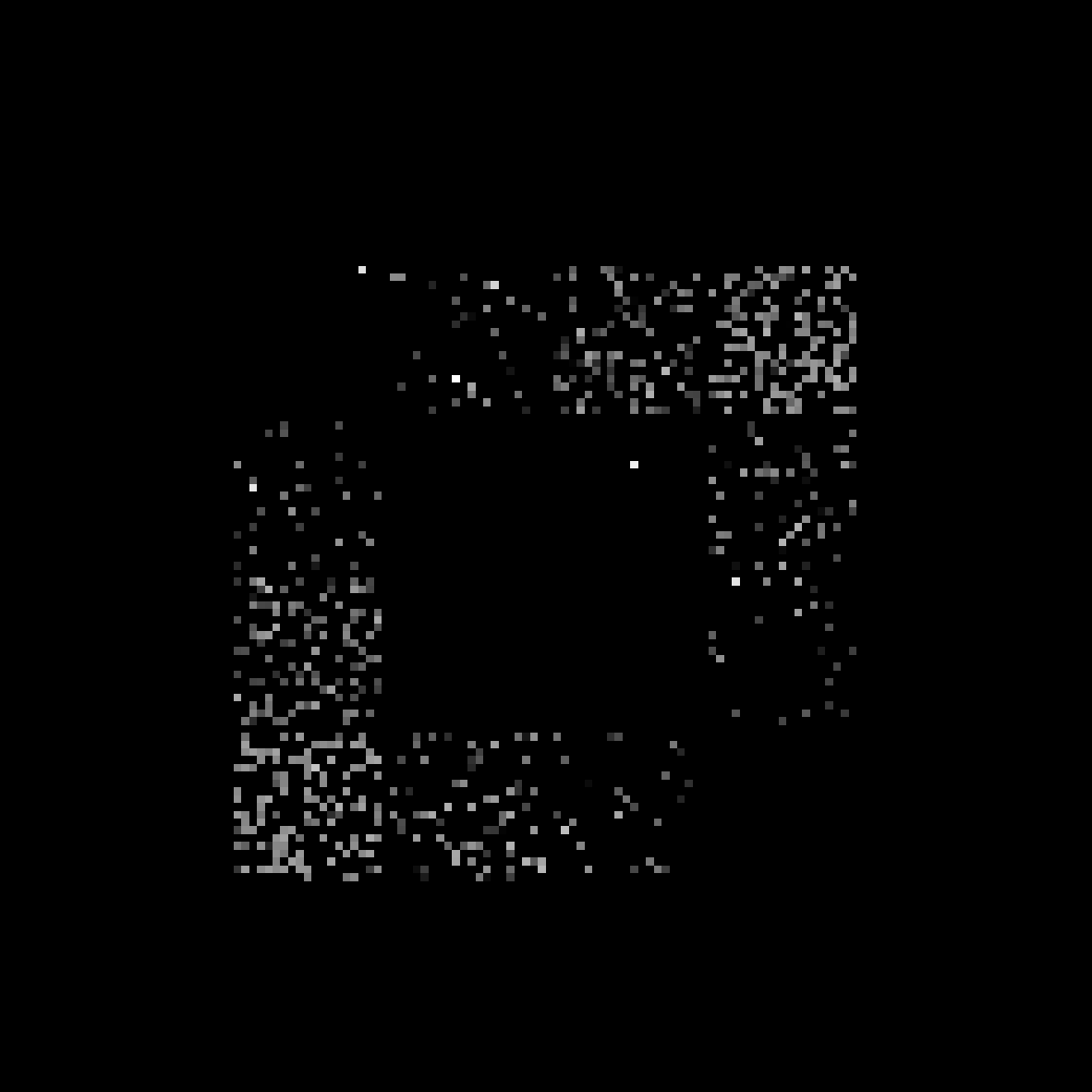}
&
\includegraphics[width = 1.5 cm]{heatmap_simulateds4n3PDcbar}
&
\rotatebox[origin=c]{90}{$PD$~(a.u.)}
\\
\end{tabular}
\caption{The depicted parameter maps were reconstructed from data simulated with an SNR = $100$ and using 16 radial k-space spokes. The proton density (PD, third row) is here equivalent to the fraction of the recovered compartments. The compartments in each voxel are sorted according to their $\ell_2$-norm distance to the origin in $T_1$-$T_2$ space. Compare to Figure~\ref{tab:grid_description}, which shows the ground truth.}
\label{fig:hms_16n2}
\end{center}
\end{figure}

Figure~\ref{fig:hms16n3} shows the reconstructed parameter maps for an $SNR = 10^3$ (the SNR is defined by the ratio between the $\ell_2$-norm of the signal and the added noise), and using 16 radial k-space spokes. The multicompartment reconstruction recovers the two compartments accurately, except for the voxels that contain the highly correlated tissues B and C. This behavior is expected due to the high correlations of the fingerprints of those tissues (c.f. Figure~\ref{fig:wm_gm}). The model also yields an accurate estimate of the proton density of the compartments that are present in each voxel (except, again, for voxels containing tissue B and C). In most voxels, the reconstruction does not detect additional compartments, even though this is is not explicitly enforced. Figure~\ref{fig:hms_16n2} shows that the performance degrades gracefully when the SNR is decreased to 100: the model still achieves similar results, only noisier, and a spurious third compartment with a relatively low proton density appears in some voxels.

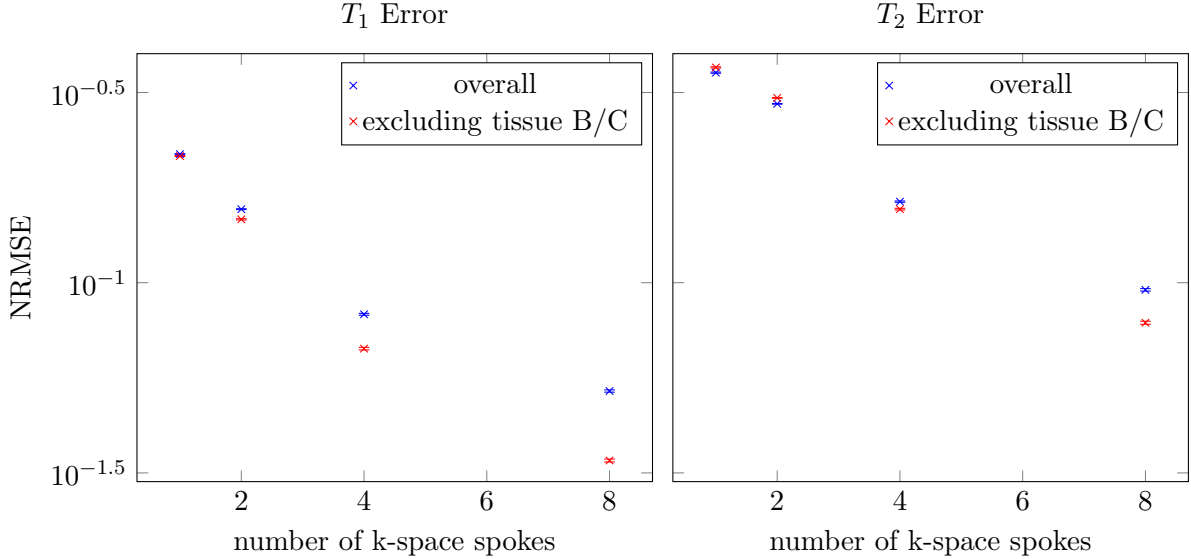
\begin{figure}[htbp]
	\centering
\begin{tikzpicture}
\begin{axis}[
ymode=log,
ymin = 0.03,
ymax = 0.4,
title = {$T_1$ Error},
xlabel = number of k-space spokes,
ylabel = NRMSE,
name = T1,
	]
\addplot+[
		mark=x,
		only marks,
        error bars/.cd,
            y dir= both,
            y explicit,
    ] table[x=ns,y=val,y error=err] {NumericalExperiment_simulated/data/errBar2/errBar_nspokes_T1_overall.dat};
\addplot+[
		mark=x,
		only marks,
        error bars/.cd,
            y dir= both,
            y explicit,
    ] table[x=ns,y=val,y error=err] {NumericalExperiment_simulated/data/errBar2/errBar_nspokes_T1_select1.dat};
\legend{overall, excluding tissue B/C}
\end{axis}

\begin{axis}[
ymode=log,
ymin = 0.03,
ymax = 0.4,
title = {$T_2$ Error},
xlabel = number of k-space spokes,
%	ylabel = Relative Error,
yticklabel = \empty,
at=(T1.right of north east),
anchor = left of north west,
	]
\addplot+[
	mark=x,
	only marks,
        error bars/.cd,
            y dir= both,
            y explicit,
    ] table[x=ns,y=val,y error=err] {NumericalExperiment_simulated/data/errBar2/errBar_nspokes_T2_overall.dat};
\addplot+[
		mark=x,
		only marks,
        error bars/.cd,
            y dir= both,
            y explicit,
    ] table[x=ns,y=val,y error=err] {NumericalExperiment_simulated/data/errBar2/errBar_nspokes_T2_select1.dat};
    \legend{overall, excluding tissue B/C}
\end{axis}
\end{tikzpicture}
\caption{The normalized root-mean-square errors (NRMSE) in the $T_1$ and $T_2$ estimates are plotted as a function of the number of radial k-space spokes at a fixed SNR $ = 1000$. The NRMSE is defined as the root-mean square of the $\Delta T_{1,2} / T_{1,2}$. The results are averaged over 10 realizations of the noise, and the hardly-visible error bars depict one standard deviation of the variation between different noise realizations. The NRMSE was also computed excluding voxels containing a combination of tissue B and C, because the corresponding fingerprints are highly correlated. }
\label{fig:err_nspokes}
%\end{center}
\end{figure}

%\caption{The relative error (the ratio of error to the ground truth value) in the $T_1$ and $T_2$ estimates obtained using different number of radial k-space spokes for a fixed $SNR = 1000$. The results are averaged over 10 repetitions. Error bars with length equal to one standard deviation are included but barely visible. % (the largest standard deviation is 0.0223).The error is also computed excluding voxels containing a combination of tissue B and C, because the corresponding fingerprints are almost indistinguishable. }

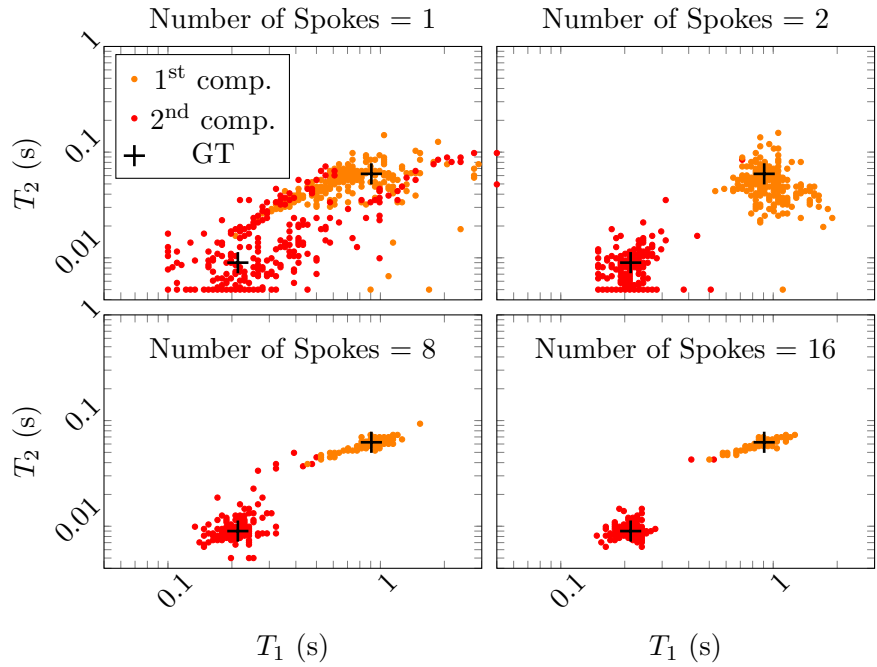
\begin{figure}[htbp]
\centering
\begin{tikzpicture}
\begin{loglogaxis}[
	title = {Number of Spokes = $1$},
	width=0.4\textwidth, height=0.3\textwidth,
	title style = {yshift = \scatterTitleYshift pt},
	xmax = \scatterToneMax,
	ymax = \scatterTtwoMax,
	xmin = \scatterToneMin,
	ymin = \scatterTtwoMin,
	major tick length = \scatterMajorTickLen,
	xticklabel style={/pgf/number format/fixed},
	xticklabel={\pgfmathfloatparsenumber{\tick} \pgfmathfloatexp{\pgfmathresult} \pgfmathprintnumber{\pgfmathresult}},
	yticklabel style={/pgf/number format/fixed},
	yticklabel={\pgfmathfloatparsenumber{\tick} \pgfmathfloatexp{\pgfmathresult} \pgfmathprintnumber{\pgfmathresult}},
   	yticklabel style = {xshift=\scatterTickShift ex, rotate=\scatterYTickRotate},
	ylabel=$T_2$ (s),
	xticklabel=\empty,
	xlabel shift = \scatterXLabelShift pt,
	ylabel shift = \scatterYLabelShift pt,
	legend pos=north west,
	name=spokes1,
]
\addplot [only marks,color=orange,mark size=1pt, mark=*]
table[x = T1, y = T2, col sep=space]{NumericalExperiment_simulated/data/num_grid/s1n3/scatter_3_1.dat};

\addplot [only marks,color=red, mark size=1pt,mark=*]
	table[x = T1, y = T2, col sep=space]{NumericalExperiment_simulated/data/num_grid/s1n3/scatter_3_2.dat};
\addplot [only marks,color=black,mark=+,mark size=4pt,mark options={line width=1pt}]
	coordinates{(0.2133,0.0090)(0.9066,0.0623)};
\legend{1\textsuperscript{st} comp., 2\textsuperscript{nd} comp., GT}
\end{loglogaxis}

\begin{loglogaxis}[
	title = {Number of Spokes = $2$},
	width=0.4\textwidth, height=0.3\textwidth,
	title style = {yshift = \scatterTitleYshift pt},
	xmax = \scatterToneMax,
	ymax = \scatterTtwoMax,
	xmin = \scatterToneMin,
	ymin = \scatterTtwoMin,
	major tick length = \scatterMajorTickLen,
	xticklabel style={/pgf/number format/fixed},
	xticklabel={\pgfmathfloatparsenumber{\tick} \pgfmathfloatexp{\pgfmathresult} \pgfmathprintnumber{\pgfmathresult}},
	yticklabel style={/pgf/number format/fixed},
	yticklabel={\pgfmathfloatparsenumber{\tick} \pgfmathfloatexp{\pgfmathresult} \pgfmathprintnumber{\pgfmathresult}},
	xticklabel=\empty,
	yticklabel=\empty,
	xlabel shift = \scatterXLabelShift pt,
	ylabel shift = \scatterYLabelShift pt,
	name=spokes2,
	at=(spokes1.north east),
	anchor=north west,
	xshift = 0.2cm,
]
\addplot [only marks,color=red, mark size=1pt,mark=*]
	table[x = T1, y = T2, col sep=space]{NumericalExperiment_simulated/data/num_grid/s2n3/scatter_3_2.dat};
\addplot [only marks,color=orange,mark size=1pt, mark=*]
	table[x = T1, y = T2, col sep=space]{NumericalExperiment_simulated/data/num_grid/s2n3/scatter_3_1.dat};
\addplot [only marks,color=black,mark=+,mark size=4pt,mark options={line width=1pt}]
	coordinates{(0.2133,0.0090)(0.9066,0.0623)};
\end{loglogaxis}

\begin{loglogaxis}[
	title = {Number of Spokes = $8$},
	width=0.4\textwidth, height=0.3\textwidth,
	title style = {yshift = -1cm},
	xmax = \scatterToneMax,
	ymax = \scatterTtwoMax,
	xmin = \scatterToneMin,
	ymin = \scatterTtwoMin,
	major tick length = \scatterMajorTickLen,
	xticklabel style={/pgf/number format/fixed},
	xticklabel={\pgfmathfloatparsenumber{\tick} \pgfmathfloatexp{\pgfmathresult} \pgfmathprintnumber{\pgfmathresult}},
	yticklabel style={/pgf/number format/fixed},
	yticklabel={\pgfmathfloatparsenumber{\tick} \pgfmathfloatexp{\pgfmathresult} \pgfmathprintnumber{\pgfmathresult}},
        xticklabel style = {yshift=\scatterTickShift ex, rotate=\scatterXTickRotate},
        	yticklabel style = {xshift=\scatterTickShift ex, rotate=\scatterYTickRotate},
	xlabel=$T_1$ (s),
	ylabel=$T_2$ (s),
	xlabel shift = \scatterXLabelShift pt,
	ylabel shift = \scatterYLabelShift pt,
	name=spokes8,
	at=(spokes1.south west),
	anchor=north west,
	yshift = -0.2cm,
]
\addplot [only marks,color=red, mark size=1pt,mark=*]
	table[x = T1, y = T2, col sep=space]{NumericalExperiment_simulated/data/num_grid/s8n3/scatter_3_2.dat};
\addplot [only marks,color=orange,mark size=1pt, mark=*]
	table[x = T1, y = T2, col sep=space]{NumericalExperiment_simulated/data/num_grid/s8n3/scatter_3_1.dat};
\addplot [only marks,color=black,mark=+,mark size=4pt,mark options={line width=1pt}]
	coordinates{(0.2133,0.0090)(0.9066,0.0623)};
\end{loglogaxis}

\begin{loglogaxis}[
	title ={Number of Spokes = $16$},
	width=0.4\textwidth, height=0.3\textwidth,
	title style = {yshift = -1cm},
	xmax = \scatterToneMax,
	ymax = \scatterTtwoMax,
	xmin = \scatterToneMin,
	ymin = \scatterTtwoMin,
	major tick length = \scatterMajorTickLen,
	xticklabel style={/pgf/number format/fixed},
	xticklabel={\pgfmathfloatparsenumber{\tick} \pgfmathfloatexp{\pgfmathresult} \pgfmathprintnumber{\pgfmathresult}},
	yticklabel style={/pgf/number format/fixed},
	yticklabel={\pgfmathfloatparsenumber{\tick} \pgfmathfloatexp{\pgfmathresult} \pgfmathprintnumber{\pgfmathresult}},
	yticklabel=\empty,
        xticklabel style = {yshift=\scatterTickShift ex, rotate=\scatterXTickRotate},
        	yticklabel style = {xshift=\scatterTickShift ex, rotate=\scatterYTickRotate},
	xlabel=$T_1$ (s),
%	label style={font=\scatterLabelFont},
	xlabel shift = \scatterXLabelShift pt,
	ylabel shift = \scatterYLabelShift pt,
	name=spokes16,
	at=(spokes8.north east),
	anchor=north west,
	xshift = 0.2cm,
]
\addplot [only marks,color=red, mark size=1pt,mark=*]
	table[x = T1, y = T2, col sep=space]{NumericalExperiment_simulated/data/num_grid/s16n3/scatter_3_2.dat};
\addplot [only marks,color=orange,mark size=1pt, mark=*]
	table[x = T1, y = T2, col sep=space]{NumericalExperiment_simulated/data/num_grid/s16n3/scatter_3_1.dat};
\addplot [only marks,color=black,mark=+,mark size=4pt,mark options={line width=1pt}]
	coordinates{(0.2133,0.0090)(0.9066,0.0623)};
\end{loglogaxis}
\end{tikzpicture}
\caption{Scatter plots showing the estimated $T_1$ and $T_2$ values in the multicompartment model for voxels containing tissue A and C. Different number of radial k-spaces spokes are tested at a fixed SNR of $10^3$. The two clusters corresponding to the two tissues concentrate more tightly around the ground truth (GT, marked with black crosses) as the number of spokes increases.}
\label{fig:FATWM_nspokes_conv}
\end{figure}

\begin{figure}[htbp]
	\centering
\begin{tikzpicture}
\begin{axis}[
	xmode=log,
	ymode=log,
	ymin = 0.02,
	ymax = 1,
	title = {$T_1$ Error},
	xlabel = SNR,
	ylabel = NRMSE,
	name = T1,
	]
\addplot+[
mark=x,
only marks,
        error bars/.cd,
            y dir= both,
            y explicit,
    ] table[x=ns,y=val,y error=err] {NumericalExperiment_simulated/data/errBar2/errBar_SNR_T1_overall.dat};
\addplot+[
mark=x,
only marks,
        error bars/.cd,
            y dir= both,
            y explicit,
    ] table[x=ns,y=val,y error=err] {NumericalExperiment_simulated/data/errBar2/errBar_SNR_T1_select1.dat};
    \legend{overall, excluding tissue B/C}
\end{axis}

\begin{axis}[
	xmode=log,
	ymode=log,
	ymin = 0.02,
	ymax = 1,
	title = {$T_2$ Error},
	xlabel = SNR,
%	ylabel = Relative Error,
	yticklabel = \empty,
	at=(T1.right of north east),
	anchor = left of north west,
	]
\addplot+[
mark=x,
only marks,
        error bars/.cd,
            y dir= both,
            y explicit,
    ] table[x=ns,y=val,y error=err] {NumericalExperiment_simulated/data/errBar2/errBar_SNR_T2_overall.dat};
\addplot+[
mark=x,
only marks,
        error bars/.cd,
            y dir= both,
            y explicit,
    ] table[x=ns,y=val,y error=err] {NumericalExperiment_simulated/data/errBar2/errBar_SNR_T2_select1.dat};
    \legend{overall, excluding tissue B/C}
\end{axis}
\end{tikzpicture}
\caption{The normalized root-mean-square errors (NRMSE) in the $T_1$ and $T_2$ estimates are plotted as a function of the SNR when acquiring 8 radial k-space spokes. The NRMSE is defined as the root-mean square of the $\Delta T_{1,2} / T_{1,2}$. The results are averaged over 10 realizations of the noise, and the hardly-visible error bars depict one standard deviation of the variation between different noise realizations. The NRMSE was also computed excluding voxels containing a combination of tissue B and C, because the corresponding fingerprints are highly correlated.}
\label{fig:err_SNR}
%\end{center}
\end{figure}
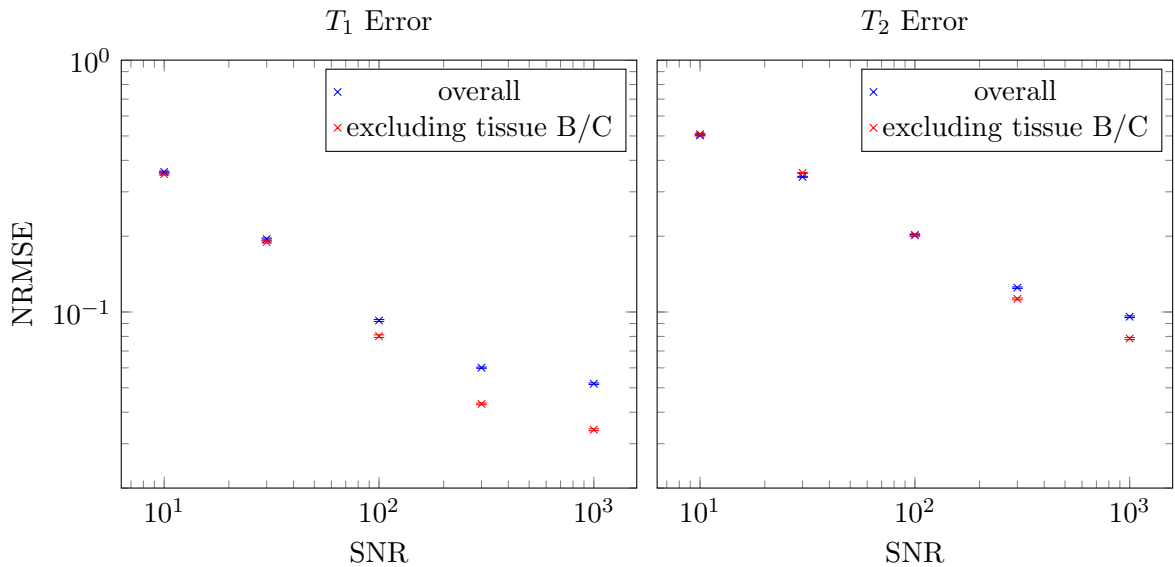

%\caption{The relative error (the ratio of error to the ground truth value) of the $T_1$ and $T_2$ estimates were obtained from 8 radial k-space spokes and with different SNRs. The results are averaged over 10 repetitions, and the error bars correspond to the standard deviation. The relative error is also computed excluding voxels containing both tissue B and C, because the corresponding fingerprints are almost indistinguishable. }

\begin{figure}[htbp]
	\centering
\begin{tikzpicture}
\begin{loglogaxis}[
	width = 0.375\textwidth,
	height = 0.375\textwidth,
	title = {SNR = $10$},
	title style = {yshift = \scatterTitleYshift pt},
	xmax = \scatterToneMax,
	ymax = \scatterTtwoMax,
	xmin = \scatterToneMin,
	ymin = \scatterTtwoMin,
	major tick length = \scatterMajorTickLen,
	xticklabel style={/pgf/number format/fixed},
	xticklabel={\pgfmathfloatparsenumber{\tick} \pgfmathfloatexp{\pgfmathresult} \pgfmathprintnumber{\pgfmathresult}},
	yticklabel style={/pgf/number format/fixed},
	yticklabel={\pgfmathfloatparsenumber{\tick} \pgfmathfloatexp{\pgfmathresult} \pgfmathprintnumber{\pgfmathresult}},
        xticklabel style = {yshift=\scatterTickShift ex, rotate=\scatterXTickRotate},
        	yticklabel style = {xshift=\scatterTickShift ex, rotate=\scatterYTickRotate},
	xlabel=$T_1$ (s),
	ylabel=$T_2$ (s),
%	label style={font=\scatterLabelFont},
	xlabel shift = \scatterXLabelShift pt,
	ylabel shift = \scatterYLabelShift pt,
	legend pos=north west,
	name = SNR10,
]
\addplot [only marks,color=orange,mark size=1pt, mark=*]
table[x = T1, y = T2, col sep=space]{NumericalExperiment_simulated/data/num_grid/s16n1/scatter_3_1.dat};

\addplot [only marks,color=red, mark size=1pt,mark=*]
	table[x = T1, y = T2, col sep=space]{NumericalExperiment_simulated/data/num_grid/s16n1/scatter_3_2.dat};
	
\addplot [only marks,color=black,mark=+,mark size=4pt,mark options={line width=1pt}]
	coordinates{(0.2133,0.0090)(0.9066,0.0623)};
	\legend{1\textsuperscript{st} comp., 2\textsuperscript{nd} comp., GT}
\end{loglogaxis}

\begin{loglogaxis}[
width = 0.375\textwidth,
height = 0.375\textwidth,
	title = {SNR = $100$},
	title style = {yshift = \scatterTitleYshift pt},
	xmax = \scatterToneMax,
	ymax = \scatterTtwoMax,
	xmin = \scatterToneMin,
	ymin = \scatterTtwoMin,
	major tick length = \scatterMajorTickLen,
	xticklabel style={/pgf/number format/fixed},
	xticklabel={\pgfmathfloatparsenumber{\tick} \pgfmathfloatexp{\pgfmathresult} \pgfmathprintnumber{\pgfmathresult}},
	yticklabel style={/pgf/number format/fixed},
	yticklabel={\pgfmathfloatparsenumber{\tick} \pgfmathfloatexp{\pgfmathresult} \pgfmathprintnumber{\pgfmathresult}},
        xticklabel style = {yshift=\scatterTickShift ex, rotate=\scatterXTickRotate},
        	yticklabel style = {xshift=\scatterTickShift ex, rotate=\scatterYTickRotate},
	xlabel=$T_1$ (s),
%	ylabel=$T_2$ (s),
	yticklabel = \empty,
%	label style={font=\scatterLabelFont},
	xlabel shift = \scatterXLabelShift pt,
	ylabel shift = \scatterYLabelShift pt,
	name=SNR100,
	at=(SNR10.right of north east),
	anchor = left of north west,
]
\addplot [only marks,color=red, mark size=1pt,mark=*]
	table[x = T1, y = T2, col sep=space]{NumericalExperiment_simulated/data/num_grid/s16n2/scatter_3_2.dat};
\addplot [only marks,color=orange,mark size=1pt, mark=*]
	table[x = T1, y = T2, col sep=space]{NumericalExperiment_simulated/data/num_grid/s16n2/scatter_3_1.dat};
\addplot [only marks,color=black,mark=+,mark size=4pt,mark options={line width=1pt}]
	coordinates{(0.2133,0.0090)(0.9066,0.0623)};
\end{loglogaxis}

\begin{loglogaxis}[
width = 0.375\textwidth,
height = 0.375\textwidth,
	title = {SNR = $1000$},
	title style = {yshift = \scatterTitleYshift pt},
	xmax = \scatterToneMax,
	ymax = \scatterTtwoMax,
	xmin = \scatterToneMin,
	ymin = \scatterTtwoMin,
	major tick length = \scatterMajorTickLen,
	xticklabel style={/pgf/number format/fixed},
	xticklabel={\pgfmathfloatparsenumber{\tick} \pgfmathfloatexp{\pgfmathresult} \pgfmathprintnumber{\pgfmathresult}},
	yticklabel style={/pgf/number format/fixed},
	yticklabel={\pgfmathfloatparsenumber{\tick} \pgfmathfloatexp{\pgfmathresult} \pgfmathprintnumber{\pgfmathresult}},
        xticklabel style = {yshift=\scatterTickShift ex, rotate=\scatterXTickRotate},
        	yticklabel style = {xshift=\scatterTickShift ex, rotate=\scatterYTickRotate},
	xlabel=$T_1$ (s),
%	ylabel=$T_2$ (s),
	yticklabel=\empty,
%	label style={font=\scatterLabelFont},
	xlabel shift = \scatterXLabelShift pt,
	ylabel shift = \scatterYLabelShift pt,
	name=SNR1000,
	at=(SNR100.right of north east),
	anchor = left of north west,
]
\addplot [only marks,color=red, mark size=1pt,mark=*]
	table[x = T1, y = T2, col sep=space]{NumericalExperiment_simulated/data/num_grid/s16n3/scatter_3_2.dat};
\addplot [only marks,color=orange,mark size=1pt, mark=*]
	table[x = T1, y = T2, col sep=space]{NumericalExperiment_simulated/data/num_grid/s16n3/scatter_3_1.dat};
\addplot [only marks,color=black,mark=+,mark size=4pt,mark options={line width=1pt}]
	coordinates{(0.2133,0.0090)(0.9066,0.0623)};
\end{loglogaxis}
\end{tikzpicture}
\caption{Scatter plots showing the estimated $T_1$ and $T_2$ values in the multicompartment model for voxels containing tissue A and C. The data was reconstructed from 16 radial k-space spokes. The two clusters corresponding to the two tissues concentrate more tightly around the ground truth (GT, marked with black crosses) as the SNR increases.}
\label{fig:FATWM_noise_conv}
\end{figure}
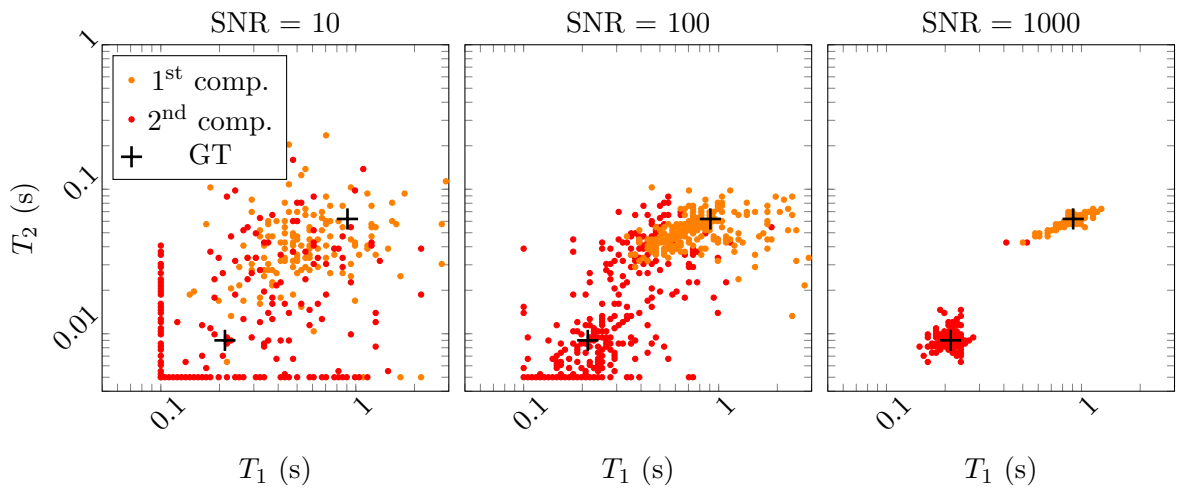

\input{fig_grid4by4}

To evaluate the performance of the method at different undersampling factors, we fit the multicompartment model to the simulated phantom data using different number of radial k-space spokes. Figure \ref{fig:err_nspokes} shows the decrease in the normalized root-mean-square-error (NRMSE) of the $T_1$ and $T_2$ estimates averaged over all voxels in the phantom. As expected, the NRMSE decreases when acquiring more k-space spokes. As explained above, the largest contributor to the NRMSE are the voxels that contain both tissue B and C, so the NRMSE excluding these voxels is also depicted. The scatter plots in Figure~\ref{fig:FATWM_nspokes_conv} show the results at each voxel containing a 50\%-50\% mixture of tissue A and C. The two compartments are separated accurately, even when only two radial k-space spokes are measured. As more samples are gathered, the two clusters corresponding to the two tissues concentrate more tightly around the ground truth. Similarly, when keeping the number of k-space spokes fixed and increasing the SNR, the NRMSE decreases as expected (Figure~\ref{fig:err_SNR}) and the estimated parameters converge to the ground truth (Figure~\ref{fig:FATWM_noise_conv}). The dependence of the reconstruction quality on the correlation of the fingerprints is visualized in Figure~\ref{fig:scatter_all_num_s16n2}, where scatter plots for all tissue combinations are shown. One can observe a dependency of the reconstruction quality on the distance of the two compartments in parameter space. In particular, the combination of tissues B and C, which have highly correlated fingerprints, results in a single compartment fit (see Figure~\ref{fig:wm_gm}). 
 
\subsection{Phantom experiment}
\label{sec:real_phantom}

\begin{figure}[tp]
\begin{center}
\begin{tabular}{ccc}
\includegraphics[width = 6 cm]{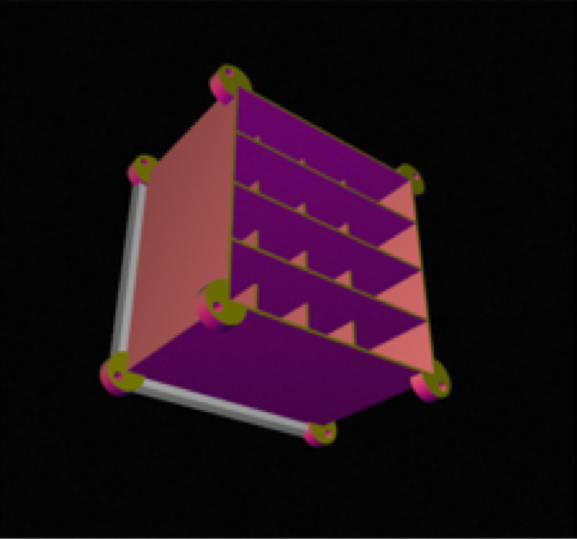}
& &
\includegraphics[width = 6.3 cm]{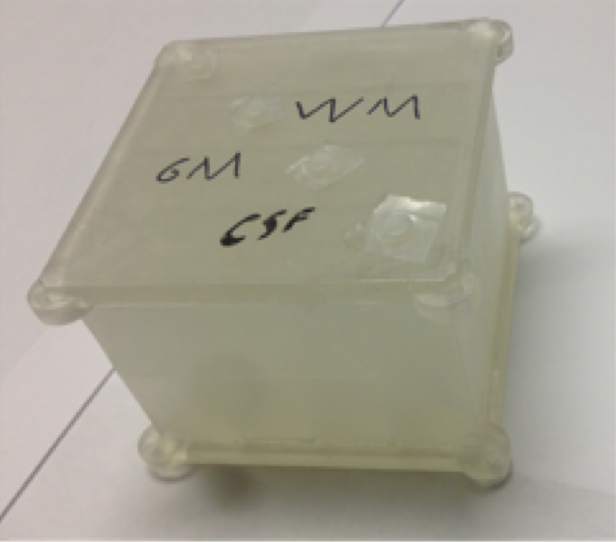}
\end{tabular}
\caption{Computer-aided design (left) and the corresponding 3D-printed phantom (right) used for the phantom evaluation. It contains two layers, one with four horizontal and one with four vertical bars. Imaging a slice that contains both layers results in the same effective structure as the numerical phantom shown in Figure~\ref{tab:grid_description}.}
\label{fig:design_grid}
\end{center}
\end{figure}

For validation on real data, we performed an imaging experiment on a clinical MR system with a field strength of 3 Tesla (Prisma, Siemens, Erlangen, Germany). We designed a dedicated phantom and manufactured it using a 3D printer (Figure \ref{fig:design_grid}). The structure of the phantom is effectively the same as that of the simulated phantom described in Section~\ref{sec:simulated_phantom} and Figure~\ref{tab:grid_description}: the diagonal elements contain only one compartment, whereas the off-diagonal elements contain two compartments. The compartments were filled with different solutions of doped water in order to achieve relaxation times in the range of biological tissues. MRF scans were performed with the sequence described in Fig.~3k of Ref.~\cite{Asslander2017arxiv} with 8 radial k-space spokes per time frame and the same parameters otherwise. 

\begin{figure}[tp]
\begin{center}
\begin{tabular}{ccc}
$T_1$ (s)
&
$T_2$ (s)
&
$PD$ (a.u.)
\\
\includegraphics[width = 5 cm]{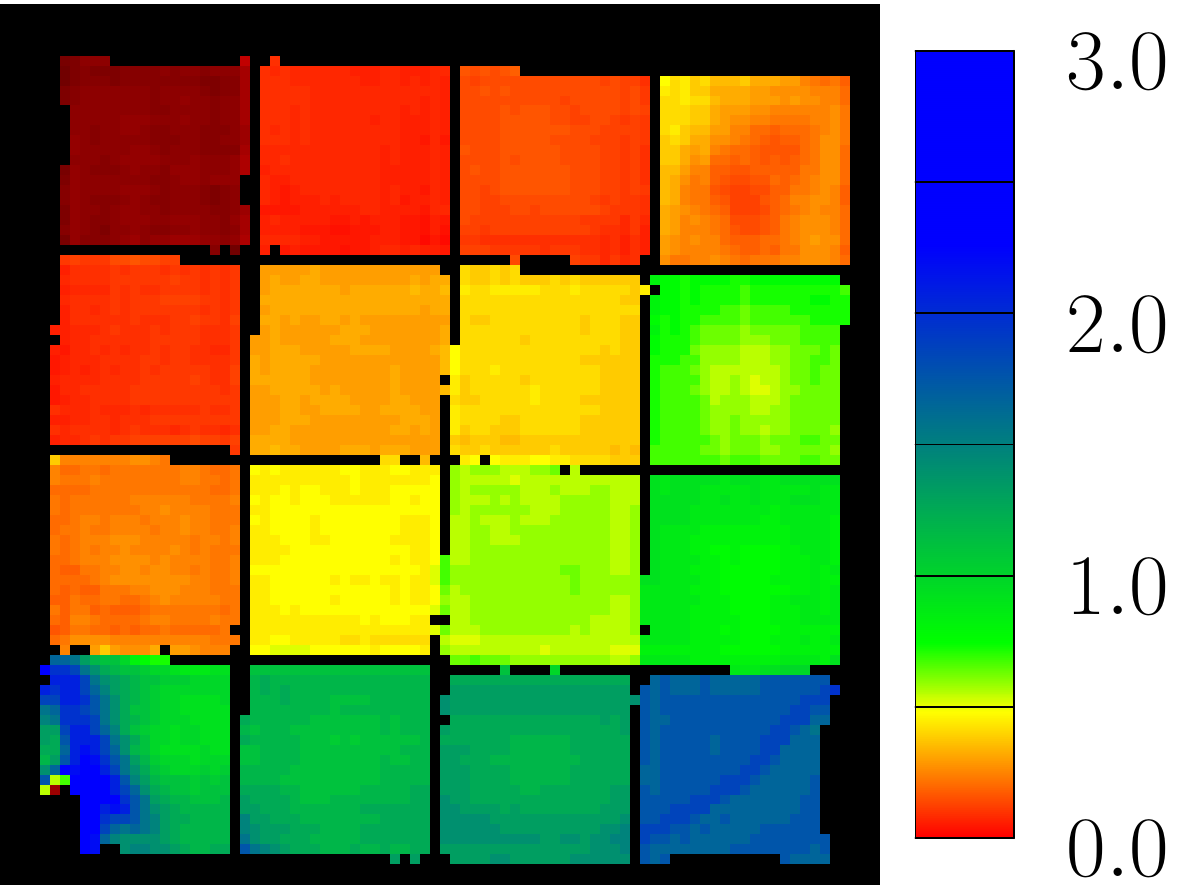}
&
\includegraphics[width = 5 cm]{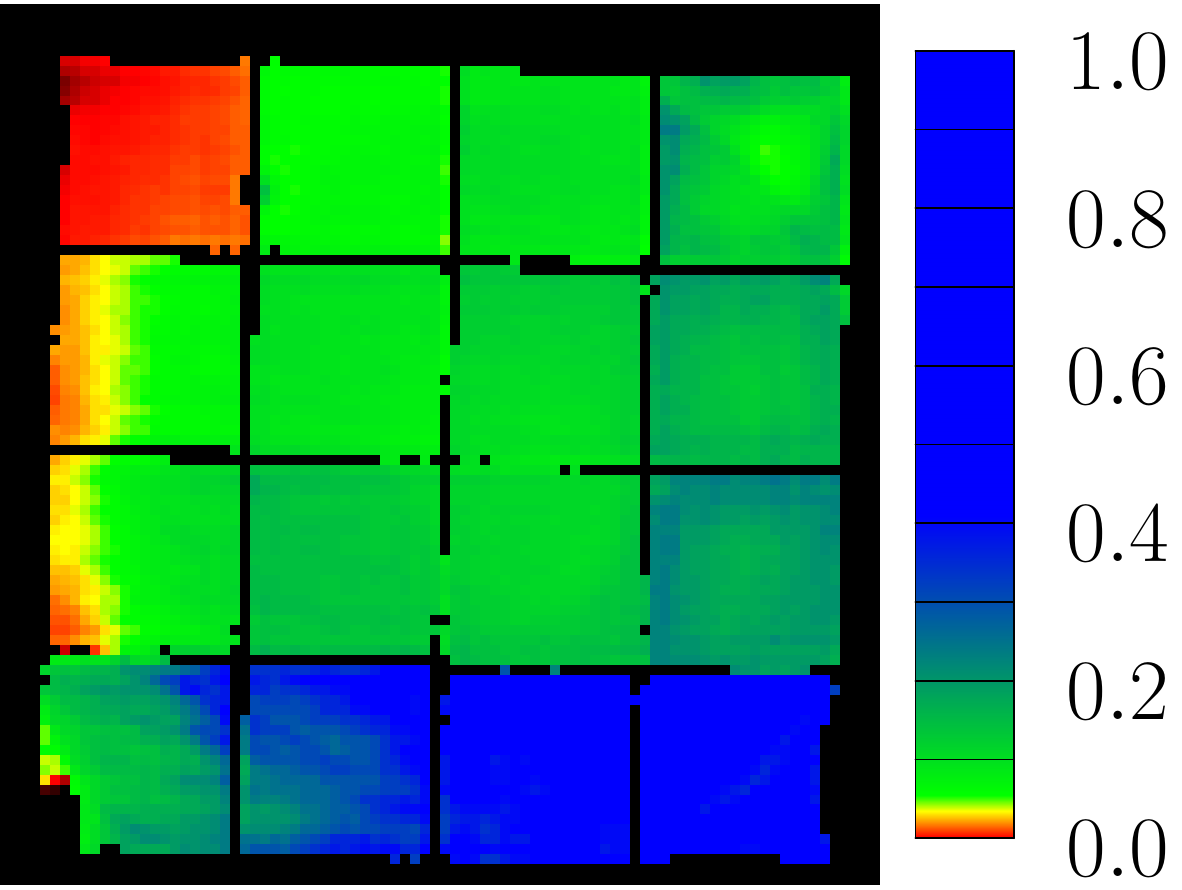}
&
\includegraphics[width = 5 cm]{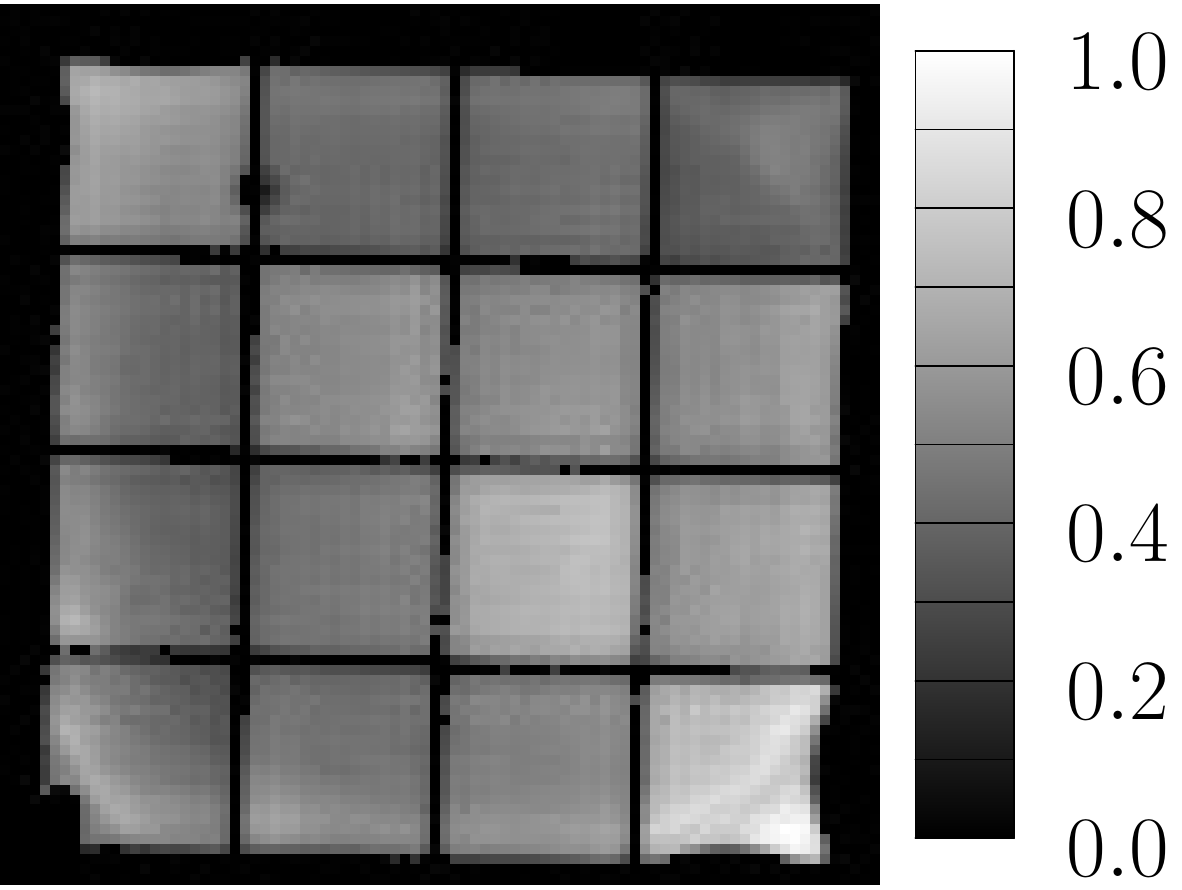}
\end{tabular}
\caption{The depicted parameter maps were reconstructed from data measured in a phantom and using a single-compartment model. The data were acquired using 8 radial k-space spokes. The colorbar is adjusted so that distinct colors correspond to the parameters of the gold-standard reference measurements: red, yellow green and blue represent the solutions A-D.}
\label{fig:hm_real_sc}
\end{center}
\end{figure}
%
%\begin{figure}[tp]
%\begin{center}
%\begin{tabular}{ |c | c | c | c | c | }
%       \hline&Fat & White matter (WM) & Gray matter (GM) & Cerebrospinal fluid (CSF) \\ \hline
%    $T_1 (s)$ & 0.1262 & 0.5075 & 0.7640  & 1.8308 \\ \hline
%    $T_2 (s)$ & $0.0174$ &  $0.1019$ & $0.1328$  &0.6573 \\ \hline
%  \end{tabular} 
%\end{center}
%   \caption{Gold standard reference measurement of the tissues present in the real phantom of Section~\ref{sec:real_phantom}. The values are estimated from the single-compartment reconstruction of areas in the phantom where only one compartment is present.}
%   \label{tab:gold_standard}
%\end{figure}

Figure \ref{fig:hm_real_sc} shows the parameter maps reconstructed with a single-compartment model and a low-rank ADMM algorithm \cite{asslander2017low}. For the diagonal elements of the grid phantom, which only contain signal from a single compartment, we consider the identified $T_1$ and $T_2$ values listed in Table~\ref{tab:gold_standard} as the reference gold standard. In the off-diagonal elements, the single-compartment model fails to account for the presence of two compartments, and the estimates are apparent relaxation times that lie in between the underlying values of the two compartments. Note that the square shape of the phantom causes significant variations in the main magnetic field towards the edges of the phantom. Since we do not account for those variations, substantial artifacts can be observed in those areas (Figure~\ref{fig:hm_real_sc}).

\begin{figure}
%\begin{tabular}{ccccc}
\centering
\begin{tabular}{ >{\centering\arraybackslash}m{0.225\linewidth}  >{\centering\arraybackslash}m{0.225\linewidth} >{\centering\arraybackslash}m{0.225\linewidth} >{\centering\arraybackslash}m{0.05\linewidth} >{\centering\arraybackslash}m{0.08\linewidth}}
First Compartment&
Second Compartment&
Third Compartment&
\\
\includegraphics[width = \heatSubPlotWidth cm]{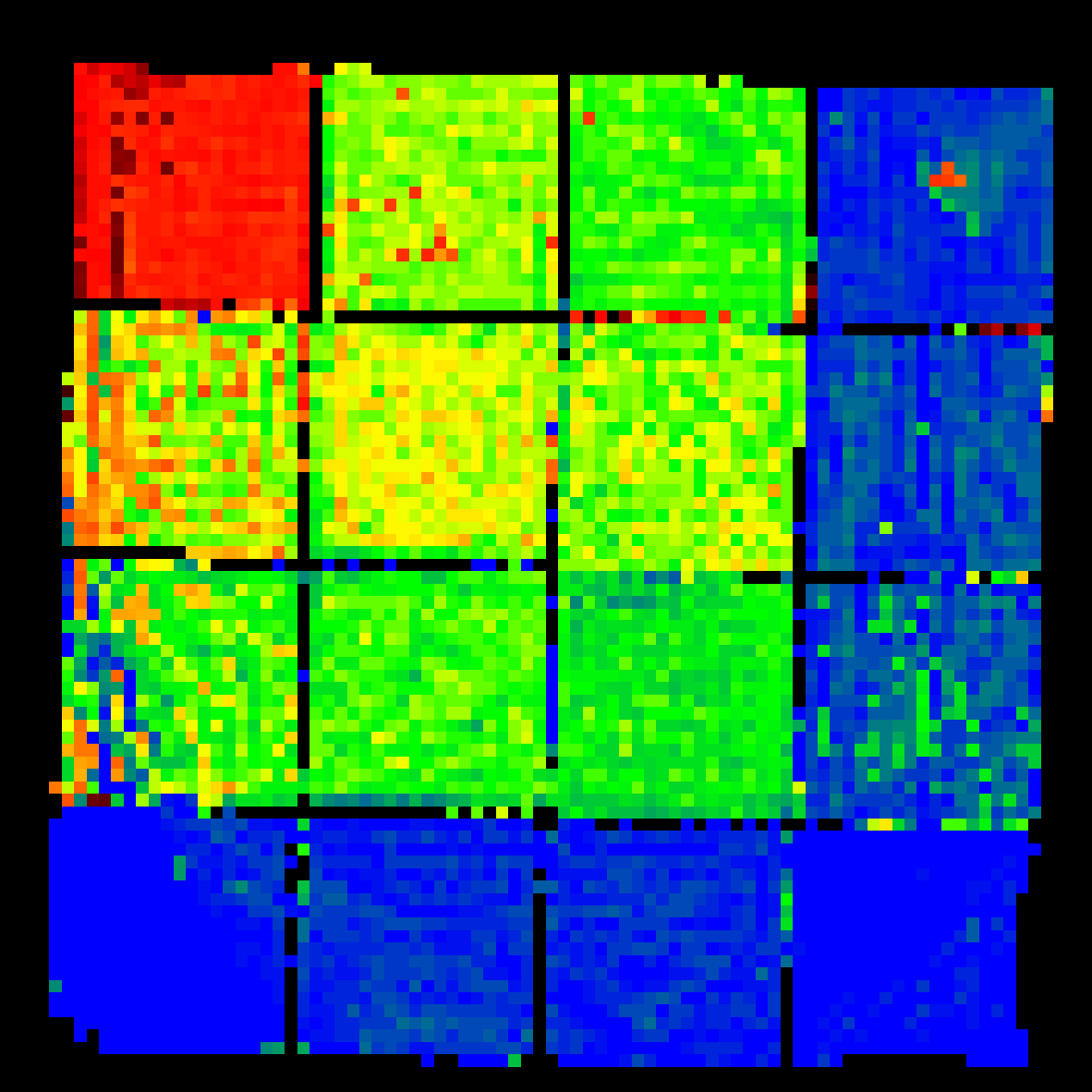}
&
\includegraphics[width = \heatSubPlotWidth cm]{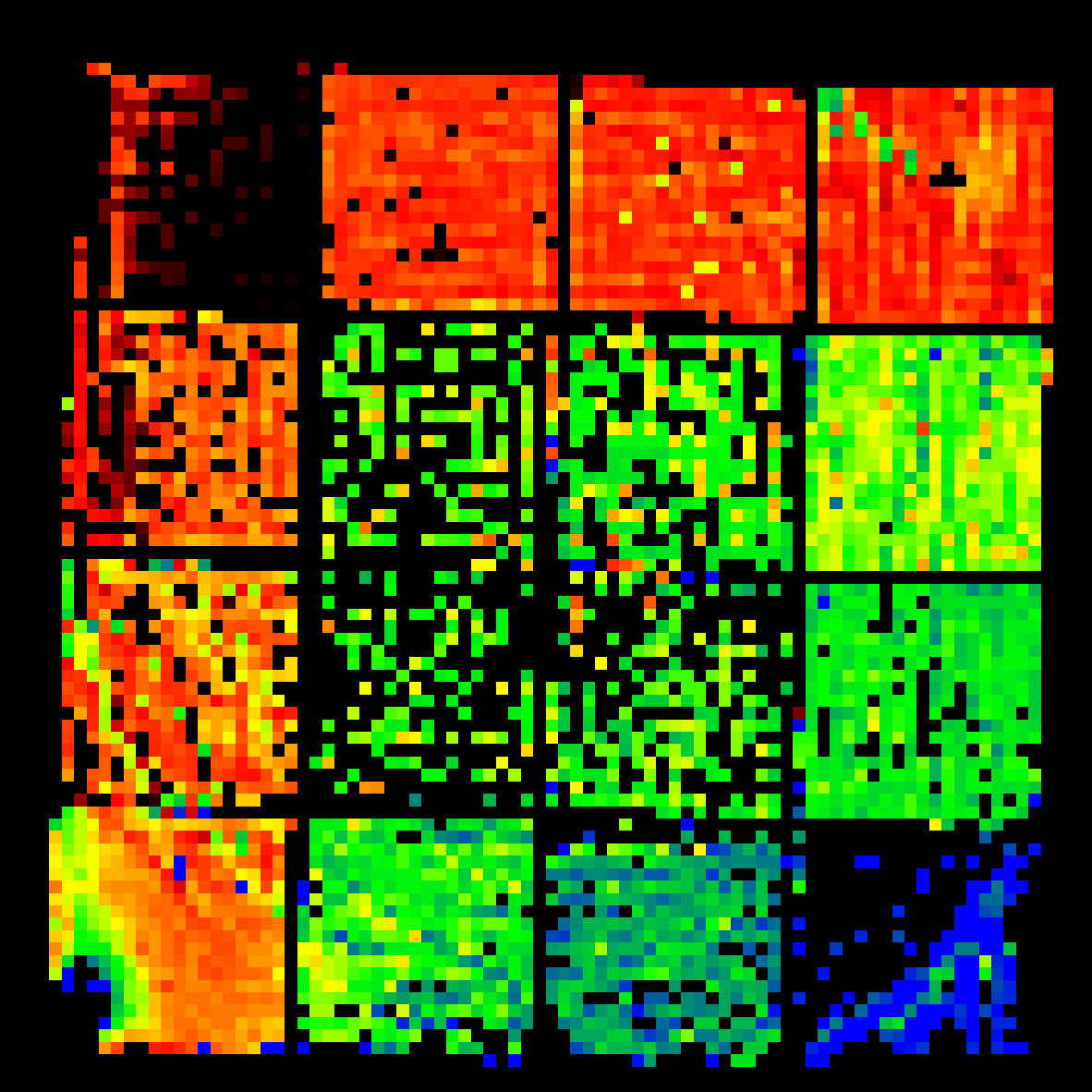}
&
\includegraphics[width = \heatSubPlotWidth cm]{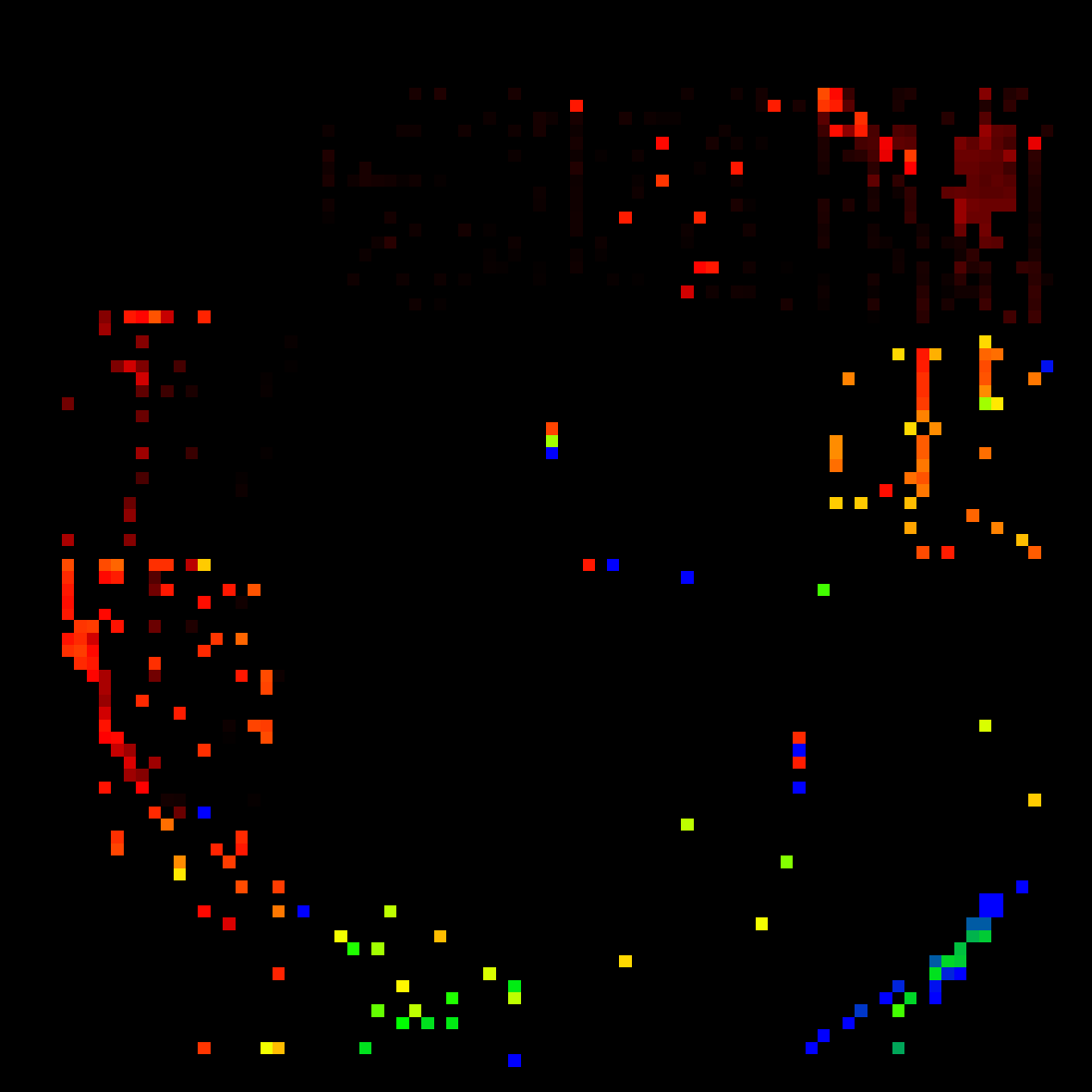}
&
\includegraphics[width = 1.5 cm]{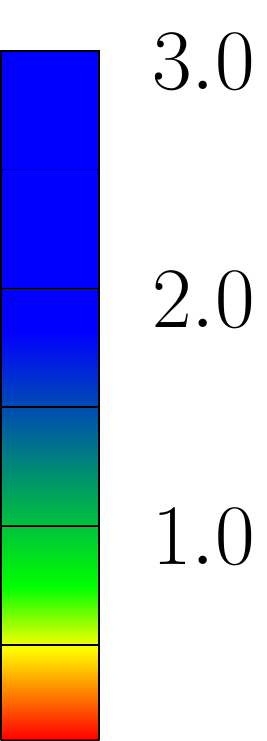}
&
\rotatebox[origin=c]{90}{$T_1$~(s)}
\\
\includegraphics[width = \heatSubPlotWidth cm]{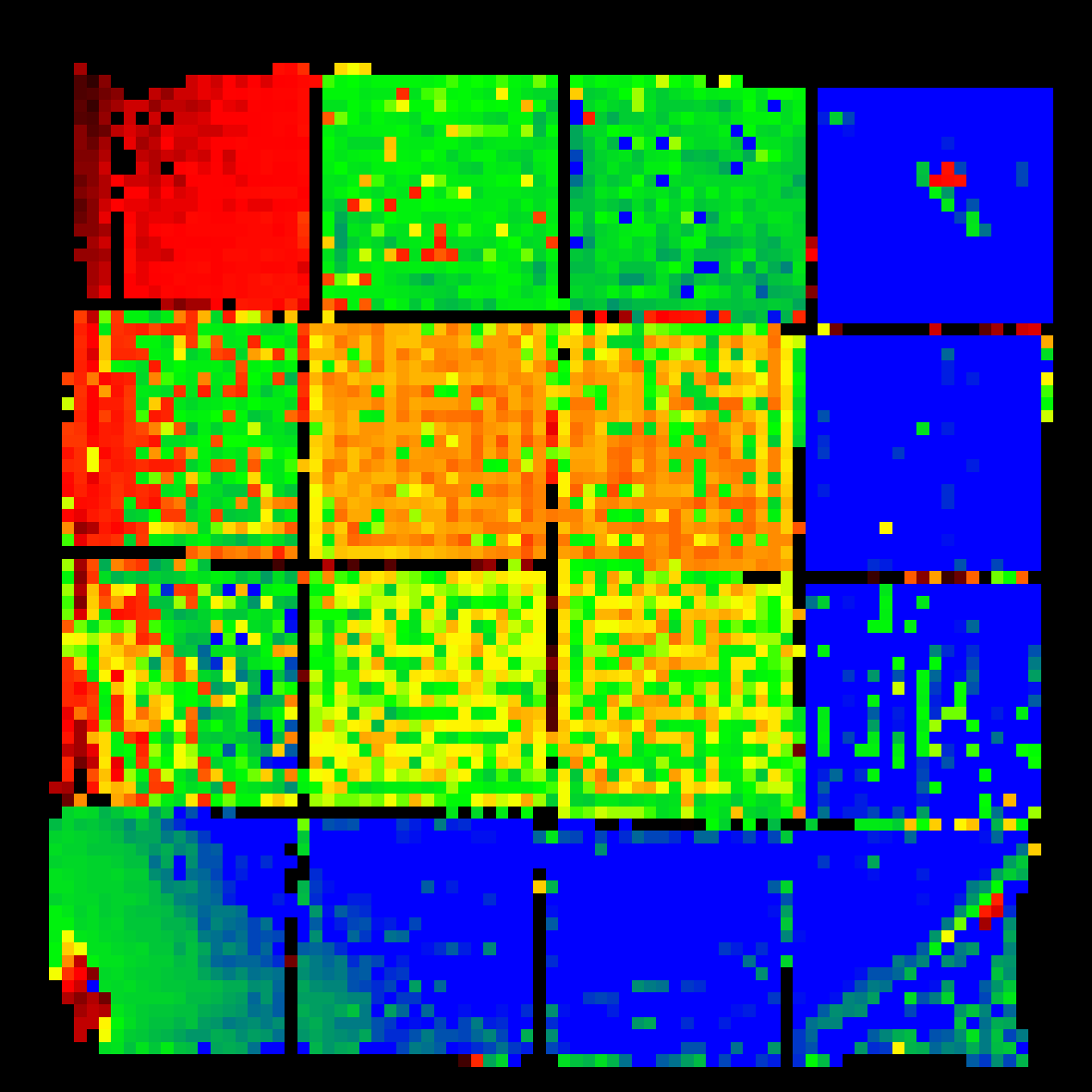}
&
\includegraphics[width = \heatSubPlotWidth cm]{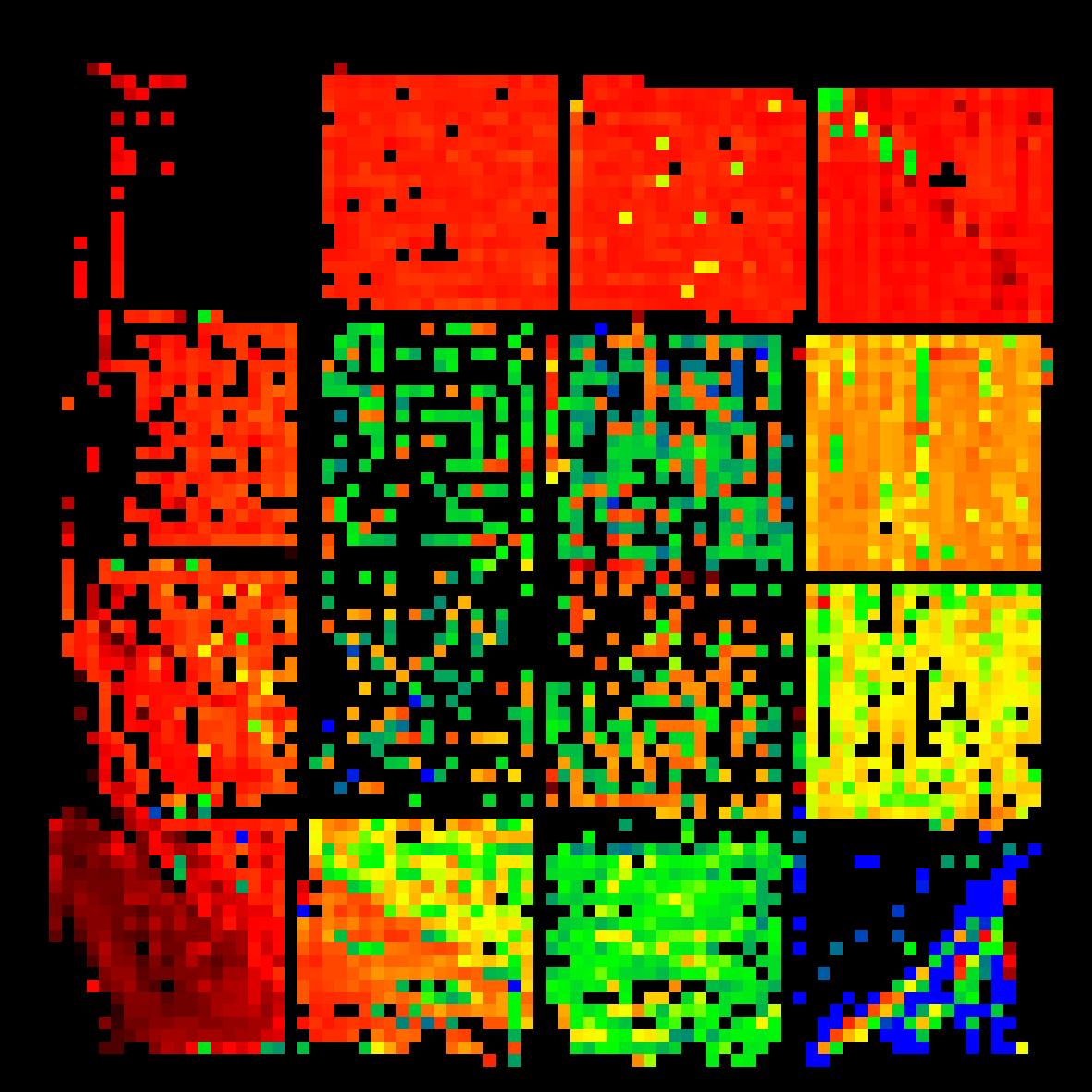}
&
\includegraphics[width = \heatSubPlotWidth cm]{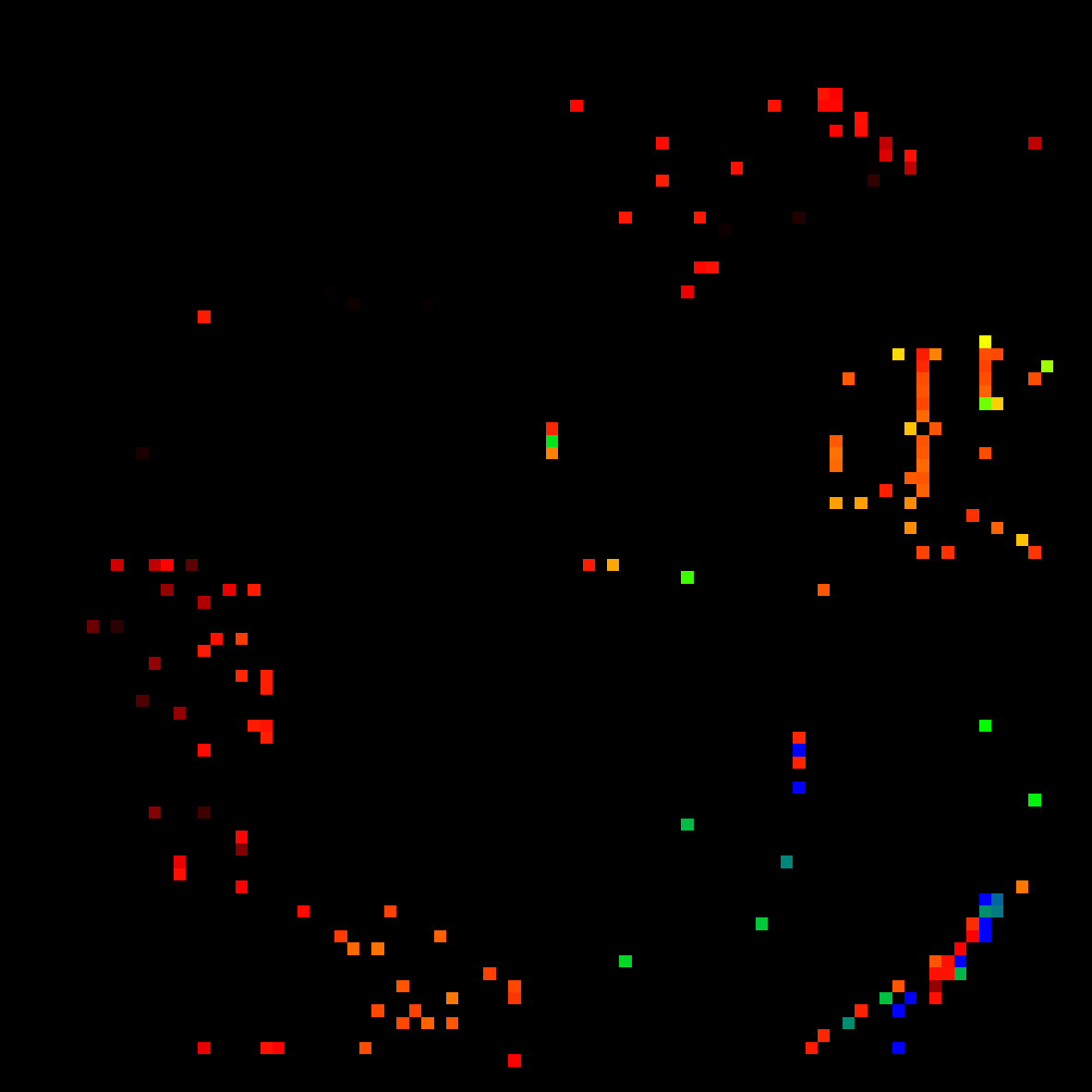}
&
\includegraphics[width = 1.5 cm]{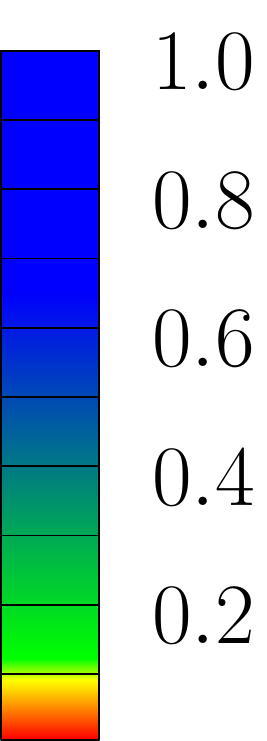}
&
\rotatebox[origin=c]{90}{$T_2$~(s)}
\\
\includegraphics[width = \heatSubPlotWidth cm]{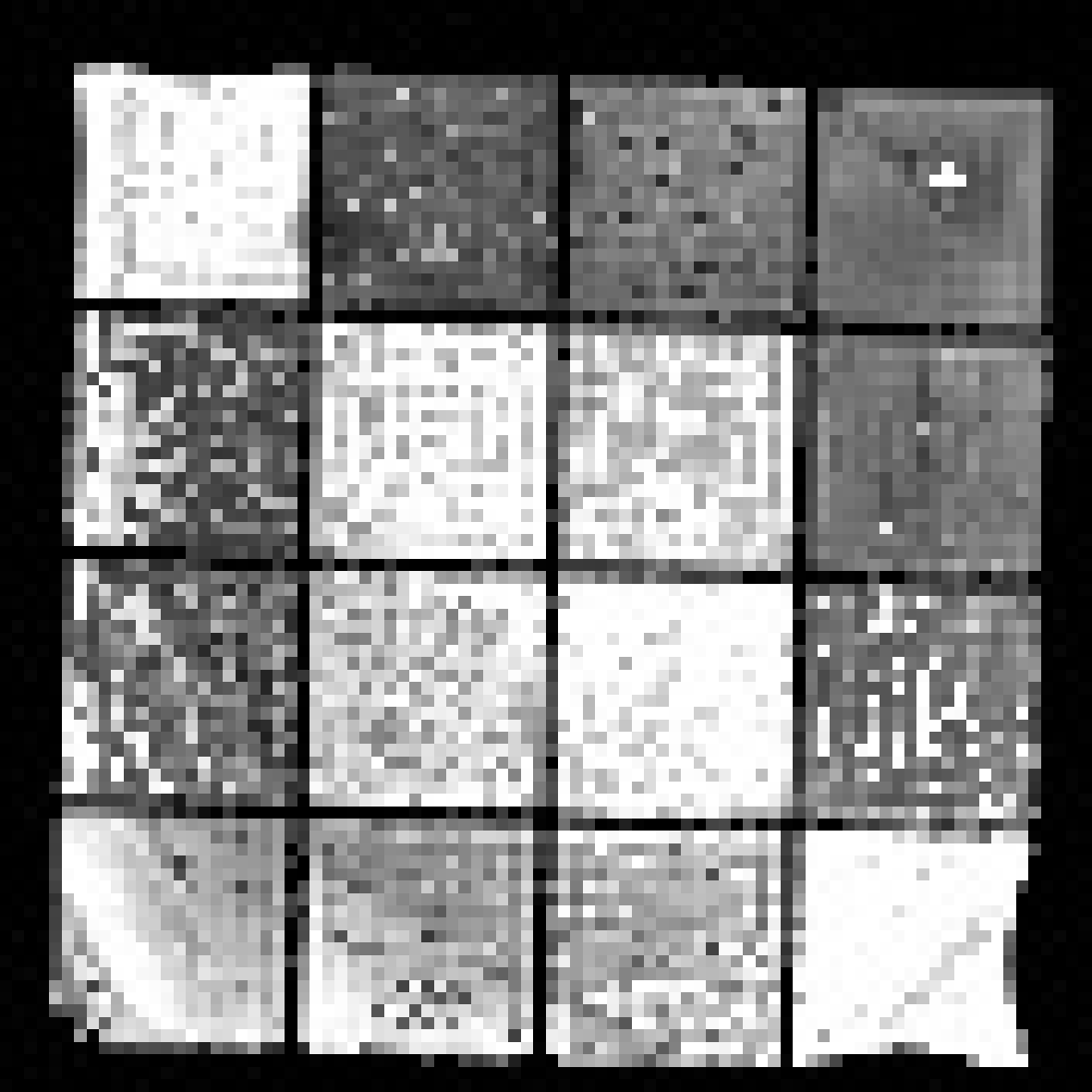}
&
\includegraphics[width = \heatSubPlotWidth cm]{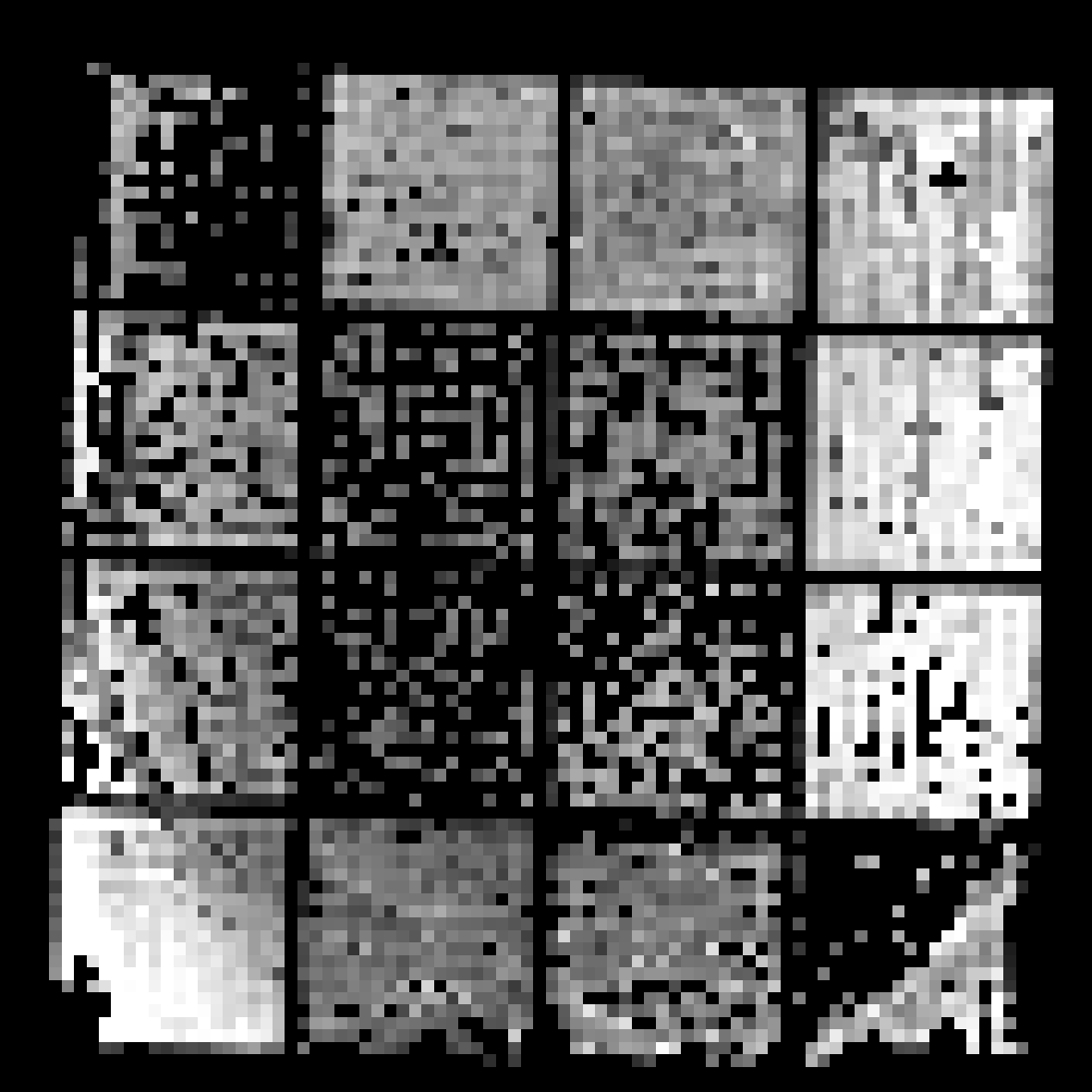}
&
\includegraphics[width = \heatSubPlotWidth cm]{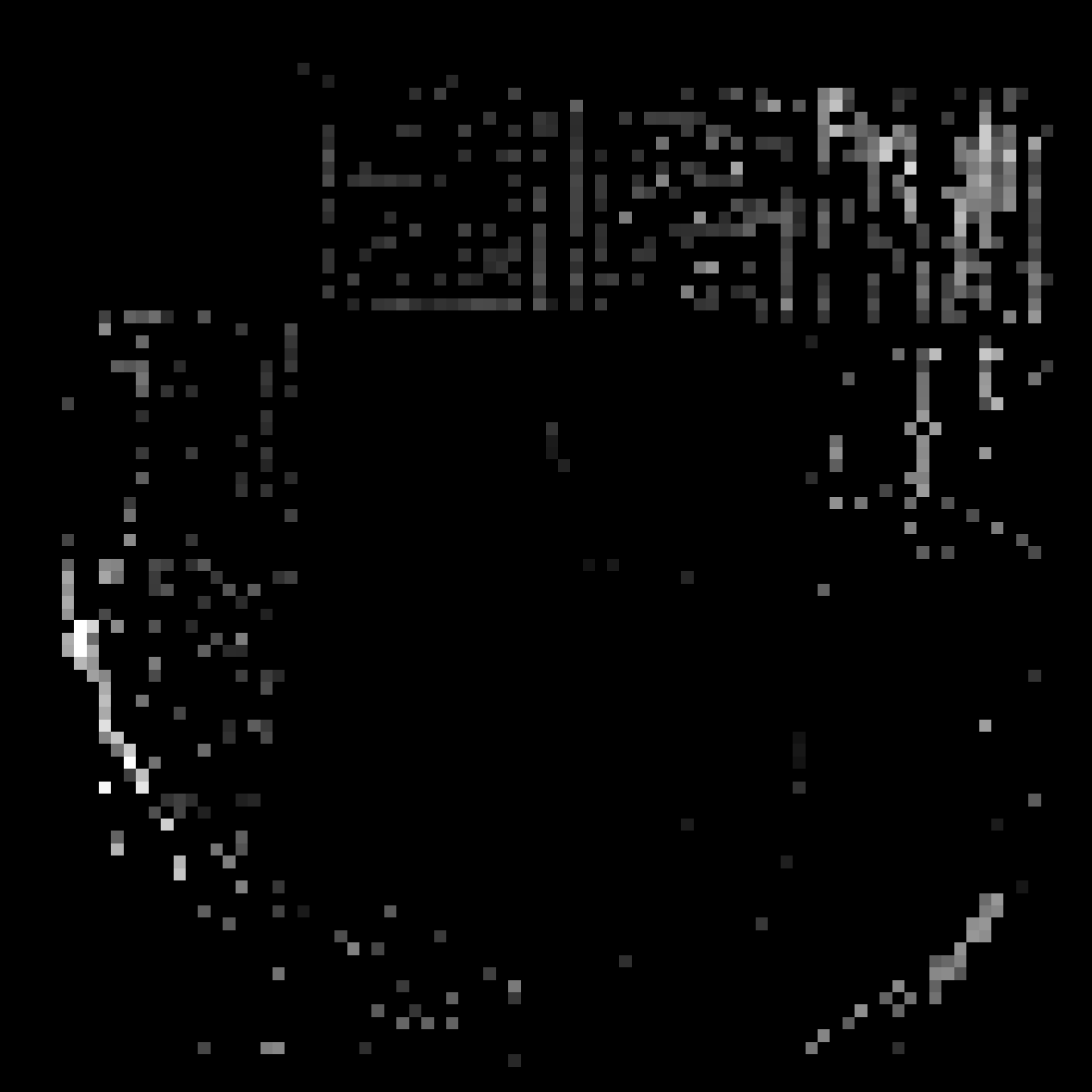}
&
\includegraphics[width = 1.5 cm]{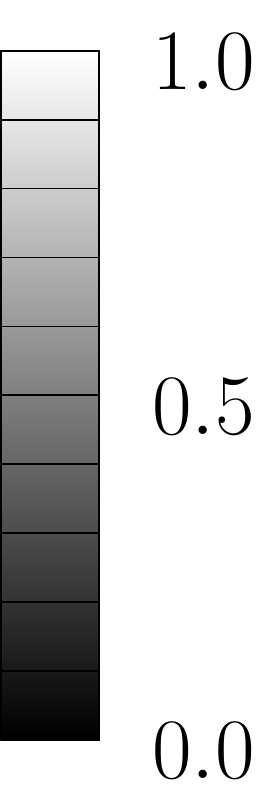}
&
\rotatebox[origin=c]{90}{$PD$~(a.u.)}
\\
\end{tabular}
\caption{The parameter maps of the scanned phantom was reconstructed using the proposed multicompartment method. The data were acquired using eight radial k-space spokes. The three compartments with the highest detected proton density are shown here, and are sorted in each voxel according to their $\ell_2$-norm distance to the origin in $T_1$-$T_2$ space. The colorbar is adjusted so that distinct colors correspond to the parameters of the gold-standard reference measurements: red, yellow green and blue represent the four solutions.}
\label{fig:hm_real_mc}
\end{figure}

Figure \ref{fig:hm_real_mc} shows the parameter maps reconstructed with the proposed multicompartment reconstruction method. At the diagonal elements, results are consistent with the single-compartment reconstruction, except for some voxels where a spurious second compartment appears with low proton density. This indicates that the method correctly identifies voxels containing only a single compartment. In the off-diagonal grid elements, two significant compartments are detected for most voxels, and the relaxation times coincide with the gold-standard within a standard deviation (Table~\ref{tab:value_real_select}). However, the noise level is quite severe at the given experimental conditions. Further, a spurious third compartment with low proton density is detected in some voxels. In contrast, many voxels containing the combination of the doped water solutions B and C are reconstructed as a single compartment with apparent relaxation times between the gold-standard values. This is expected, since the corresponding fingerprints are almost indistinguishable, as in the case of the simulated tissues B and C (c.f. Fig.~\ref{fig:noiseless}). Errors are larger for $T_2$ than for $T_1$, as expected from the Cram\'er-Rao bound estimation performed in~\cite{Asslander2017arxiv}. The artifacts towards the edge of the phantom are due to variations in the main magnetic field (c.f. to the single compartment reconstruction in Figure~\ref{fig:hm_real_sc}).  

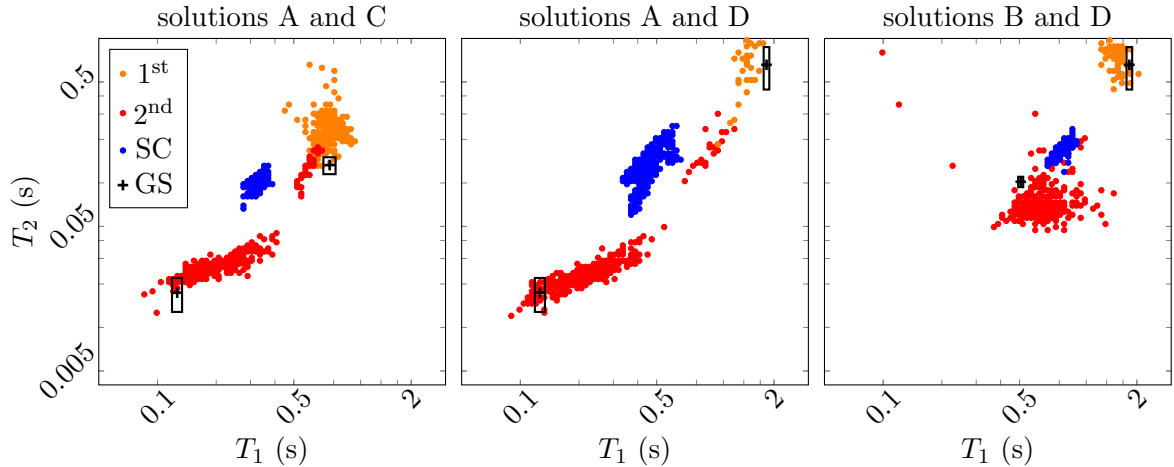
\begin{figure}[tp]
\begin{tikzpicture}
\begin{loglogaxis}[
	width = 0.375\textwidth,
	height = 0.375\textwidth,
	title = solutions A and C,
	title style = {yshift = \scatterTitleYshift pt},
	xmax = \scatterToneMax,
	ymax = \scatterTtwoMax,
	xmin = \scatterToneMin,
	ymin = \scatterTtwoMin,
	major tick length = \scatterMajorTickLen,
	xtick={0.1,0.5,2},
	xticklabels={0.1,0.5,2},
	extra x ticks = {.1,.2,.3,.4,.5,.8,1.1,1.4,1.7,2},extra x tick labels={},
	ytick={.005, .05, .5,1},
	yticklabels={0.005, 0.05, 0.5},
	extra y ticks = {.005, .01,.02,.03,.04,.05, .1,.2,.3,.4, .5}, extra y tick labels={},
    xticklabel style = {yshift=\scatterTickShift ex, rotate=\scatterXTickRotate},
   	yticklabel style = {xshift=\scatterTickShift ex, rotate=\scatterYTickRotate},
	xlabel=$T_1$ (s),
	ylabel=$T_2$ (s),
%	label style={font=\scatterLabelFont},
	xlabel shift = \scatterXLabelShift pt,
	ylabel shift = \scatterYLabelShift pt,
	legend pos=north west,
	name=left,
	clip mode=individual,
	set layers
]
\addplot [only marks,color=orange,mark size=1pt, mark=*,on layer=axis background]
table[x = T1, y = T2, col sep=space]{NumericalExperiment_simulated/data/real_grid/mc/scatter_3_1.dat};
\addplot [only marks,color=red, mark size=1pt,mark=*,on layer=axis background]
	table[x = T1, y = T2, col sep=space]{NumericalExperiment_simulated/data/real_grid/mc/scatter_3_2.dat};
\addplot [only marks,color=blue, mark size=1pt,mark=*,on layer=axis background]
	table[x = T1, y = T2, col sep=space]{NumericalExperiment_simulated/data/real_grid/sc/scatter_3.dat};
\addplot [only marks,color=black,mark=+,mark size=2pt,mark options={line width=1pt}]
	coordinates{(0.1262,0.0174)(0.7640,0.1328)};
%\addplot [very thick] (0.1262-0.0075,0.0174-0.0046) rectangle (0.1262+0.0075,0.0174+0.0046);
\draw  [black,thick,on layer = axis foreground]
	(axis cs: 0.1262-0.0075,0.0174-0.0046) rectangle (axis cs: 0.1262+0.0075,0.0174+0.0046);
\draw  [black,thick,on layer = axis foreground]
	(axis cs: 0.7640-0.0550,0.1328-0.0178) rectangle (axis cs: 0.7640+0.0550,0.1328+0.0178);
\legend{1\textsuperscript{st}, 2\textsuperscript{nd}, SC, GS}
\end{loglogaxis}

\begin{loglogaxis}[
	width = 0.375\textwidth,
	height = 0.375\textwidth,
	title = solutions A and D,
	scale=\subScatterPlotScale,
	title style = {font = \scatterTitleFont,yshift = \scatterTitleYshift pt},
	xmax = \scatterToneMax,
	ymax = \scatterTtwoMax,
	xmin = \scatterToneMin,
	ymin = \scatterTtwoMin,
	major tick length = \scatterMajorTickLen,
	xtick={0.1,0.5,2},
	xticklabels={0.1,0.5,2},
	extra x ticks = {.1,.2,.3,.4,.5,.8,1.1,1.4,1.7,2},extra x tick labels={},
	ytick={.005, .05, .5,1},
	yticklabels={0.005, 0.05, 0.5},
	extra y ticks = {.005, .01,.02,.03,.04,.05, .1,.2,.3,.4, .5}, extra y tick labels={},
        xticklabel style = {font=\scatterTickFont,yshift=\scatterTickShift ex, rotate=\scatterXTickRotate},
        	yticklabel style = {font=\scatterTickFont,xshift=\scatterTickShift ex, rotate=\scatterYTickRotate},
	xlabel=$T_1$ (s),
%	ylabel=$T_2$ (s),
	yticklabel=\empty,
	label style={font=\scatterLabelFont},
	xlabel shift = \scatterXLabelShift pt,
	ylabel shift = \scatterYLabelShift pt,
	name=middle,
	at=(left.right of north east),
	anchor=left of north west,
	clip mode=individual,
	set layers
]
\addplot [only marks,color=red, mark size=1pt,mark=*]
	table[x = T1, y = T2, col sep=space]{NumericalExperiment_simulated/data/real_grid/mc/scatter_4_2.dat};
\addplot [only marks,color=orange,mark size=1pt, mark=*]
	table[x = T1, y = T2, col sep=space]{NumericalExperiment_simulated/data/real_grid/mc/scatter_4_1.dat};
\addplot [only marks,color=blue, mark size=1pt,mark=*]
	table[x = T1, y = T2, col sep=space]{NumericalExperiment_simulated/data/real_grid/sc/scatter_4.dat};
\addplot [only marks,color=black,mark=+,mark size=2pt,mark options={line width=1pt}]
	coordinates{(0.1262,0.0174)(1.8308,0.6573)};
\draw  [black,thick,on layer = axis foreground]
	(axis cs: 0.1262-0.0075,0.0174-0.0046) rectangle (axis cs: 0.1262+0.0075,0.0174+0.0046);
\draw  [black,thick,on layer = axis foreground]
	(axis cs: 1.8308-0.0612,0.6573-0.2143) rectangle (axis cs: 1.8308+0.0612,0.6573+0.2143);
\end{loglogaxis}

\begin{loglogaxis}[
	width = 0.375\textwidth,
	height = 0.375\textwidth,
	title = solutions B and D,
	scale=\subScatterPlotScale,
	title style = {font = \scatterTitleFont,yshift = \scatterTitleYshift pt},
	xmax = \scatterToneMax,
	ymax = \scatterTtwoMax,
	xmin = \scatterToneMin,
	ymin = \scatterTtwoMin,
	major tick length = \scatterMajorTickLen,
	xtick={0.1,0.5,2},
	xticklabels={0.1,0.5,2},
	extra x ticks = {.1,.2,.3,.4,.5,.8,1.1,1.4,1.7,2},extra x tick labels={},
	ytick={.005, .05, .5,1},
	yticklabels={0.005, 0.05, 0.5},
	extra y ticks = {.005, .01,.02,.03,.04,.05, .1,.2,.3,.4, .5}, extra y tick labels={},
        xticklabel style = {font=\scatterTickFont,yshift=\scatterTickShift ex, rotate=\scatterXTickRotate},
        	yticklabel style = {font=\scatterTickFont,xshift=\scatterTickShift ex, rotate=\scatterYTickRotate},
	xlabel=$T_1$ (s),
%	ylabel=$T_2$ (s),
	yticklabel=\empty,
	label style={font=\scatterLabelFont},
	xlabel shift = \scatterXLabelShift pt,
	ylabel shift = \scatterYLabelShift pt,
	name=right,
	at=(middle.right of north east),
	anchor=left of north west,
	clip mode=individual,
	set layers
]
\addplot [only marks,color=orange,mark size=1pt, mark=*]
	table[x = T1, y = T2, col sep=space]{NumericalExperiment_simulated/data/real_grid/mc/scatter_8_1.dat};
\addplot [only marks,color=red, mark size=1pt,mark=*]
	table[x = T1, y = T2, col sep=space]{NumericalExperiment_simulated/data/real_grid/mc/scatter_8_2.dat};
\addplot [only marks,color=blue, mark size=1pt,mark=*]
	table[x = T1, y = T2, col sep=space]{NumericalExperiment_simulated/data/real_grid/sc/scatter_8.dat};
\addplot [only marks,color=black,mark=+,mark size=2pt,mark options={line width=1pt}]
	coordinates{(0.5075,0.1019)(1.8308,0.6573)};
\draw  [black,thick,on layer = axis foreground]
	(axis cs: 0.5075-0.0152,0.1019-0.0084) rectangle (axis cs: 0.5075+0.0152,0.1019+0.0084);
\draw  [black,thick,on layer = axis foreground]
	(axis cs: 1.8308-0.0612,0.6573-0.2143) rectangle (axis cs: 1.8308+0.0612,0.6573+0.2143);
\end{loglogaxis}
\end{tikzpicture}
\caption{The scatter plot shows the estimated relaxation times of the measured phantom at the example of voxels containing different combinations of compartments. For comparison, the estimates obtained from the single-compartment (SC) are shown, as well as the gold-standard (GS) values measured on the diagonal, where only a single compartment in present. The rectangles represent the standard deviation of the gold-standard measurement. }
\label{fig:sc_real_select}
\end{figure}

Further insights are provided by the scatter plots depicted in Figure~\ref{fig:sc_real_select}, which combines the estimated relaxation times of all voxels containing respective combination of doped water solutions. The multicompartment reconstruction correctly detects the presence of both tissues and the estimated relaxation times cluster around the gold-standard values. In contrast, the single-compartment algorithm results in a single cluster of apparent relaxation times at a position in parameter space that provides little information about the underlying tissue composition. 

\begin{table}
	\centering
	\begin{tabular}{ |c | c | c | c | c | }
		\hline  & A & B & C & D \\ \hline
		%$T_1 (s)$ & 0.126171848960461 & 0.507467423232209 & 0.764011098056793  & 1.830782684810722 \\ \hline
		%$T_2 (s)$ & $0.017437642716351$ &  $0.101936662952149$ & $0.132780857415372$  &0.657337069217406 \\ \hline
		$T_1 \text{ (s)}$ & $0.1262 \pm 0.0075$ & $0.5075 \pm 0.015$ & $0.7640 \pm 0.055$  & $1.831 \pm 0.061$ \\ \hline
		$T_2 \text{ (s)}$ & $0.0174 \pm 0.0046$ &  $0.1019 \pm 0.0084$ & $0.1328 \pm 0.018$  & $0.66 \pm 0.21$ \\ \hline
	\end{tabular} 
	\caption {The gold-standard (GS) values measured by single-compartment reconstruction on the diagonal grids of phantom described in Figure~\ref{fig:design_grid}, where only a single compartment in present. }
	\label{tab:gold_standard}
\end{table}

\begin{table}
\centering
\begin{tabular}{ |c | c | c | c | c | c | c |  }
       \hline  & \multicolumn{2}{c|}{A  and C} & \multicolumn{2}{c|}{A and D} & \multicolumn{2}{c|}{B and D} \\ \hline
    %$T_1 (s)$ & 0.1262 & 0.5075 & 0.7640  & 1.8308 \\ \hline
    %$T_2 (s)$ & $0.0174$ &  $0.1019$ & $0.1328$  &0.6573 \\ \hline
    & A & C &A & D & B & D \\ \hline
    $T_1 \text{ (s)}$ & $0.35 \pm 0.25$ & $0.58 \pm 0.26$ 
    & $0.24 \pm 0.26$ & $1.40 \pm 0.57$ 
    & $0.66 \pm 0.18$ & $1.49 \pm 0.36$\\ \hline
    $T_2 \text{ (s)}$ & $0.068 \pm 0.071$ & $0.20 \pm 0.20$  
    &  $0.063 \pm 0.24$ &  $1.20 \pm 0.58$ 
    & $0.080 \pm 0.11$  & $1.25 \pm 0.44$\\ \hline

    %$T_1 \text{ (s)}$ First Comp. & $0.3501 \pm 0.2535$ & $0.2361 \pm 0.2634$ & $0.6582 \pm 0.1843$ \\ \hline
    %$T_2 \text{ (s)}$ First Comp. & $0.0683 \pm 0.0710$ &  $0.0626 \pm 0.2406$ & $0.0803 \pm 0.1084$ \\ \hline
    %$T_1 \text{ (s)}$ Second Comp. & $0.5789 \pm 0.2627$ & $1.3981 \pm 0.5740$ & $1.4922 \pm 0.3556$   \\ \hline
    %$T_2 \text{ (s)}$ Second Comp. & $0.1998 \pm 0.2045$ &  $1.2015 \pm 0.5789$ & $1.2548 \pm 0.4389$  \\ \hline
  \end{tabular} 
  \caption {Mean values and standard deviations of the multicompartment estimates for the selected combinations appearing in Figure~\ref{fig:sc_real_select}.}
    \label{tab:value_real_select}
\end{table}

\section{Conclusions and Future Work}
In this work we propose a framework to perform quantitative estimation of parameter maps from magnetic resonance fingerprinting data, which accounts for the presence of more than one tissue compartment in each voxel. This multi-compartment model is fitted to the data by solving a sparse linear inverse problem using reweighted $\ell_1$-norm regularization. Numerical simulations show that an efficient interior-point method is able to tackle this sparse-recovery problem, yielding sparse solutions that correspond to the parameters of the tissues present in the voxel. In contrast, alternative approaches based on $\ell_2$-norm reweighting~\cite{McGivney2017} yield locally dense solutions, which are interpreted as a probability density of the tissue parameters.

The proposed method is validated through simulations, as well as with a controlled phantom imaging experiment on a clinical MR system. The results indicate that incorporating a sparse-estimation procedure based on reweighed $\ell_1$-norm regularization is a promising avenue towards achieving multicompartment parameter mapping from MRF data. However, the proposed method still requires a high SNR, which entails long scan times for in vivo applications. A promising research direction is to incorporate additional prior assumptions on the structure of the parameter maps in order to leverage the method in the realm of clinically-feasible scan times. Future work will further include the application of the proposed framework to the quantification of the myelin-water fraction in the human brain, which is an important biomarker for neurodegenerative disease.

\subsection*{Acknowledgements}
C.F. is supported by NSF award DMS-1616340. J.A., F.K., R.L. and M.C. are supported by NIH/NIBIB R21 EB020096, NIH/NIAMS R01 AR070297 and NIH P41 EB017183. S. L. is supported by a seed grant from the Moore-Sloan Data-Science Environment at NYU. S. T. is supported in part by the Office of Naval Research under Award N00014-17-1-2059.

\newpage
\begin{small}
\bibliographystyle{abbrv}
\bibliography{refs}
\end{small}

\appendix

\end{document}